# DISTRIBUTION AND KINEMATICS OF O VI IN THE GALACTIC HALO

B. D. Savage[1], K. R. Sembach[2], B. P. Wakker[1], P. Richter[1], M. Meade[1],

E. B. Jenkins[3], J. M. Shull[4], H. W. Moos,[5] and G. Sonneborn[6]

## ABSTRACT

Far-Ultraviolet Spectroscopic Explorer (FUSE) spectra of 100 extragalactic objects and two distant halo stars are analyzed to obtain measures of O VI $\lambda\lambda1031.93$, 1037.62 absorption along paths through the Milky Way thick disk/halo. Strong O VI absorption over the approximate velocity range from -100 to 100 km s$^{-1}$ reveals a widespread but highly irregular distribution of thick disk O VI, implying the existence of substantial amounts of hot gas with T ~3x10$^5$ K in the Milky Way halo. The integrated column density, log [N(O VI) cm$^{-2}$] , ranges from 13.85 to 14.78 with an average value of 14.38 and a standard deviation of 0.18. Large irregularities in the distribution of the absorbing gas are found to be similar over angular scales extending from less than one to 180 degrees implying a considerable amount of small and large scale structure in the absorbing gas. However, the inferred small scale structures must be quite common in order to provide a large sky covering factor of thick disk O VI along the paths to extragalactic objects. The overall distribution of Galactic O VI is not well described by a symmetrical plane-parallel layer of patchy O VI absorption. The simplest departure from such a model that provides a reasonable fit to the observations is a plane-parallel patchy absorbing layer with an average O VI mid-plane density of $n_o$(O VI) = 1.7x10$^{-8}$ cm$^{-3}$, a scale height of ~2.3 kpc, and a ~0.25 dex excess of O VI in the northern Galactic polar region. The distribution of O VI over the sky is poorly correlated with other tracers of gas in the halo, including low and intermediate velocity H I, Hα emission from the warm ionized gas at ~10$^4$ K, and hot X-ray emitting gas at ~10$^6$ K. The O VI has a velocity dispersion, b ≈ 30 to 99 km s$^{-1}$ with an average value of 60 km s$^{-1}$ and standard deviation of 15 km s$^{-1}$. Thermal broadening alone cannot explain the large observed profile widths, since gas at T ~ 3x10$^5$ K, the temperature at which O VI is expected to peak in abundance, has b(O VI ) = 17.7 km s$^{-1}$. The average O VI absorption velocities toward high latitude objects (|b| > 45°) range from -46 to 82 km s$^{-1}$, with a high-latitude sample average of 0 km s$^{-1}$ and a standard deviation of 21 km s$^{-1}$. O VI associated with the thick disk moves both toward and away from the plane with roughly equal frequency. High positive velocity O VI absorbing wings extending from ~100 to ~250 km s$^{-1}$ observed along 22 lines of sight may be tracing the flow of O VI into the halo. A combination of models involving the radiative cooling of hot fountain gas, the cooling of supernova bubbles in the halo, and the turbulent mixing of warm and hot halo gases is required to explain the presence of O VI and other highly ionized atoms found in the halo. The preferential venting of hot gas from local bubbles and superbubbles into the northern Galactic polar region may explain the enhancement of O VI in the North. If a fountain flow dominates, a mass flow rate of approximately 1.4 M$_\odot$yr$^{-1}$ of cooling hot gas to each side of the Galactic plane with an average density of 10$^{-3}$ cm$^{-3}$ is required to explain the average value of log [N(O VI)sin|b|] observed in the southern Galactic hemisphere. Such a flow rate is comparable to that estimated for the Galactic intermediate-velocity clouds.

*Subject headings:* Galaxy: halo - ISM: Abundances - ISM:clouds -Ultraviolet: ISM

[1] Department of Astronomy, University of Wisconsin, 475 N. Charter Street, Madison, WI 53706

[2] Space Telescope Science Institute, 3700 San Martin Drive, Baltimore, MD 21218

[3] Princeton University Observatory, Peyton Hall, Princeton, NJ 08544

[4] Center for Astrophysics and Space Astronomy, Department of Astrophysical and Planetary Sciences, University of Colorado, Boulder, CO 80309-0389

[5] Department of Physics and Astronomy, Johns Hopkins University, Baltimore, MD 21218

[6] Laboratory for Astronomy and Solar Physics, NASA Goddard Space Flight Center, Code 681, Greenbelt, MD 20771

# 1. INTRODUCTION

The lithium-like $(1s^2 2s)$ $^2S_{1/2}$ -> $(1s^2 2p)$ $^2P_{1/2,\ 3/2}$ electronic transitions of the highly ionized atoms of O VI, N V, and C IV provide fundamental information about gas in interstellar and intergalactic space with temperatures ranging from approximately $1 \times 10^5$ K to $5 \times 10^5$ K. Table 1 gives the basic spectroscopic information about these transitions including wavelengths, f-values, energy required for the creation and destruction of the ion, temperatures of peak abundance assuming collisional ionization equilibrium, and references to the literature describing absorption line studies of these ions for gas in the Galactic disk and halo. Gas with $1 \times 10^5$ K $< T < 5 \times 10^5$ K cools very rapidly and may exist in cooling bubbles, cooling flows, or in the transition temperature interface regions between warm $T \sim 10^4$ K and hot $T \sim 10^6$ K gas.

Among the three lithium-like high ion states accessible to ultraviolet (UV) observatories, O VI is especially important because of the large cosmic abundance of oxygen and the large energy (113.9 eV) required to convert O V into O VI. With such a high energy for production, it is unlikely that O VI will be produced by photoionization from extreme UV radiation from normal stars since most hot stars have strong $He^+$ absorption edges at 54.4 eV. In contrast, interstellar C IV has a production energy of 47.9 eV and can be found in photoionized gas with $T \sim 10^4$ K as well as in transition temperature gas with $T \sim 10^5$ K. Although N V has a production energy of 77.5 eV and mostly traces hot gas, its lower cosmic abundance often makes it difficult to detect. If subjected to hard radiation (extreme UV or soft X-rays), it is possible to find O VI in warm photoionized gas where the level of ionization depends on the ratio of the ionizing photon density to the gas density. Therefore, O VI can be produced by photoionization in very low density gas or very high radiation density environments.

Studies of interstellar O VI have been hampered by the difficulties involved in making observations in the far-UV wavelength region of its resonance doublet at 1031.93 and 1037.62 Å. Instruments operating efficiently at wavelengths shortward of ~1150 Å require reflecting optics with special coating (such as LiF or SiC) and windowless detectors. Although the Copernicus satellite successfully observed interstellar O VI toward many stars in the Galactic disk (Jenkins 1978a,b), the studies were limited to stars with visual magnitudes $m_V < 7$ and therefore did not yield useful information about the extension of O VI away from the Galactic plane. Therefore, the first observations of the extension of highly ionized interstellar gas into the Galactic halo were based on observations of N V, C IV, and Si IV obtained with the International Ultraviolet Explorer (IUE) satellite and with the Hubble Space Telescope (HST) (see references in Table 1).

Except for brief observing programs with the Hopkins Ultraviolet Telescope (HUT; Davidsen 1993) and the spectrographs in the Orbiting and Retrievable Far and Extreme Ultraviolet Spectrometers (ORFEUS; Hurwitz & Bowyer 1996; Hurwitz et al. 1998; Widmann et al. 1998; Sembach, Savage, & Hurwitz 1999), the study of O VI in the Milky Way halo has required the high throughput capabilities of the Far-Ultraviolet Spectroscopic Explorer (FUSE) satellite launched in 1999 (Moos et al. 2000; Sahnow et al. 2000). The first FUSE studies of O VI in the Galactic halo toward 11 AGNs revealed the widespread but irregular distribution of O VI with an extension away from the Galactic plane approximately described by an exponential layer with a scale height of 2.7 kpc (Savage et al. 2000). Observations with FUSE also recorded the presence of O VI absorption in Galactic high velocity clouds (Sembach et al. 2000) and Galactic O VI emission from four high latitude directions (Shelton et al. 2001; Shelton 2002; Dixon et al. 2001). FUSE has also been used to study the widespread presence of O VI in the Large Magellanic Cloud (LMC; Howk et al. 2002a) and in the Small Magellanic Cloud (SMC; Hoopes et al. 2002). These LMC and SMC observations contain information about the irregular distribution of Galactic O VI over small angular scales (Howk et al. 2002b).

The possible existence of hot gas in the Milky Way halo was first proposed by Spitzer (1956) and is considered in a number of more recent theoretical papers. Several recent review papers include Spitzer (1990), McKee (1993), and Savage (1995). The hot gas can be detected through UV high ionization line absorption, UV high ionization line emission, and X-ray line emission. The UV absorption and emission observations sample gas in the $1 \times 10^5$ K to $5 \times 10^5$ K temperature range, while the X-ray emission observations sample hotter gas with $T > 10^6$ K. High ionization state UV absorption measurements provide direct information on the column density of the absorbing ion integrated along the line of sight (i.e. $N(ion) = \int n(ion)\, dx$). In contrast, the intensity of high ionization line emission depends on the product of the ion and electron density times the line excitation rate integrated along the line of sight, but allowance must be made for the scattering of emitted resonance-line photons out of and into the line of sight. Interpretations of the observed X-ray emission, which also depends on the product of the ion and electron density, requires modeling the excitation processes, assumptions about collisional ionization equilibrium (or lack thereof), and the energy-dependent foreground attenuation of the emitted photons.

This paper is organized as follows: The observations and basic data reductions are discussed in §2. The interstellar measurements are presented in §3. The difficulty of separating low- and high-velocity O VI is discussed in §4. We associate the low-velocity O VI absorption from approximately -100 to +100 km s$^{-1}$ with O VI in the thick disk of the Milky Way. The column density distribution of the O VI thick disk/halo absorption is discussed in §5. The extension of O VI into the halo is discussed in §6. The kinematic character of the thick disk O VI is described in §7. Sight lines of particular interest are discussed in §8. The relationships among O VI and other Galactic ISM species are discussed in §9. The implications of these observations for the origin of highly ionized gas in the Milky Way are discussed in §10. A summary of the principal results appears in §11.

## 2. OBSERVATIONS AND REDUCTIONS

The FUSE O VI catalog paper (Wakker et al. 2002) contains the full details of the FUSE observations, the data handling, the basic ISM measurements, and plots of the O VI absorption line profiles for all of the objects in our study. Here we review the most essential aspects of the observations and reductions of the spectra for the 100 extragalactic objects and two stars listed in Table 2.

The objects are named according to the convention adopted by Wakker et al. (2002) and are ordered by increasing Galactic longitude. We list object name, Galactic longitude (l), Galactic latitude (b), redshift (z) or recession velocity (v in km s$^{-1}$), and other information relevant to the measurements. The new observations include the star vZ 1128 in the globular cluster M3 at a distance of 10 kpc and a distance z = 9.8 kpc away from the Galactic plane (Howk et al. in preparation) and PG 0832+675, a post-asymptotic giant branch star with V = 14.49 in the direction of HVC Complex A with an estimated distance of 8.1 kpc and z = 4.7 kpc (Ryans et al. 1997).

The distribution of the objects on the sky is illustrated in Figure 1 in views centered on the Galactic center (Fig. 1a) and on the Galactic anticenter (Fig. 1b). Most objects were observed as part of the FUSE science team guaranteed observing time for programs P101, P107, P108, and P207. To provide more complete sky coverage, we have supplemented the observations with FUSE guest investigator observations of additional extragalactic objects. These observations are publicly available through the FUSE data archive, and in all cases the guest investigator spectra were obtained for unrelated scientific investigations.

Full details of the FUSE instrument and its in-flight performance are found in Moos et al. (2000) and Sahnow et al. (2000). The spectral integrations were obtained in the time-tagged photon address mode with the objects centered in the large (30"x30") aperture of the LiF1 channel. Our analysis has been restricted to data from the two FUSE LiF channels since they provide most of the effective area of FUSE at the wavelengths of the O VI doublet. The data processing employed the standard CALFUSE (version 1.8.7) reduction software available at the Johns Hopkins University in November 2000.

The zero point in the wavelength scale in the region of the O VI $\lambda\lambda$1031.93, 1037.62 doublet was determined by registration of the atomic lines of Ar I $\lambda$1048.22 and Si II $\lambda$1020.70 to the H I 21 cm emission observed in or near the direction of each object. The final adopted velocity scale includes an additional positive correction of 10 km s$^{-1}$ to the O VI $\lambda$1037.62 line in order to bring it into agreement with the velocity observed for the O VI $\lambda$1031.93 line (see Wakker et al. 2002). The resulting velocity registration of the O VI data should be accurate to ~ 10 to 20 km s$^{-1}$ (2$\sigma$) depending on the signal-to-noise ratio (S/N) of the spectra. FUSE spectra are oversampled by the detectors with ~0.0065 Å pixel$^{-1}$ ( ~2.0 km s$^{-1}$ pixel$^{-1}$) before additional binning at wavelengths near the O VI absorption. This pixel size can be compared to the instrumental spectral line spread function, which is approximated by a Gaussian with FWHM ~ 20-25 km s$^{-1}$ at the O VI doublet for LiF1A spectra. Therefore, depending on the S/N of the observation, the final spectra were rebinned by factors of 5, 10, or 15 to binned pixel widths of ~10, 20, or 30 km s$^{-1}$.

For the brightest objects with excellent quality LiF1A observations, our analysis is based on data from that channel alone because of its superior resolution and S/N. For the lower quality data, the spectra from the two LiF channels were combined into a single spectrum. Combined spectra have a slightly lower resolution than spectra based only on LiF1 observations. In Table 2 a numerical data quality assessment is listed in column 5, where quality 4 implies excellent data with the observed S/N > 14 in the continuum adjacent to the O VI $\lambda$1031.93 line in a 10-pixel bin which is equivalent to one resolution element. Quality 3 implies good data with S/N = 9 to 14; quality 2 implies fair data with S/N = 5 to 9; and quality 1 implies poor data with S/N = 3 to 5.

Sample FUSE spectra from the LiF1A channel extending from 1020 to 1045 Å are shown for Mrk 421, Mrk 1095, and PKS 2005-489 in Figure 2. Galactic atomic absorption lines are identified and Galactic H$_2$ lines are indicated with the tick marks above the spectra. Terrestrial emission lines are also identified below each spectrum. The O VI $\lambda$1037.62 line is often confused by blending with C II* $\lambda$1037.02 and the H$_2$ (5-0) R(1) and P(1) lines at 1037.15 and 1038.16 Å. The O VI $\lambda$1031.93 line is usually relatively free of blending, since the H$_2$ (6-0) P(3) and R(4) lines are often relatively weak and are displaced in velocity by -214 and 125 km s$^{-1}$ from the rest of O VI velocity. H$_2$ is strong in the spectrum of Mrk 1095, moderate in strength toward PKS 2005-489, and absent in the spectrum of Mrk 421. The O VI absorption toward Mrk 1095 is very weak but is very strong toward PKS 2005-489.

# 3. ABSORPTION LINE MEASUREMENTS

Information about O VI in the Milky Way halo has been derived from the relatively blend-free O VI $\lambda$1031.93 line. For 20 lines of sight it is possible to obtain reliable information from the O VI $\lambda$1037.62 line over a restricted velocity range near the line center typically from LSR velocities of ~ -70 to 100 km s$^{-1}$. For the other lines of sight severe blending or low S/N limits the value of the O VI $\lambda$1037.62 observations.

Wakker et al. (2002) describe in detail the steps taken to remove the effects of blends with other absorbers and to measure the Galactic O VI lines. Below we summarize important aspects of the process. Usually, low-order Legendre polynomials (n < 3) were fitted to regions free of absorption within 1000 km s$^{-1}$ of the Galactic interstellar O VI lines. In 10 cases a higher order fit (n = 3-5) was needed. Absorption profiles for the O VI $\lambda$1031.93 line are shown versus LSR velocity in Figure 3 for a sample of 12 objects. The continuum placement shown is normally quite reliable since most of the objects are AGNs, which usually have well-defined power-law continua.

Along the sight lines where the H$_2$ (6-0) P(3) and R(4) lines at 1031.19 Å and 1032.35 Å blend with O VI, the expected strength of the contamination is based on an analysis of other H$_2$ J = 3 and 4 absorption lines in the spectrum (see Wakker et al. 2002). The estimated H$_2$ absorption line profiles have been plotted on the O VI absorption measurements shown in Figure 3. The H$_2$ absorption features near -214 and 125 km s$^{-1}$ are usually not a problem for measurements of O VI absorption at velocities more than ±20 km s⁻¹ away from these two absorption lines. When the H₂ lines blend with broad O VI absorption, the blending can usually be reliably corrected for.

With the exception of NGC 7469, the (6-0) R(0) line of HD at 1031.91 Å makes a negligible contribution to the O VI λ1031.93 absorption because the total H₂ column densities are small for these high-latitude lines of sight through the low-density halo. For NGC 7469 we applied a correction to the HD absorption in the core of the O VI λ1031.93 line by fitting the extra narrow absorption feature (see Wakker et al. 2002). The line of Cl I λ1031.51 with log (λf) = 2.19 (Morton 1991) is also weak and contributes negligible absorption because it requires a large column density of H₂ to be detectable.

In some cases, blending from intergalactic medium (IGM) Lyβ λ1025.72 absorption occurs in the spectral region of the Milky Way O VI λλ1031.93, 1037.62 doublet. The validity of the identification of a line as Lyβ can often be determined with reference to the weaker Lyγ λ972.54 line in the FUSE spectra or to the stronger Lyα λ1215.67 line if Hubble Space Telescope spectra exist. Only for PG 1048+342, PG 1216+69, and HE 1228+0131 were the Milky Way O VI λ1031.93 observations not usable because of blending with very strong IGM Lyβ absorption. Redshifted absorption lines from the neutral, low, and intermediate-ionization ISM in the background objects is also present near the Milky Way O VI λ1031.93 line in a number of cases. The offending lines include O I λ988.77, Si II λ1020.70, S III λ1012.50, C III λ977.02, Lyβ λ1025.72, Lyδ λ949.74, and Lyε λ937.80. For 5 sight lines one of these contaminates the Galactic O VI λ1031.93 absorption.

 Equivalent widths for the O VI λ1031.93 line are given by Wakker et al. (2002) for the absorption we associate with the thick disk/halo O VI of the Milky Way based on integrations over the LSR velocity range from $v_-$ to $v_+$ listed in columns 6 and 7 of Table 2. The rationale for the selected integration range, which is usually from approximately -90 to 90 km s⁻¹, is discussed in §4. The O VI λλ1031.93 equivalent widths range from 82±12±18 mÅ for Mrk 1095 to 429±12±59 mÅ for PKS 2005-489. The first set of errors for each object allows for statistical noise in the data and for continuum placement uncertainties. The second set of errors represents the systematic error due to fixed-pattern noise, the uncertainty associated with deciding how to separate in velocity O VI absorption in the disk and low halo from O VI absorption in high velocity gas, and for selected cases systematic errors from the deblending of H₂ absorption (see Wakker et al. 2002 for detailed discussions).

The O VI absorption lines are relatively broad and are generally resolved by the FUSE observations. For O VI at $3\times10^5$ K, the temperature at which O VI peaks in abundance under conditions of collisional ionization equilibrium, the thermal Doppler contribution to the O VI line width corresponds to b = 17.7 km s⁻¹ (FWHM = 29.4 km s⁻¹), where b = $(2kT/m_o)^{1/2}$ is the standard Doppler spread parameter. The expected thermal width of the O VI line is therefore comparable to the FUSE resolution of 20-25 km s⁻¹ at the O VI doublet. Since the O VI lines are usually resolved, reliable column densities can be obtained using the apparent optical depth method (Savage & Sembach 1991), where the apparent column density per unit velocity is given by

$$N_a(v) \text{ [ions cm}^{-2} \text{ (km s}^{-1})^{-1}] = (m_ec/\pi e^2) (f\lambda)^{-1} \tau_a(v) = 3.768\times10^{14} (f\lambda)^{-1} \tau_a(v). \qquad (1)$$

In Eq.(1), f is the oscillator strength of the line, λ is the wavelength of the line in Å, and $\tau_a(v)$ is the observed or apparent absorption optical depth defined by

$$\tau_a(v) = \ln [F_c(v) / F(v)], \qquad (2)$$

where F(v) and $F_c(v)$ are the observed and estimated continuum fluxes at velocity v. If an absorption line is fully resolved, then $N_a(v) = N(v)$, where N(v) is the actual column density per unit velocity. If the lines are only partially resolved, a comparison of $N_a(v)$ for the weak and strong line of a doublet allows an assessment of the possible presence of unresolved saturated absorption in the instrumentally blurred recording of the absorption. When the values of $N_a(v)$ for the two doublet absorption lines agree, the simulations by Savage & Sembach (1991) imply that unresolved saturated absorption is not likely affecting the derived column densities (also see Jenkins 1996).

In Figure 4 we compare the O VI doublet values of $N_a(v)$ for two extragalactic lines of sight for cases where the spectral line blending of the O VI $\lambda1037.62$ line is not severe. The comparison of the values of $N_a(v)$ for the weak and strong lines of the doublet is generally only possible over a limited velocity range near the line center of the O VI $\lambda1037.62$ absorption. To illustrate this blending problem, we display in Figure 4 for PG 2005-489 and Mrk 279 the absorption line profiles for the O VI $\lambda\lambda1031.93$, 1037.62 lines and the derived $N_a(v)$ profiles, with the strong line profile shown as the heavy line and the weak line profile shown as the thin line. The extra features on the $N_a(v)$ profile for the weak line are from the contaminating $H_2$ and C II* lines identified on the panels displaying the O VI $\lambda1037.62$ absorption profiles. The good agreement between the $N_a(v)$ curves for the weak and strong lines over the velocity range from -50 to 100 km s$^{-1}$ for PG 2005-489 and Mrk 279 implies that there is no unresolved saturated absorption in the observed profiles for these objects and that the $N_a(v)$ profiles for the relatively uncontaminated O VI $\lambda1031.93$ line should give a reliable measure of the true column density per unit velocity blurred to the FUSE resolution.

Larger values of $N_a(v)$ for the weak O VI line compared to the strong line (deviating by > $1\sigma$) are measured for 5 of 20 O VI survey objects that have been observed with high enough S/N to make the comparison meaningful (see Wakker et al. 2002). Footnote b to Table 2 provides more information about the results for the five objects where the measurements imply some saturation. In that footnote we list the object name, the ratio of the weak line and strong line column densities integrated over the uncontaminated portions of the line profiles, $N_W/N_S$, and the error. In addition we list the logarithmic saturation correction, $\Delta\log N(corr)$, to the measured column density for the strong line using the saturation correction method recommended by Savage & Sembach (1991; see their Table 4). An improved estimate for the column density for these six objects is therefore $\log N = \log N_S + \Delta\log N(corr)$, where $\log N_S$ is the integrated logarithmic column density for the strong line from column 13 of Table 2. Since Table 2 only contains the strong line results, the correction we have determined from Savage & Sembach (1991) has been adjusted so that it can be applied to the strong rather than the weak absorption line. The saturation correction is large for only one object (0.57 dex for Mrk 421). For the other four objects modest saturation corrections ranging from 0.11 to 0.24 dex are implied.

Our ability to reliably study saturation effects in the O VI $\lambda1031.93$ line through a comparison with the O VI $\lambda1037.62$ line is limited to only 20 objects. We have therefore chosen to base our scientific analysis of these observations on the O VI strong line values of the apparent column density listed in Table 2. Based on finding evidence for saturation in 5 of 20 cases where saturation information is available, we estimate that unresolved saturation may be affecting about 25% of the values of $\log N$ listed in Table 2. However, large (~0.5 dex) saturation corrections are probably limited to ~5% of the lines of sight.

In Table 2 columns 6-16 we list the basic interstellar absorption line measurements for each object. The quantities listed include $v_-$, $v_+$, $\overline{v}_{exp}$, $\overline{v}_{obs}$, $\sigma_v$, b, $\sigma_b$, $\log N(O\ VI)$, $\sigma_{sc}$, $\sigma_{sys}$, and

$\log [N(O\ VI)\sin|b|]$. Here $\log N(O\ VI) = \log [\int N_a(v)dv]$, where we set limits to the column density integration for gas associated with the thick O VI disk of the Milky Way from $v_-$ and $v_+$ (see § 4), and the integration is over $N_a(v)$ for the strong line of the doublet at 1031.93 Å. The values of $v_-$ and $v_+$ with attached colons identify those cases where blending with high velocity gas absorption causes the velocity extent of the lower velocity absorption to be uncertain (see Wakker et al. 2002).

To study the effects of Galactic rotation on the absorption line profiles, we have calculated the quantity, $\overline{v}_{exp}$, which is the column density weighted average LSR velocity expected for O VI absorption toward each object assuming that the O VI arises in a co-rotating plane-parallel exponential gas layer with a scale height of h = 2.5 kpc and a velocity dispersion of b = 60 km s$^{-1}$. The co-rotation assumption uses the Galactic rotation curve of Clemens (1985) and requires gas at large distances from the Galactic plane to have the same rotational speed as gas in the underlying disk. Our estimate of $\overline{v}_{exp}$ ignores the effects of gas flowing onto or away from the Galactic plane.

The O VI scale height is quite uncertain because of the large irregularity in the distribution of Galactic O VI and the asymmetry between the Galactic north and south polar regions (see §6). We take h = 2.5 kpc as a representative value. When computing the expected average velocity, we assumed the gas layer ends at z = 10 kpc.

Table 2 also contains the average LSR velocities and velocity dispersions for the thick disk absorption. These were computed from the expressions

$$\bar{v}_{obs} = \int v \, N_a(v) dv \, / \int N_a(v) dv \qquad (3)$$

$$b = \sqrt{2} \, [ \int (v - \bar{v}_{obs})^2 \, N_a(v) dv \, / \int N_a(v) dv \, ]^{1/2}. \qquad (4)$$

The velocity dispersion, b, is similar to the Doppler spread parameter commonly used in the Voigt profile fitting of interstellar absorption lines. The calculation of b does not allow for instrumental blurring or unresolved saturated structure within the lines. Both effects are relatively small for the broad O VI lines considered here. This velocity dispersion measures the combined effects of thermal Doppler line broadening, the multiple component nature of the absorption, interstellar turbulence, Galactic flows, and Galactic rotation.

The various errors listed ($\sigma_v$ and $\sigma_b$) are the errors in $\bar{v}_{obs}$ and b. The error in $\bar{v}_{obs}$ is based on a quadrature addition of the statistical error and the 10 km s$^{-1}$ velocity alignment error. The error for b is the statistical error. Except for cases of overlapping low and high velocity absorption, the systematic error in b should be small, since the value is measured relative to the value of $\bar{v}_{obs}$. For the overlapping cases we have attached a colon to the listed value of b to indicate additional uncertainty from the blending. Statistical (photon counting) errors combined with continuum placement errors, $\pm\sigma_{sc}$, and systematic errors, $\pm\sigma_{sys}$, are listed for log N(O VI). The systematic error includes our estimate for a combination of fixed-pattern noise errors, velocity cutoff errors, and uncertainties associated with deblending H$_2$ in several cases. The derivation of these errors is discussed in Wakker et al. (2002).

At the end of Table 2, we include for comparison the average values of log N(O VI) and standard deviations for the Milky Way halo gas absorption along lines of sight to the Large and Small Magellanic Clouds (LMC and SMC) from Howk et al (2002b). The relatively large standard deviations of 0.13 and 0.17 dex for Milky Way halo gas absorption toward stars in the LMC and SMC reveals that the irregularity in the distribution of O VI extends to small angular scales (see Howk et al. 2000b and §5).

## 4. SEPARATING THE THICK O VI DISK ABSORPTION FROM HIGH VELOCITY CLOUD ABSORPTION

The lines of sight to the objects observed in this program pass through Milky Way disk gas, thick disk/halo gas, intermediate-velocity clouds (IVCs), high-velocity clouds (HVCs), and may even sample intergalactic O VI in the Local Group of galaxies. At high Galactic latitude the dividing lines in velocity between high, intermediate, and low velocity are generally at |$v_{LSR}$| = 90 and 30 km s$^{-1}$, although the effects of Galactic rotation must also be considered. The O VI absorption and $N_a(v)$ profiles shown in Figures 3 and 4, and in Wakker et al. (2002) trace a complex set of processes and phenomena involving hot gas in the Milky Way disk, thick disk/halo, and beyond. Clearly separating the absorption produced by the various phenomena based on the kinematical imprints is a major challenge. We refer to O VI in the Milky Way disk-halo interface extending several kpc away from the Galactic plane as the "thick disk O VI."

The O VI profiles exhibit a diversity of strengths and kinematical behavior. Disk, and thick disk O VI is clearly detected within the general velocity range from approximately -90 km s$^{-1}$ to 90 km s$^{-1}$ toward 91 of the 102 objects listed in Table 2. In addition, the absorption profiles reveal O VI at high velocities with |$v_{LSR}$| ranging from ~100 to 400 km s$^{-1}$ in ~60 % of the lines of sight. The O VI absorption at high-velocity is considered by Sembach et al. (2002). We consider aspects of the O VI high velocity sky in this paper because the absorption it produces sometimes blends with O VI absorption occurring in the thick disk region.

Whenever a known H I HVC is present along the line of sight to an object, we usually see associated high-velocity O VI absorption that spans the approximate velocity range of the H I HVC. The best examples include the following: (1) The clear detection of strong O VI absorption associated with Complex C in the directions to Mrk 279, Mrk 290, Mrk 501, Mrk 506, Mrk 817, Mrk 876, PG 1259+593, PG 1351+640, and PG 1626+554. (2) The detection of strong O VI absorption associated with the Magellanic Stream in the direction of Fairall 9, NGC 7714, NGC 7469, and PG 2349-15. (3) The detection of O VI in the warped outer regions of the Milky Way toward H 1821+643 and 3C 382.0. (4) The presence of O VI absorption at velocities that suggests an association with known H I HVCs, but in directions situated several degrees beyond the traditional H I boundaries of the HVCs as seen in 21 cm emission. These include NGC 1705, PKS 0558-304, Mrk 509, ESO 265-G23, PG 0052+251, and Mrk 357.

The high-velocity O VI also reveals other phenomena. Toward a number of objects, a very broad positive velocity wing of O VI absorption is sometimes seen extending from v ~ 100 to 200 km s$^{-1}$ or even to 300 km s$^{-1}$. There are also O VI HVCs isolated in velocity that do not have an associated H I HVC.

In order to study the thick disk O VI, it is necessary to separate the absorption it produces from the absorption arising from other phenomena. In the case of Complex C and the Magellanic Stream, the origin of the high-velocity O VI absorption is clearly due to processes unrelated to those associated with OVI in the thick disk. Complex C traces the infall of low-metallicity gas falling onto the Milky Way (Wakker et al.1999; Gibson et al. 2001; Richter et al. 2001a), while the Magellanic Stream likely traces tidal matter pulled from the SMC (Gardiner & Noguchi 1996; Lu et al. 1998; Gibson et al. 2000). However, in other cases the high-velocity O VI absorption may be related to the processes that cause matter to move from the Galactic disk into the halo. For example, the high positive velocity wings of O VI absorption could represent the detection of outflowing hot gas participating in a Galactic fountain flow. The O VI absorption associated with gas in the outer warped region of the Milky Way probably has an origin more closely related to that for thick disk O VI than for other HVC phenomena. It therefore appears reasonable to consider the absorption associated with the warp to be part of the thick disk O VI absorption.

The velocity ranges we have adopted for reporting column densities of the O VI absorption at low, intermediate, and high velocities are illustrated in Figure 5. The velocity ranges plotted for the objects in the figure are ordered by Galactic longitude. They are also separated into four different latitude intervals: b < -50°, -50° < b < 0°, 0° < b < 50°, and b > 50°. The thickness of the line is proportional to the column density measured over the velocity extent of the line.

For ~75% of the sight lines, we find that either the high-velocity absorption is isolated from the low-velocity absorption or there is no high velocity O VI down to an equivalent width limit of 65 mÅ. The thick disk O VI absorption for these cases is found to be mostly limited to the velocity range from -120 to 120 km s$^{-1}$, although it might extend to ±135 km s$^{-1}$ in some cases, or be limited to ±70 km s$^{-1}$ in others. Toward HVC Complex C, the O VI absorption always extends to more negative velocities, and the separation of absorption due to the thick disk and Complex C is usually not clear. In these cases we found it necessary to split the HVC and thick disk absorption at some reasonable intermediate velocity based on the shape of the O VI profile in combination with the 21 cm data. The systematic logarithmic column density error associated with changing the integration range of the thick disk absorption by ±15 km s$^{-1}$ is included in the value of $\sigma_{sys}$ listed in Column 15 of Table 2.

The IVCs have gas-phase abundances similar to the abundances of gas found in the disk (Richter et al. 2001a, 2001b) and are known to be situated in the low halo at distances away from the Galactic plane ranging from 0.5 to 2 kpc (Wakker 2001). The IVC phenomenon is probably connected to the processes that create the thick gaseous disk. We have therefore included the velocity range of the IVCs in our integration range for the O VI we associate with the thick disk. In columns 6 and 7 of Table 2, we list the velocity integration range we have adopted for measuring equivalent widths and log N(O VI) associated with O VI in the thick disk of the Milky Way. Additional comments about separating thick disk and HVC absorption are found in Wakker et al. (2002) and Sembach et al. (2002).

## 5. COLUMN DENSITY DISTRIBUTION FOR THE THICK DISK ABSORPTION

Relatively strong O VI absorption associated with the thick disk of Galactic O VI is observed toward 91 of the 102 objects in our program. Various O VI column density averages are given in Table 3. The values of log N(O VI) range from 13.85 (for NGC 7714) to 14.78 (for PKS 2005-489). The $3\sigma$ limits on log N(O VI) for the 11 non-detections range from 13.75 to 14.23. The non-detections are from the lower quality observations (quality 1 and 2), and the $3\sigma$ column density limits are generally larger than the smallest column densities recorded by the higher quality observations (quality 3 and 4). The equivalent column density perpendicular to the Galactic plane, log [N(O VI)sin|b|], ranges from 13.45 (for Mrk 1095) to 14.68 (for 3C 273). Figures 6a and 6b show histograms of the distribution of log N(O VI) and log [N(O VI)sin|b|] for the 91 objects. The median, average, and standard deviation of log N(O VI) are 14.38, 14.36, and 0.18 respectively, while for log[N(O VI)sin|b|] the values are 14.23, 14.21, and 0.23, respectively. These numbers were calculated by omitting the column-density limits for the 11 non-detections. Including the non-detections would lower the averages slightly. However, that effect would be compensated by an increase in the averages if we could include an allowance for saturation in the O VI $\lambda$1031.93 profiles that may be affecting ~25% of the lines of sight (see §3).

The distribution of values of log N(O VI) on the sky is shown in Aitoff projections (Figures 7a and 7b) and in polar projections (Figures 8a and 8b). The values of the O VI column densities are coded as colored circles $12^\circ$ in radius surrounding the small data points representing the direction to each survey object. In directions where objects are closer together than $12^\circ$, the color-coding extends halfway to the adjacent object. Upper limits are denoted with the small, $2.5^\circ$ radius, colored circles. The choice of $12^\circ$ radius circles provides the visual impression that the covering factor for O VI is large, which is consistent with the high detection frequency of thick disk O VI toward the 102 survey objects. Figures 9a and 9b contain Aitoff projections of the O VI column densities coded as circles with sizes proportional to log N(O VI). Column density limits are displayed as downward pointing triangles with sizes proportional to the $3\sigma$ limit. In Figure 9a the O VI measurements are plotted along with the contours of the non-thermal Galactic Radio Loops I-IV as defined by Berkhuijsen et al. (1971). In Figure 9b the O VI measurements are superposed on a grey scale representation of the 0.25 keV X-ray background as recorded by ROSAT (Snowden et al. 1997). The relationships between O VI, the soft X-ray background, and the Radio Loops are discussed in §9.3 and 9.4.

The distribution of O VI on the sky is quite irregular. The directions in the sky with low values of log N(O VI) include the region in the direction of $l \sim 64^\circ$ to $172^\circ$ and $b \sim -41^\circ$ to $-59^\circ$ sampled by Mrk 926, PG 2349-014, NGC 7714, NGC 7469, Mrk 1502, Mrk 335, NGC 1068, NGC 588, NGC 595, and UGC 12163 with log N(O VI) <14.23, <13.97, 13.85, 13.96, 14.00, 14.07, 13.87, <14.07, <13.75, and 14.00 respectively, and the region around $l \sim 201^\circ$ to $231^\circ$ and $b \sim -21^\circ$ to -$37^\circ$ sampled by Mrk 1095, PKS 0405-012, HE 0450-2958, and SBS 0335-052 with log N(O VI) = 13.89, 14.10, < 14.00, and <13.99, respectively.

Isolated directions to individual objects with low values or limits for log N(O VI) adjacent to objects with relatively large O VI column densities include HE 1115-1753 ($l = 273.65^\circ$, b = $39.64^\circ$) with log N(O VI) < 13.92 which is $6.6^\circ$ from IRAS F11431-1810 with log N(O VI) = 14.56 and PG 0052+251 (l = $123.91^\circ$, b = $-37.44^\circ$) with log N(O VI) <13.97, which is $6.8^\circ$ from Mrk 357 with log N(O VI) = 14.40.

Directions with the highest values of log N(O VI), ranging from 14.60 to 14.78, include PKS 2005-489, Mrk 509, and Ton 1924-416 near the Galactic center to the south, Ton S210 near the South Galactic Pole, 3C 273 inside the edge of Galactic Radio Loop IV, and HS 1102+3441 at high latitude toward the Galactic anti-center.

The large values of log N(O VI) toward PKS 2005-489, Tol 924-416, and Mrk 509 suggest an excess of O VI absorption in directions extending toward the Galactic center at negative latitudes, where we have suitable number of background objects. The excess in directions toward the southern Galactic center is apparent in Figures 7a and 9a and also includes the direction to ESO 141-G55. The average value of log N(O VI) is 14.63 for these four southern Galactic center objects with l between 330° and 36° and b between -24° and -33°. This can be compared to an average of 14.23 for the six other objects in the south with b between -21° and -35° and l ranging from 63° to 258°. The possible local and distant origins of this excess are discussed in §8.7.

Unfortunately, we can not search for a similar excess in the amount of O VI for directions over the northern Galactic center for latitudes in the range from b = 20° to 40°. For the northern Galactic hemisphere there are no FUSE observations of extragalactic objects in the region of generally high interstellar dust extinction associated with the Sco-Oph region in directions extending from l = 310° to 40° and b = 0° to 40°.

The irregular distribution of O VI in the Galaxy is clearly revealed in Figure 10, where values of log N(O VI) and log [N(O VI)sin|b|] are shown plotted against Galactic longitude and latitude. Figure 10 also shows that there are substantial differences in the values of log N(O VI) and log [N(O VI)sin|b|] in the northern Galactic hemisphere compared to the southern Galactic hemisphere. Table 3 contains averages over various ranges of Galactic latitude for the measured O VI column densities.

At low latitudes (15° < |b| <45°) the average values of log [N(O VI)sin|b|] are similar in the north and south.
North (15° < b < 45°): <log [N(O VI)sin|b|]> = 14.10±0.17(stdev) averaged over 22 objects.
South (-45° < b | <-15°): <log [N(O VI)sin|b|]> = 14.06±0.26 (stdev) averaged over 16 objects.
However, these values are affected by incomplete sampling of the sky in directions with high interstellar obscuration from dust, particularly in the north in the region 15° < b < 45° and 330° < l < 30°.

For |b| > 45° the observations should not be seriously impacted by incomplete sky coverage bias introduced by local obscuration effects. It is therefore interesting to directly compare the average values of log [N(O VI)sin|b|] over the north and south Galactic polar regions:
North (45° < b < 90°) : <log [N(O VI)sin|b|]> = 14.35 ± 0.15 (stdev) averaged over 40 objects.
South (-90° < b < -45°) : <log [N(O VI)sin|b|]> = 14.12 ± 0.25 (stdev) averaged over 13 objects.

There is a large excess of 0.20 to 0.30 dex in the average value of log [N(O VI)sin|b|] for the north Galactic Polar region with (45° < b < 90°) compared to lower latitude directions in the north and to the entire southern Galactic sky. While incomplete sky coverage at the lower latitudes in the north is probably affecting the derived value for this excess, the difference between the averages for the north and the south are quite striking. The excess in log [N(O VI)sin|b|] in the north is mostly responsible for the difference seen in Figure 6b between the histogram for the entire sky (thin line) and for |b| > 45° (thick line).

The observations displayed in Figures 7, 8, 9, and 10 reveal considerable variation in the value of log N(O VI) from one sight line to the next. At angular separations from 0.5° to 5° toward stars observed in the LMC (Howk et al. 2002a) and from 0.05° to 3.3° for stars observed in the SMC (Hoopes et al. 2002), Howk et al. (2002b) have found star to star differences in log N(O VI) for gas in the Milky Way halo with average differences of 0.21 and 0.20 dex, respectively. The standard deviations in the values of logN(O VI) for Milky Way halo gas toward the LMC and SMC of 0.13 and 0.17 dex, respectively, are comparable to the standard deviation of 0.18 found for the FUSE full sample of 91 objects listed in Table 3. The standard deviation of logN(O VI) in the LMC sample increases to 0.18 dex if the velocity range of the integration is restricted to –50 to 50 km s$^{-1}$ (see footnote c to Table 2).

The variation we see in Figures 7-9 reveals that the levels of small angular scale variation in logN(O VI) observed toward the LMC and SMC also extend to larger angular scales. In the top panel of Figure 11 we show the relative variation in N(O VI), $\Delta$ N(O VI)$_{ij}$ / N(O VI)$_{ij}$ , for object pairs plotted against the angular separation of the pair, $\Delta\theta_{ij}$ , in degrees. Here
$\Delta$N(O VI)$_{ij}$ = N(O VI)$_i$ – N(O VI)$_j$ and N(O VI)$_{ij}$ = [N(O VI)$_i$ +N(O VI)$_j$]/2. Only pairs with errors in $\Delta$ N(O VI)$_{ij}$ / N(O VI)$_{ij}$ of $\leq$ 0.25 are included in this analysis. The dashed lines in the middle panel show <$\Delta$ N(O VI)$_{ij}$ / N(O VI)$_{ij}$> and the dispersion, $\sigma$[$\Delta$ N(O VI)$_{ij}$ / N(O VI)$_{ij}$], versus <$\Delta\theta_{ij}$>, with the average taken over a 15° range in of $\Delta\theta_{ij}$ but displayed every 5°. The solid lines in the middle panel show the 90, 75, 50, 25, and 10% levels in the distribution of $\Delta N(O\ VI)_{ij}$ / $N(O\ VI)_{ij}$ over $15^o$ versus $<\Delta\theta_{ij}>$, with the 50% (median) level plotted as the heavy solid line. The average of $\Delta N(O\ VI)_{ij}$ / $N(O\ VI)_{ij}$ hardly changes as a function of $<\Delta\theta_{ij}>$. It ranges from an average of 0.35 at small $<\Delta\theta_{ij}>$ to a maximum of 0.55 in the range of $<\Delta\theta_{ij}>$ from 80 to $120^o$ and declines to 0.4 at $<\Delta\theta_{ij}> \sim 170^o$. This behavior places constraints on the spatial properties of the O VI absorbing structures in the Galaxy (see §10).

The lower panel of Figure 11 shows the relative variation in the H I column density $<\Delta N(H\ I)_{ij}$ / $N(H\ I)_{ij}>$ as a function of $<\Delta\theta_{ij}>$ calculated in exactly the same way as for O VI. The H I column densities are from the sources discussed by Wakker et al. (2002) and cover the same velocity range as for O VI except for a few cases where such an integration would break an H I IVC component into two. The H I observations refer to the column density averaged over radio antenna beams having an extent of typically 20'. The behavior of the variations found for H I as a function of $<\Delta\theta_{ij}>$ is roughly similar to that found for O VI except the peak amplitude of the variation in H I for $<\Delta\theta_{ij}>$ between 60 and $130^o$ is somewhat larger (~0.80 versus ~0.55 for O VI).

## 6. THE EXTENSION OF OVI AWAY FROM THE GALACTIC PLANE

The large O VI column densities measured along extragalactic sight lines imply a substantial amount of O VI is situated in the Galactic thick disk at large distances away from the plane of the Galaxy. In the Galactic disk, the average density of O VI determined from Copernicus satellite data for stars with d < 2 kpc is found to be $2x10^{-8}$ cm$^{-3}$ (Jenkins 1978b; Savage et al. 2000). New FUSE observations of more distant stars imply a somewhat smaller average of $1.7x10^{-8}$ cm$^{-3}$ (Jenkins et al. 2001). Assuming a plane-parallel stratification for the patchy distribution of O VI absorbing structures in an exponential atmosphere with a scale height, h, and an average O VI mid-plane density of $<n_o(O\ VI)> = 2x10^{-8}$ cm$^{-3}$, Savage et al. (2000) estimated an O VI scale height of 2.7 kpc based on the median value of log $[N(O\ VI)sin|b|] = 14.21$ toward the 11 objects in the first FUSE extragalactic sample. The scale height follows from $N_\perp(O\ VI) = N(O\ VI)sin|b| = n_o(O\ VI)$ h and allowed for the irregular distribution of the absorbing structures through the inclusion of a 0.21 dex patchiness parameter (see Savage et al. 1990 and 2000). This first estimate of the Galactic O VI scale height from FUSE observations is much larger than the first estimate from the Copernicus survey data of h = 0.3 (+0.2, -0.1) kpc (Jenkins 1978b) and smaller than the value h = 5.50 (+2.37, -2.09) kpc estimated by Widmann et al. (1998) based on ORFEUS high resolution spectrometer data.

With the large increase in the size of the sample of objects along extragalactic lines of sight we re-evaluate the nature of the extension of O VI away from the Galactic plane. Figure 12 shows log $[N(O\ VI)sin|b|]$ versus log $|z(kpc)|$ for various data samples including: (1) The Copernicus sample of stars in the Galactic disk from Jenkins (1978a), adopting the distances listed in his Table 2 (open squares); (2) The Milky Way absorption toward stars in the LMC and SMC (upward pointing open triangles) from Howk et al. (2002b) and Hoopes et al. (2002); and (3) The observations from Table 2 (plotted as open circles with the 100 extragalactic objects plotted on the right hand side of the figure in the region beyond the break in the log $|z(kpc)|$ axis and the two stars plotted near log $|z| \sim 1.0$). The extragalactic data from Table 2 are organized into two groups for the south and north Galactic hemispheres, with the ordering in latitude going from $-90^o$ on the left to $90^o$ on the right. Upper limits shown in Figure 12 are plotted as downward pointing triangles. There is a considerable spread in the values of log $[N(O\ VI)]sin|b|]$ at all values of $|z|$. The data points for the LMC and SMC directions at $|z| = 50$ and 70 kpc, respectively, refer to the spread in values of log$[N(O\ VI)sin|b|]$ extending over the relatively small angular extents of these two galaxies. In contrast the spread for the Copernicus observations and the extragalactic observations illustrate the variation in the values of log $[N(O\ VI)sin|b|]$ extending over most of the sky.

The three solid lines in Figure 12 show the expected behavior of the log $[N(O\ VI)sin|b|]$ versus log $|z(kpc)|$ distribution for a smoothly distributed, exponentially stratified plane-parallel Galactic atmosphere with an O VI mid-plane density $n_o(O\ VI) = 1.7x10^{-8}$ cm$^{-3}$ and exponential scale heights of 1.0, 2.5, and 10 kpc. The use of a mid-plane density of $1.7 \times 10^{-8}$ cm$^{-3}$ appears well justified since the value represents an average extending over ~220 stars in the combined Copernicus and FUSE disk star sample (Jenkins et al. 2001). The large irregularity in the distribution of O VI and the enhancement in the amount of O VI over the northern Galactic hemisphere introduces a complication when trying to estimate a Galactic scale height for O VI. The data points in Figure 12 are consistent with the O VI scale height lying in the range between 1.0 and 10 kpc.

One can imagine a wide range of possible models to fit the observed values of log [N(O VI)sin|b|] that are more complicated than a simple symmetrical plane-parallel exponential model that only allows for the patchy distribution of the absorbing structures. The simplest model that improves the fit is a superposition of a plane-parallel patchy absorbing layer with an exponential scale height of ~2.3 kpc and a ~0.25 dex excess of O VI at the higher northern Galactic latitudes. The cross-hatched region in Figure 10e shows the distribution of data points expected for a simple plane-parallel distribution of O VI absorbing clouds that fits the observations for the southern Galactic hemisphere where <log [N(O VI)sin|b|]> = 14.09±0.25. The model has a scale height of 2.3 kpc and a midplane density of $1.7 \times 10^{-8}$ cm$^{-3}$. The patchiness of 0.25 dex in the model is indicated as the crosshatched region in Figure 10e. While this simple model roughly describes the observations for the southern Galactic hemisphere and for b < 45° in the North, it underestimates the observed values of log N(O VI) from b ~ 45° to 90°. Averages of log [N(O VI)sin|b|] over various ranges in b are listed in Table 3. The ~0.25 dex excess in the northern Galactic hemisphere for b > 45° could be the result of the imprint of local or distant Galactic structures on the distribution of O VI in the sky as viewed from our position in the Galaxy.

Some local Galactic phenomena that may affect our view of the distribution of O VI in the sky include: (1) The extension of the Local Bubble into the halo; (2) Nearby superbubbles such as Galactic Radio Loops I, IV, and the North Polar Spur; (3) An enhancement of the amount of O VI in directions of the sky with substantial amounts of intermediate velocity gas; (4) Galactic fountain activity over nearby spiral arms such as the Perseus arm, the local spiral arm, and the Sagittarius Arm.

Distant phenomena that might affect our view of O VI in the sky include: (1) A possible enhancement in the amount of O VI associated with a very high z-extension of hot gas from the central regions of the Galaxy; (2) An interaction between thick disk gas and the gas associated with infalling high velocity clouds such as Complex C, which covers a substantial portion of the northern Galactic polar region; (3) The warp of the outer Galaxy to positive z in the general direction 60° < l < 150° and to negative z in the general direction of 240° < l < 300°; (4) The motion of the Milky Way through Local Group gas.

Without knowing more about the true z-distance of the O VI gas producing the excess in the northern Galactic hemisphere, it is difficult to establish which of the various possible causes actually dominates. However, a reasonable possibility appears to be the preferential venting of hot gas into the northern Galactic hemisphere from the Local Bubble and nearby superbubbles. This interpretation receives some support from the Na I observations of Sfeir et al. (1999), which were designed to map the structure of the local hot bubble. The Na I measurements imply that the local bubble opens in a chimney-like structure to high latitude in directions roughly defined by enhanced intensity in the 0.25 keV X-ray background.

The rather complicated view of the distribution of O VI in the thick disk of the Milky Way causes difficulties when considering the possible existence of a highly extended (tens of kpc) halo of hot gas containing O VI surrounding the Milky Way. The best information we have regarding the existence of O VI very far from the disk of the Milky Way is for those directions where the effects of Galactic rotation separate nearby and distant absorption. In the direction to H 1821+643 with l = 94.00° and b = 27.42°, the high negative velocity extent of the O VI profiles implies the detection of O VI in the outer warped regions of the Milky Way at a Galactocentric distance of ~24 kpc and a |z| distance of ~ 7 kpc (see §8.5).

## 7. KINEMATICS OF THE THICK DISK O VI

The kinematic properties of the O VI absorption provide important information about the various physical processes controlling the distribution of gas at T ~ $3 \times 10^5$ K in the Galactic halo. However, clearly isolating the different origins of the observed kinematics is difficult. The two simple measures of the kinematic properties of the O VI absorption we have chosen to study are the average LSR velocity of the absorption, $\bar{v}_{obs}$ (see Eqn. 3) and the velocity dispersion of the absorption as measured by b (see Eqn. 4). The errors associated with these quantities are discussed in §3 and by Wakker et al. (2002). When there is substantial blending bewteen thick disk absorption and high velocity absorption, the systematic errors are large.

### 7.1. Average O VI Velocity and Velocity Width

The distributions of the values of $\bar{v}_{obs}$ and b are shown in Figures 6c and 6d. In these figures the upper (light line) histogram is for the complete sample, while the lower (heavy line) histogram is for the directions to objects at high latitude (|b| > 45°). The values for $\bar{v}_{obs}$ and b averaged over various samples of the 91 sightlines for which measures of $\bar{v}_{obs}$ and b have been obtained are listed in Table 4. For the 91 objects we find: $\bar{v}_{obs}$(min) = -46 km s$^{-1}$, $\bar{v}_{obs}$(max) = 82 km s$^{-1}$, $\bar{v}_{obs}$(median) = 3 km s$^{-1}$, $\bar{v}_{obs}$(average) = 2 km s$^{-1}$, and a standard deviation of 21 km s$^{-1}$. The medians, averages, and standard deviations do not change much for the various sub-samples of the 91 objects including one at high latitude (|b| > 45°), low latitude (|b| < 45°), high latitude in the north (b > 45°), and high latitude in the south (b < -45°).

Turning to the O VI absorption-line velocity dispersion, b (km s$^{-1}$), we find for the full sample b(min) = 30 km s$^{-1}$, b(max) = 99 km s$^{-1}$, b(median) = 59 km s$^{-1}$, b(average) = 61 km s$^{-1}$, and a standard deviation of 15 km s$^{-1}$. The values of b also show very little difference from one sub-sample to the next. For example, the values of b(km s$^{-1}$) for the low-latitude sample (15° < |b| < 45°), b(average) = 60 km s$^{-1}$ and standard deviation = 17 km s$^{-1}$, are similar to the values for the high-latitude sample, b(average) = 61 km s$^{-1}$ and standard deviation = 14 km s$^{-1}$. This implies that the effects of Galactic rotation on the lower-latitude sample are no larger than the effects of inflow and outflow on the higher-latitude sample. The observed line dispersions are substantially larger than the value b = 17.7 km s$^{-1}$ expected for thermal Doppler broadening alone in gas at T = $3 \times 10^5$ K. The O VI profiles are therefore likely shaped by other processes such as inflow, outflow, turbulence, and Galactic rotation.

The values of b for the extragalactic lines of sight observed by FUSE are much larger than those derived for directions to stars in the Galactic disk. Jenkins (1978a) lists values of $<(v - \bar{v}_{obs})^2>$ for O VI absorption toward 62 stars observed by the Copernicus satellite with distances ranging from ~ 0.1 to 3.2 kpc in the Galactic disk corresponding to b = $\sqrt{2} <(v - \bar{v}_{obs})^2>^{0.5}$ ranging from 10.7 to 56 km s$^{-1}$ with a median of 27 km s$^{-1}$. The median value of b for the FUSE extragalactic sample of 60 km s$^{-1}$ is 2.2 times larger than the median value measured for the disk sample. The extra pathlength through O VI-bearing halo gas for the FUSE extragalactic sample may explain much of the difference.

We examine the behavior of $\bar{v}_{obs}$ with direction in the Galaxy in Figure 13 where values of $\bar{v}_{obs}$ are displayed as colored circles in an Aitoff projection with the Galactic anticenter at the center of the figure and longitude increasing to the left. Increasing positive velocities are color-coded, with yellow to red 12° radius circles surrounding each object, while decreasing negative velocities are color-coded from light to dark blue. At higher Galactic latitudes ( |b| > 45°), the effects of Galactic rotation should be small and the measurements should be most sensitive to the effects of flows perpendicular to the Galactic plane. The gas at high latitudes in each Galactic hemisphere is moving with positive and negative velocity with nearly equal frequency. The visual impression one gets from Figure 13 is consistent with the average values and standard deviations listed in Table 4. For $|b| > 45^{\circ}$, $\bar{v}_{obs}$ (average) = 0 km s$^{-1}$ with a standard deviation = 21 km s$^{-1}$. These numbers hardly change for other sub-samples of the observations, including one for the northern Galactic polar region ($b > 45^{\circ}$) and southern Galactic polar region ($b < -45^{\circ}$) (see Table 4).

At low Galactic latitudes ($|b| < 45^{\circ}$) where the kinematic effects of inflow and outflow are reduced, it should be possible to see the effects of Galactic rotation on the measured values of $\bar{v}_{obs}$ to each object, particularly those at the lowest latitudes in directions where Galactic rotation effects should be large. An inspection of Figure 13 at positive latitudes reveals an excess of values of $\bar{v}_{obs}$ with positive velocity (yellow and red) for directions with $l \sim 240^{\circ}$ to $300^{\circ}$ and with negative velocity (green and blue) for directions with $l \sim 60^{\circ}$ to $120^{\circ}$. Such an effect is not seen at negative latitudes.

Another way of searching for the effects of Galactic rotation is to plot the observed values of $\bar{v}_{obs}$ against $\bar{v}_{exp}$, where $\bar{v}_{exp}$ is the average expected LSR velocity for the O VI absorption to an object according to a model of Galactic rotation including assumptions for the stratification of the gas away from the Galactic plane and the turbulent velocity in the gas (see §3). In Figures 14a and 14c we display $\bar{v}_{obs}$ versus $\bar{v}_{exp}$ for objects with $45^{\circ} < b < 90^{\circ}$ and $-90^{\circ} < b < -45^{\circ}$, respectively. In Figures 14b and 14d we display $\bar{v}_{obs}$ versus $\bar{v}_{exp}$ for objects with $0^{\circ} < b < 45^{\circ}$ and $-45^{\circ} < b < 0^{\circ}$, respectively. Filled circles are for measurements where there is no serious contamination between thick disk and high velocity O VI absorption. Open circles are for cases with serious contamination between thick disk and high velocity absorption. Half filled circles are for cases where there are possible contamination problems.

The displays of Figures 14a and 14c for high-latitude directions are consistent with the color-coded view seen in Figure 13. At high latitudes, where the effects of Galactic rotation are small ($\bar{v}_{exp} < 20$ km s$^{-1}$), the values of $\bar{v}_{obs}$ exhibit a large spread in velocity centered on $\bar{v}_{obs} = 0$ km s$^{-1}$. The displays in Figures 14b and 14d for objects at low latitude ($|b| < 45^{\circ}$), where the effects of inflow and outflow should be reduced provide the greatest chance of seeing the effects of Galactic rotation. Rotation effects are difficult to see at low latitude in the southern Galactic hemisphere (Fig. 14d). For the northern hemisphere at low latitude (Fig. 14b) there is a rough correlation (correlation coefficient ~0.32 or ~15% probability of being random). The correlation between $\bar{v}_{obs}$ and $\bar{v}_{exp}$ is affected by the large dispersion in velocity of the observed values of $\bar{v}_{obs}$. This dispersion is evidently produced by a combination of turbulence, inflow/outflow, and the general spatial patchiness of the O VI absorption. The effects of Galactic rotation are more easily seen in the O VI velocity channel maps found in Wakker et al. (2002).

The two objects showing the largest Galactic rotational effects are H 1821+643 in the direction $l = 94.00^{\circ}$ and $b = 27.42^{\circ}$ with $\bar{v}_{exp}$ = -31 km s$^{-1}$ and $\bar{v}_{obs}$ = -46 km s$^{-1}$, and ESO 0265+G23 in the direction $l = 285.91^{\circ}$ and $b = 16.59^{\circ}$ with $\bar{v}_{exp}$ = 30 km s$^{-1}$ and $\bar{v}_{obs}$ = 48 km s$^{-1}$.

The kinematic behavior of O VI in Figure 14 shows a higher degree of irregularity than a similar plot shown for C IV by Savage et al. (1997), where the effects of Galactic rotation were more clearly visible. O VI may trace a more disturbed and turbulent phase of halo gas than C IV. However, to be sure, it will be necessary to obtain more measures of C IV absorption along the same lines of sight for which O VI measurements now exist.

The distribution of the O VI line width, b, on the sky is shown in Figure 15 where the circle size is proportional to b. The filled circles are for lines of sight where the measures of b are reliable. The open circles are for cases where the measurements of b are uncertain because of blending between thick disk absorption and high velocity absorption. While large values of b are found in both the northern and southern Galactic hemispheres, directions with small values of b are mostly found in the southern hemisphere where 6 of 29 directions have $b < 40$ km s$^{-1}$ compared to 1 of 62 in the northern hemisphere. This difference is also reflected in the distribution of the O VI column density on the sky (see Figs. 8a and 9a) with low values of logN(O VI) mostly occurring in the southern Galactic hemisphere. The correlation between logN(O VI) and b is discussed in §7.3.

## 7.2. O VI and H I Intermediate Velocity Clouds

The IVCs in the Solar neighborhood represent an important component of the neutral interstellar gas (Kuntz & Danly 1996; Wakker 2001). They contain much of the kinetic energy of the neutral medium (Kulkarni & Fich 1985), have metallicities of 0.5-1.0 Solar (Richter et al. 2001a, 2001b; Wakker 2001), and have distance limits in the range from z = 0.5 to 2 kpc. Their motions, distances, and abundances suggest that the IVCs are probably tracing the flow of gas in a Galactic fountain. The IVCs cover a substantial portion of the north Galactic polar region of the sky and are less abundant in the south Galactic polar region.

It is reasonable to wonder if the IVCs are somehow connected to the distribution and kinematics of O VI. In Figures 16a and 16b we display Aitoff projections of the (H I) IVCs, with l = 120° at the center of the figures and Galactic longitude increasing to the left based on the Leiden-Dwingeloo H I Survey (Hartmann & Burton 1997). The IVCs are represented with the grey scale proportional to the H I column density in the IVCs in the velocity range from $v = -30$ to $-90$ km s$^{-1}$ in Figure 16a and from $v = 30$ to $90$ km s$^{-1}$ in Figure 16b. The IVC names are not noted on this figure; full discussions of the names and properties of the IVCs are found in Kuntz & Danly (1996) and Wakker (2001). Figures 16a and 16b also show open circles denoting the direction and column density of O VI integrated over the same velocity range of the IVCs (-30 to -90 km s$^{-1}$ for Fig. 16a; 30 to 90 km s$^{-1}$ for Fig. 16b) for each object listed in Table 2.

Many of the objects observed in our program lie in the direction of IVCs. When an IVC lies in the direction to an object, we have listed information about the name of the IVC and possible association of O VI absorption at the velocity of the IVC in Table 5. H I IVCs exist toward 59 of the 102 objects in our program. The IVCs include the Low Latitude IV Arch (LLIV Arch), the IV Arch, the IV Spur, Complex K, and a number of isolated IVCs. O VI is detected at the velocities of the IV Arch in 6 of 8 cases, at the velocity of Complex K in 2 of 3 cases, and at the velocities of the LLIV Arch in 4 of 6 cases (also see Richter et al. 2001b). O VI is not detected at the velocity of the IV Spur in 3 of 3 cases. We conclude that O VI absorption occurs near the velocities of IVCs in most, but not all cases.

Since O VI absorption seems to be associated with IVCs, it is possible that the significant variation in the O VI column density across the sky might be in part produced by an enhancement of O VI absorption in directions of the sky having IVCs. In Table 3 we have listed separate averages for log N(O VI) and log [N(O VI)sin|b|] for directions with IVCs and for directions with no IVCs. The averages for directions with IVCs and without IVCs differ by only 0.01 dex. Therefore, the presence or absence of H I IVCs is not affecting the total column density of Galactic O VI through the entire halo. Similarly, the presence or absence of an IVC does not affect the average observed dispersion, <b(km s$^{-1}$)>, obtained for the sample of O VI profiles (see Table 4). A more careful comparison is provided in Figures 16 and 17, where the integration range for the O VI column densities exactly matches that for the IVCs. There is no obvious enhancement in the values of log N(O VI) in Figures 16a and 16b for directions on or off IVCs. That lack of correlation is displayed in Figure 17 where the O VI column density over the velocity range of the IVCs is plotted against the IVC H I column density for all the data points shown in Figures 16a and 16b. Over a logarithmic range in the IVC H I column density extending from 18.5 to 20.2 there is no relationship between the amount of O VI and the amount of IVC H I. We conclude that the presence or absence of an H I IVC has no significant effect on the total amount of O VI observed along an extragalactic line of sight through the halo. This result based on a large number of sight lines through IVCs disagrees with the low spectral resolution ORFEUS measurements of O VI absorption toward 14 halo stars, which provided weak evidence for an excess of O VI along sight lines characterized by intermediate negative velocity gas (Hurwitz & Bowyer 1996). It is possible that the averaging over long paths through the halo has reduced the size of the effect.

## 7.3. High Positive Velocity O VI Absorption Wings

The presence of O VI absorption at high velocity ( $|v_{LSR}| > 100$ km s$^{-1}$) and its association with H I HVCs is discussed extensively by Sembach et al. (2002). The presence of O VI absorption near the velocities of many of the H I HVCs indicates that the HVCs are probably interacting with an unseen highly extended ( R > 70 kpc) hot corona to the Milky Way or with a hot Local Group medium. Much of the O VI absorption at high velocity appears to be tracing phenomena in regions of the Galaxy far from the Galactic thick disk of O VI.

However, there is one aspect of the high velocity O VI absorption that could be related to phenomena occurring in and near the Galactic thick disk of O VI. The O VI profiles for 22 objects exhibit very interesting high positive velocity absorption wings extending from approximately 100 to 250 km s$^{-1}$. Since the wings are often weak, they are most readily detected in spectra with good S/N. The objects are listed in Table 6 where we give object name, l, b, the velocity range of the wing, and the value of log N(O VI) in the wing integrated over the listed velocity range. Seven of the O VI line profiles shown in Figure 3 show the positive velocity wings including the profiles for Mrk 817, PG 0953+414, Mrk 421, PG 0844+349, IRAS F11432-1810, 3C 273, and PKS 2005-489.

The distribution on the sky of objects with these wings can be seen in Figure 17 of Sembach et al (2002). Objects with the high positive velocity absorption wings are prevalent in the northern Galactic region with $180^\circ < l < 360^\circ$ and $30^\circ < b < 80^\circ$ where 18 objects with wings are found. Positive velocity wings are also seen toward 5 objects near the center of HVC Complex C (see Sembach et al. 2002). These include PG 1415+451, PG 1626+554, Mrk 876, Mrk 817, and PG 1351+640.

Two objects in the general direction of the Galactic center at a latitude around -30° have high positive velocity absorption wings, including Mrk 509 and PKS 2005-489. These wings may trace outflowing gas from the central regions of the Galaxy.

Seven objects exhibiting wings including 3C 273 lie in directions bounded by Galactic Radio Loops I and/or IV (see footnote b in Table 6). In the case of 3C 273, Sembach et al. (2001) have proposed that the O VI wing that extends from 105 to 260 km s$^{-1}$ may be tracing the outflow of hot gas into the halo from the Sco-Cen OB association. Such an outflow process may be responsible for the wings for the other objects in the general direction of Radio Loops I and IV.

Based on a study of the distribution of extreme-UV sources and Na I absorbing clouds at high latitude, Welsh et al. (1999) have found evidence that the local interstellar bubble cavity may open into the northern Galactic halo with a shape resembling an elongated chimney with a wide funnel. The large number of positive velocity O VI wings seen in the north and the 0.25 dex excess of the thick disk O VI in the north could be tracing the hot gas related to this proposed chimney.

Various possible extragalactic origins for the positive velocity wings are discussed by Sembach et al. (2002).

## 7.4. N(O VI) versus b (km s$^{-1}$)

We searched the O VI data for relationships among various observed quantities. In most cases, the result was simply a scatter diagram (see §9). However, a corrrelation exists between log N(O VI) and the line-width parameter, b (km s$^{-1}$) (see Fig. 18). The observed column density of O VI is larger when the line width is larger. The effect holds for the measurements from Table 2 (filled circles) and includes the Copernicus satellite (Jenkins 1978a) sample of disk stars (open circles). The trend also holds for gas in the LMC (upward pointing open triangles) and SMC (downward pointing open triangles), even though the metallicity of the LMC and SMC are reduced by ~0.3 and ~0.7 dex compared to the Milky Way.

The relationship between N(O VI) and b shown in Figure 18 was first noticed by Heckman et al. (2002). At higher columns, this relation flattens out slightly. They showed that this general relation between N and b is independent of the oxygen abundance over the range O/H = 0.1 to 2.0 solar. The observed trend at high columns is similar to that predicted for radiatively cooling gas if it is assumed that the cooling flow velocity and the line breadth are strongly correlated. It has been argued by Heckman et al. (2002) that the column density/line width correlation is strong evidence for collisional ionization of O VI in a variety of systems. Photoionization models do not reproduce the large line widths observed in many of the O VI absorbers, including those we observe in the Milky Way thick disk/halo and high velocity clouds (see Sembach et al. 2002).

## 8. INTERESTING SIGHT LINES

In this section detailed comments are provided for particularly interesting cases where the O VI absorption associated with the thick disk gas of the Milky Way is unusual in some way. We discuss the lines of sight with especially strong and weak O VI absorption. We also discuss cases where the O VI profiles are kinematically interesting, including narrow profiles and multi-component profiles.

### 8.1. Very Strong O VI absorption.

The sources with the largest O VI column densities in our sample of 102 objects are PKS 2005-489 and 3C 273.

PKS 2005-489, with log N(O VI) = 14.78±0.02 and log [N(O VI)sin|b|] = 14.51, lies under the Galactic center in the direction $l$ = 350.37° and b = -32.60°. The O VI absorption and $N_a(v)$ profiles are shown in Figures 3 and 4. The absorption in the strong O VI $\lambda$1031.93 line extends from -100 to +225 km s$^{-1}$, with maximum absorption at ~ 60 km s$^{-1}$. The absorption from -100 to +120 km s$^{-1}$ is attributed to the thick disk of the Milky Way. The extended wing of absorption from +120 to +225 km s$^{-1}$ with log N(O VI) = 13.90±0.06 is discussed by Sembach et al. (2002) and in §7.3. The great strength of the O VI absorption in this direction may be related to an enhanced abundance of hot gas over the central regions of the Galaxy (see §8.7). The nearby directions to ESO 141-G55 and Ton 1924-416 are also among the 23 objects with the largest N(O VI).

3C 273, with log N(O VI) = 14.73±0.01 and log [N(OVI)sin|b|] = 14.68, lies in the direction $l$ = 289.95°, b = 64.36° that passes near Galactic Radio Loops I, IV, and the North Polar Spur (see Figs. 9a and 9b). The 0.25 keV X-ray emission in this direction (Snowden et al. 1997) is quite strong (see Fig. 9b). In a detailed study of 3C 273, Sembach et al. (2001) found that the main portion of the O VI absorption profile that extends from -100 to 100 km s$^{-1}$ is similar in appearance to the profiles for N V and C IV. However, the wing in O VI absorption extending from 100 to 240 km s$^{-1}$ is not detected in other species. The gas in this direction has evidently been heated to high temperatures by multiple supernova explosions in the Sco-Cen OB association. It is quite possible that the direction to 3C 273 provides us with a direct view of the flow of gas from the disk into the halo through a chimney-like structure. Similar absorption wings observed toward other objects are discussed in §7.3.

### 8.2. Weak O VI absorption

The 6 lines of sight with the smallest measured values of log N(O VI) are NGC 7714, Mrk 1095, NGC 7469, NGC 1068, Mrk 1502, and UGC 12163. With O VI column densities ranging from log N(O VI) = 13.85 to 14.00, these measured column densities are similar to the smallest 3σ limits to log N(O VI) determined for 6 objects including NGC 595, HE 0405-2958, HE 1115-1735, PG 0052+251, PG 2349-014, and SBS 0355-052. Column density limits for 6 additional objects exceed logN(O VI) = 14.00. In the following, we discuss several of these interesting cases of low O VI column density.

Mrk 1095 lies toward the Galactic anti-center in the direction $l$ = 201.69° and b = -21.13°. The FUSE spectra for Mrk 1095 are of good quality (see Figs. 2 and 3). The O VI absorption is weak, with log N(O VI) = 13.89±0.07 and log [N(O VI)sin|b|] = 13.45. The absorption extends from -35 to 50 km s$^{-1}$, has $\bar{v}_{obs}$ = 5 km s$^{-1}$ and b = 31 km s$^{-1}$. This is also one of the narrowest profiles found in the FUSE O VI sample. The direction passes just below the Orion Association and continues through the outer Galaxy. The low abundance of O VI in a direction passing through the halo over this active region of hot stars with strong stellar winds is surprising, since the creation of hot gas in the ISM is thought to occur preferentially in OB associations (MacLow, McCray & Norman 1989). Possibly the Orion association has not yet evolved long enough to produce superbubbles extending far from the Galactic plane. The four other objects sampling this general direction of the Galaxy are HE 0450-2958 [$l$ = 231.13°, b = -37.59°, log N(O VI) < 14.00 (3σ)], Mrk 618 [ $l$ = 206.72°, b = -34.66°, log N(O VI) = 14.27±0.07], PKS 0405-12 [$l$ = 204.93°,

b = -41.76°, log N(O VI) = 14.10±0.04], and SBS 0335-052 [ l = 191.34°, b = -44.69°, log N(O VI) < 13.99].

NGC 7469 (l = 83.10°, b = -45.47°) with log N(O VI) = 13.96±0.06  lies near the direction to NGC 7714 (l = 88.22°,  b = 55.56°) with  log N(O VI) = 13.85±0.14.  These two objects are also near other objects with low values or limits for log N(O VI) including Mrk 926 [log N(O VI) < 14.23(3σ)] , PG 0052+251 [log N(O VI) < 13.97(3σ)], PG 2349-014 [log N(O VI) < 13.97(3σ)], Mrk 1502 [log N(O VI) = 14.00±0.10] , Mrk 335 [log N(O VI) = 14.07±0.03], Mrk 304 [log N(O VI) = 14.20±0.08], NGC 1068 [log N(O VI) = 13.98±0.06], NGC 588 [log N(O VI) < 14.07], NGC 595 [log N(O VI) < 13,75], and UGC 12163 [log N(O VI) = 14.00±0.12].   An inspection of Figure 7b reveals that the general region in the sky around l = 80° to 175° and b = -30° to –60° has a deficiency of O VI.  This is also a Galactic direction where the soft 0.25 keV X-ray background is low (see Fig. 9b and §9.3).

### 8.3. Narrow O VI absorption Profiles

The three objects with the narrowest Galactic O VI λ1031.93 profiles are Mrk 618, Mrk 1095 and Mrk 1502 with b = 30, 31, and 33 km s$^{-1}$, respectively.   All three objects are in directions with low values of log N(O VI) (see §8.2).  Even though the O VI lines are narrow,  they are broader than the value of b = 18  km s$^{-1}$ expected for pure Doppler broadening of O VI at T = $3\times10^5$K.

### 8.4. Multi-component O VI absorption

Most  of the O VI absorption profiles appear relatively smooth, making it difficult to clearly identify substructure in the absorption.  However, even the smooth profiles exhibit asymmetries, which suggests that an overlapping superposition of absorbing structures may be responsible for each profile.  In several cases the component nature of the absorption is more clearly defined.  Some of these include:

ESO 141-G55.  The principal O VI absorption is centered near 10 km s$^{-1}$ and a weaker component appears at –50 km s$^{-1}$.

H 1821+643.  A strong narrow O VI absorption near 13 km s$^{-1}$ likely is due to gas in the outer regions of the planetary nebula K1-16 (see §8.6).  Several additional components span the full velocity range of the remaining absorption from -145 to 115 km s$^{-1}$.

HE 0226-4110.  At least three O VI components appear be required to describe the absorption from -120 to 110 km s$^{-1}$.

MRC 2251-178.  Two or perhaps three components would be required to represent the complex absorption from –145 to 70 km s$^{-1}$.

Mrk 9. At least three components are contributing to the O VI absorption extending from -100 to  90 km s$^{-1}$.

### 8.5. O VI Absorption in the Outer Arm of the Milky Way

O VI absorption associated with the outer spiral arm of the Milky Way, where gas warps to large positive latitudes, can be studied in the spectra of several objects.  These include 3C 382.0 (l = 61.31°, b = 17.45°),  H 1821+643 (l = 94.00°, b = 27.42°) , and HS 0624+6907 (l = 145.71°, b = 23.25°).  The O VI profiles for both 3C 382.0 and H 1821+643 extend from –130 to 110 km s$^{-1}$ and from – 160 to 115 km s$^{-1}$, respectively.  In these directions the outer arm H I emission is found at –87 km s$^{-1}$ for 3C 382.0 and  at –87 and –128 km s$^{-1}$ for H 1821+643.  In both cases it appears that the O VI absorption extends well beyond the range of the H I emission and that O VI likely exists in the outer arm of the Milky Way.  However, for HS 0624+6907, the O VI absorption extends from -85 to 90 km s$^{-1}$,  while the H I associated with the outer arm is centered near -100 km s$^{-1}$ implying no O VI in the outer arm gas in that direction.

For the direction to H1821+643 Savage et al. (1995) have analyzed HST GHRS observations of Galactic UV absorption produced by  S II, Si II, Mg II, C IV, and N V.  Their measurements of C IV reveal a strong absorbing component at $v_{LSR}$ = -120 km s$^{-1}$ with  b = 25 km s$^{-1}$ and N(C IV) = (6.1±1.1)x10$^{13}$ cm$^{-2}$.  The absorption spans the velocity range from -80 to

-160 km s$^{-1}$, which corresponds well to the negative velocity extension of the FUSE O VI absorption.  Assuming a flat Galactic rotation curve with an outer Galaxy rotational speed of 220 km s$^{-1}$, a component velocity of $-120$ km s$^{-1}$ for C IV and O VI implies that the absorption may be occurring at a Galactocentric distance of ~24 kpc and a distance away from the Galactic plane of ~7 kpc (see Fig. 2 in Savage et al. 1995).  The detection of strong O VI absorption in this gas along with the C IV casts serious doubts on the possibility advanced by Savage et al. (1995) that the C IV is produced in low density warm gas ($n_{total}$ ~ $6x10^{-3}$ cm$^{-3}$) in the outer Galaxy photoionized by the extragalactic background.  To explain the O VI,  a hot collisionally ionized gas origin appears more likely.

The Galactic warp interpretation is consistent with the absence of such a high negative velocity wing toward UGC 12163 (l = 92.14°, b = -25.34°) where the negative velocity extension of the O VI absorption is to $-100$ km s$^{-1}$.  In the general direction of l = 80 to 120° , the Galaxy warps to positive latitudes rather than negative  latitudes.

### 8.6. O VI Absorption Probably Associated with the Planetary Nebula K1-16

The line of sight to H 1821+643 passes through the local bubble, close to a nearby planetary nebula (K1-16), over the Perseus and Outer spiral arms, and through the outer Galactic Warp.  For a detailed discussion of Goddard High Resolution Spectrograph (GHRS) observations of Galactic absorption along this line of sight see Savage et al. (1995).  The line of sight passes 96" away from the central star of the planetary nebula K1-16, which has an optical radius of 57" (Kohoutek 1963; Acker et al. 1992). The central star of the nebula is a He rich PG 1159 star with an effective temperature of $1.4x10^5$ K, a high mass loss rate estimated to be $8x10^{-9}$ M$_\odot$ yr$^{-1}$, and a terminal velocity of 4000 km s$^{-1}$ (Rauch & Werner 1995; Kruk & Werner 1998; Koesterke & Werner 1998).

The FUSE observations of H 1821+643 reveal strongly saturated O VI absorption at $v_{LSR}$ ~ 13 km s$^{-1}$ with both lines  of the doublet having relatively narrow components of similar strength (Wakker et al. 2002).  The O VI $\lambda$1031.93 absorption profile for H 1821+643 is shown in Figure 3. Very  strong absorption components are also seen in N V and C IV (Savage et al. 1995) at $v_{LSR}$ ~ -7 km s$^{-1}$. The 20 km s$^{-1}$ velocity difference between O VI and N V+ C IV could be a real difference or the result of velocity calibration errors in the FUSE and/or GHRS observations. These very strong highly ionized absorption components could be associated with hot gas near the boundary of Galactic Radio Loop III or with the planetary nebula.  FUSE observations of the central star of the nebula also reveal interstellar O VI absorption with a strength somewhat greater than for the 13 km s$^{-1}$ O VI component observed toward H 1821+643.  Possibly both lines of sight are tracing highly ionized gas that is not associated with the planetary nebula.  However, an association with the nebula appears more likely.  The central star is so hot ($1.4x10^5$ K) that the O VI might be created by photoionization provided the nebula is clumpy enough to allow O VI ionizing radiation from the star to penetrate to regions beyond its optical boundary.   An alternate possibility is that the mechanical interaction of the stellar wind with the surrounding nebula creates hot gas and X-ray photons which photoionize a region extending beyond the boundaries of the optical nebula. Because of the possible association of the 13 km s$^{-1}$ O VI absorption component with the planetary nebula, we have removed the  component before proceeding with the  analysis of the broader aspects of the O VI absorption toward H 1821+643 (see Wakker et al. 2002).

### 8.7. Galactic Center Lines of Sight

The large values of log N(O VI) toward PKS 2005-489, Tol 1924-416, Mrk 509, and ESO 141-G55, which lie in directions toward the Galactic center between l = 330° and 36° and b = -24° and -33° (see Figs. 7a and 9), could be the result of an excess of hot gas associated with processes occurring in the central region of the Milky Way.  However,  in the directions to PKS 2005-489, Tol 1924-416, and ESO 141-G55, the absorption could also be associated with energetic events responsible for the origin of Radio Loop I (see Fig. 9a and §9.4). The clear separation of absorption occurring in nearby versus distant regions of the Galaxy will require observations of  O VI absorption toward nearer stars located in the general direction of these extragalactic sources.

# 9. O VI VERSUS OTHER ISM TRACERS

In this section we discuss the relationships between the O VI absorption and other tracers of interstellar gas, including H I, Hα, soft X-ray emission, non-thermal radio emission, and C IV absorption. The measurements for H I, Hα, and the soft X-ray emission are listed in Table 7.

## 9.1. N(O VI) vs N(H I)

Although we would not necessarily expect a good relationship between the O VI column density and the column density of H I toward our sample of objects, it is of interest to see if the amount of cool and warm neutral gas as traced by H I is somehow affected by the amount of hot gas traced by O VI. It is possible that there might be a displacement effect where cool and warm gas is missing when there is an excess of hot gas. Therefore, in a comparison of N(H I) versus N(O VI), we might expect to see an anti-correlation rather than a correlation.

We compare log N(O VI) for the thick Galactic disk to log N(H I) in Figure 19a. The values of N(H I) are obtained from the H I observations discussed and displayed in Wakker et al. (2002). The H I column densities have been obtained from integrations over the same velocity range as for O VI (see the values of $v_-$ to $v_+$ listed in Table 2), except for a few cases where this would break an IVC component into two. Then, the H I for the IVC was fully included in the integration (see notes to Table 7). Although the column densities of H I and O VI both span more than a decade, we see no relationship between N(O VI) and N(H I) for the sample of objects included in our program. Comparisons of O VI emission with estimates for the absorption in the same direction imply that the O VI occupies a very small fractional volume of the ISM (Shelton et al. 2001). We should therefore not be too surprised that we do not see an anticorrelation.

## 9.2. N(O VI) versus Hα Emission from the Warm Ionized Medium

In Figure 19b we compare the column density of O VI with the intensity of Hα emission from the warm ionized medium. Values of log N(O VI) toward each survey object are plotted against log I(Hα), where I(Hα) is the velocity integrated surface brightness of diffuse Hα emission in Rayleighs ($10^6 / 4\pi$ photons cm$^{-2}$ s$^{-1}$ sr$^{-1}$) measured by the Wisconsin Hα mapper experiment (Haffner et al. 2002) over a 1° diameter field of view in a pointing direction containing the extragalactic object but not centered on the object. One Rayleigh is equivalent to an ionized gas emission measure, EM = $\int n_e^2$ dx , of 2.25 cm$^{-6}$ pc in ionized gas with T = 8000 K. The column density of O VI and the emission measure of diffuse Galactic Hα are poorly correlated. This is in marked contrast to the relatively good correlation between I(Hα) and N(H I) (see Fig. 19d).

## 9.3. N(O VI) versus the Soft X-ray Background Emission

O VI and X-rays are apparently associated with hot gas in the ISM, so we checked to see if N(O VI) and the brightness of X-ray emission are correlated. Since O VI peaks in abundance at T ~ $3 \times 10^5$ K under conditions of equilibrium collisional ionization, we have chosen to compare the O VI absorption to the 0.25 keV soft X-ray background radiation, which likely is produced in hot gas with T ~ $(1-3) \times 10^6$ K. For 0.25 keV X-rays, the photoelectric absorption optical depth is unity for a total hydrogen column density (neutral and ionized) of ~$1 \times 10^{20}$ cm$^{-2}$. For most of the high-latitude lines of sight in the FUSE O VI program N(H I) ~ (1 - 4) $\times 10^{20}$ cm$^{-2}$ (see Table 7). Therefore, the photoelectric absorption of distant X-ray emitting gas along these lines of sight should result in an attenuation of approximately 1 to 4 optical depths.

In Figure 9b we display an all-sky Aitoff projection of the 0.25 keV soft X-ray background count rate from Snowden et al. (1997) at a resolution of 36'. The X-ray data are represented by the grey scale image, and the column densities of O VI toward each object in the FUSE survey are represented by the open circles, with the circle size proportional to log N(O VI). The 11 3σ upper limits to log N(O VI) are displayed as triangles with a size proportional to the limit.  We note a possible correlation between the value of log N(O VI) and the soft X-ray brightness. For example, the high O VI column density toward 3C 273 (l = 289.95°, b = 64.36°) traces gas where the 0.25 keV X-ray emission is particularly large while the O VI column densities are small toward 8 of 12 objects in the region of generally low X-ray emission from l ~ 60° to 150° and b ~ -30° to -60°.

In Figure 19c we directly compare values of log N(O VI) with values of the 0.25 keV X-ray count rate from the ROSAT survey,  log I(X).  The values  of log N(O VI) are for the thick disk (see Table 2) while the values of  I(X) ($10^{-6}$ counts s$^{-1}$ arcmin$^{-2}$) listed in column 10 of Table 7 represent the sum of  the ROSAT  (R1+R2 band)  0.25 keV counts in  8 ROSAT  12' pixels lying closest to the direction of each object. This sum did not include the central ROSAT pixel because it could  be affected by an X-ray point source associated with the background AGN or QSO.   To assess  the possible effects  of photoelectric attenuation of the X-rays, we have plotted the observations in Figure 19c with different symbols to  distinguish different values of N(H I).  The open and filled symbols are for values of N(H I) greater than or less than   $2 \times 10^{20}$ cm$^{-2}$ , respectively.

As is well known, the intensity of 0.25 keV X-ray observations of the diffuse background displays an anti-correlation with Galactic N(H I) (Snowden et al. 1998, 2000).  The anti-correlation of N(H I) and I(X) for our survey directions  is shown in Figure 19f. That anti-correlation also shows up in Figure 19c as a clear separation of data points sorted according to N(H I). The X-ray emission is large when N(H I) < $2 \times 10^{20}$ cm$^{-2}$  (filled symbols) and small when N(H I)  > $2 \times 10^{20}$ cm$^{-2}$  (open symbols).   In directions where N(H I)  is relatively small and where foreground photoelectric attenuation should also be small (filled symbols) there is no obvious correlation between log N(O VI) and log I(X).

The  diffuse X-ray background is a complex superposition of a non-uniform local bubble component and attenuated halo and extragalactic background components (Snowden et al. 1998, 2000).  In order to properly compare N(O VI) with the soft X-ray diffuse background, we  need to correct for the photoelectric absorption produced by foreground neutral and ionized gas. While we have good information about the total amount of neutral gas we do not know its line of sight distribution very well.  Also, we have very little knowledge about the amount of absorbing ionized gas along our high latitude lines of sight.  The warm ionized gas could provide a considerable amount of  X-ray absorption since its column density  is typically 30 to 50% of the H I column density along high latitude lines of sight.  Given all these uncertainties, we do not pursue the relationships between O VI absorption and X-ray emission any further.

An interesting result that emerges from our O VI correlation study is that, while the correlations of N(O VI) versus N(H I) and  N(O VI)  versus  I(Hα) are poor,  the correlation I(Hα) versus N(H I) is relatively good (see Fig. 19d). The latter correlation is perhaps not too surprising, since many tracers of gas in the ISM are relatively well correlated.  The good correlation of I(Hα) versus N(H I)  and the good anti-correlation of I(X) and N(H I) explains why I(Hα) and I(X)  also have a relatively good anti-correlation (see Fig. 19e).

### 9.4. O VI and Galactic Non-thermal Radio Loops

Galactic non-thermal radio loops trace the sites of energetic events in the Galaxy and may therefore contain an excess of hot interstellar gas (see Sembach et al. 1997).  In Figure 9a we show the boundaries of Galactic Radio Loops I-IV  as defined by the observations of Berkhuijsen et al. (1971) along with the total O VI  column density toward each survey object, displayed as the filled circles with size related to log N(O VI).   For Radio Loops II and III there is no obvious enhancement in the O VI column density when moving from directions off the radio loops to directions inside the loops. In fact, the region defined by Radio Loop II represents a region generally deficient in O VI.

In the direction of Radio Loop IV we have previously noted the substantial enhancement in the O VI column density toward 3C 273 (l = 289.95°, b = 64.36°, log [N(O VI)sin|b|] = 14.67), which is in the direction of a bright soft X-ray emitting filaments near the edge of Radio Loop IV

(see Figs. 9a and 9b). The values of log [N(O VI)sin|b|] toward 7 other objects inside of Radio Loop IV including Mrk 1383, Tol 1247-232, PG 1302-102, PG 1307+085, HE 1326-516, ESO 572-G34, and IRAS F11431-1810 range from < 14.06 for HE 1326-0516 to 14.49 for Mrk 1383, respectively.

Similarly, for the 7 objects in directions within the boundaries of Radio Loop I but outside of Radio Loop IV , we find two objects (PKS 2005-489, PG 1352+183) with large values of log [N(O VI)sin|b|], three objects with average values (ESO 141-G55, NGC 4649, and Tol 1924-416), and two objects with low values (HE 1115-1735, ESO 265-G23). Determining if the excesses in log [N(O VI)sin|b|] toward PKS 2005-489 and PG 1352+183 are due to an association with Radio Loop I and not caused by a more distant phenomena (e.g., hot gas associated with the Galactic center) will require measures of O VI toward nearer stars in the general directions of these three objects. We conclude that the amount of O VI is not strongly enhanced in the directions of Galactic Radio Loops.

## 9.5. O VI, N V, and C IV

A detailed intercomparison of the relationships among the various highly ionized species found in the halo is beyond the scope of this paper since it will require determining reliable values of the column density of Si IV, C IV, and N V for many of the objects listed in Table 2 from existing or future observations with the spectrographs aboard the HST. However, a preliminary study of the amount of N V and C IV compared to O VI through the entire halo is valuable since the ratios of N V and C IV to O VI provides constraints on the ionization processes in the halo. In Table 8 we compare values of log N(O VI), log N(N V), and log N(C IV) for paths through the entire halo for cases where values of log N(N V) and log N(C IV) have been obtained with GHRS or STIS at resolutions, $\lambda/\Delta\lambda$, exceeding 10, 000. Results for H 1821+643 are omitted from this table because of the strong probable contamination to C IV, N V, and O VI from the planetary nebula K1-16 (see §8.6).

The values of N(C IV)/N(O VI) range from 0.49 to 1.74. Ignoring the upper limit for Mrk 205 and the one high value of the ratio for ESO 141-G55, which traces gas over the Galactic center (see Sembach et al. 1999), we find that the values of N(C IV)/N(O VI) range from 0.49 to 0.79 with an average of 0.62 and a standard deviation of 0.10. The values of N(N V)/N(O VI) range from 0.13 to 0.24. Ignoring the upper limit for Fairall 9, we find a range from 0.14 to 0.24 with an average of 0.18 and a standard deviation of 0.05. These ionic ratios are compared to theoretical expectations in §10.2.

The average value of N(C IV)/N(O VI) = 0.62±0.10 for 6 complete paths through the Galactic halo can be compared to the average 0.62 (+0.68, -0.32) found by Jenkins & Bowen (private communication) based on a preliminary analysis of O VI absorption toward 54 stars in the Galactic disk with |z| < 0.4 kpc for which values of N(C IV) are available from the IUE high ion survey of Savage et al. (2001). The fact that the average is similar for disk and halo lines of sight suggests that the processes creating O VI and C IV may not change significantly from the disk to the halo. This result is in marked contrast to the conclusion reached by Spitzer (1996) based on a sample of 6 relatively near-by stars in the disk for which N(O VI) was measured by the Copernicus satellite and N(C IV) was measured by HST or IUE. The 6 star average of N(C IV)/N(O VI) = 0.17± 0.08 suggested a large increase in the value of the C IV to O VI ion ratio from the disk into the low halo. It is likely that the smaller value of the high ion ratio for disk gas determined by Spitzer is dominated by conditions in the ISM relatively close to the Sun. In contrast the preliminary result from Jenkins & Bowen provides a more global assessment of the C IV to O VI ratio in the disk of the Galaxy.

## 10. ORIGIN OF O VI AND OTHER HIGHLY IONIZED SPECIES IN THE GALACTIC HALO

The FUSE O VI survey observations reported here provide important new insights into the distribution and kinematics of highly ionized gas in the Milky Way halo. The FUSE observations increase by an order of magnitude the number of lines of sight for which absorption-line information is now available for extragalactic paths through the entire halo. By obtaining information on O VI, we are able to gain information about one of the best tracers of hot, collisionally ionized gas available in the entire electromagnetic spectrum. The observations confirm the basic validity of Spitzer's (1956) prediction that the ISM of the Galaxy contains a hot gas phase that extends well away from the Galactic plane.

Although the FUSE O VI observations greatly increase the number of paths through the halo for which absorption line data is now available, it is important to understand how the new O VI observations fit into the broad context of information about the distribution and kinematics of various other species in the halo of the Galaxy. In Table 9 we compare the distributions of various components of the Galactic ISM from H I to O VI. Because of the irregular distribution of the ISM it is difficult to determine reliable estimates for the extension of the gas away from the Galactic plane. However, the highly ionized species (Si IV, C IV, N V, and O VI) are ~ 10 times more extended than the warm neutral medium and ~ 3 times more extended than the warm ionized medium traced by electrons and Al III. Among the high ions, it appears that the two species that are more easily photoionized (Si IV and C IV) may have a greater extension than the two species that mostly trace collisionally ionized gas (N V and O VI). If this result is confirmed through a more careful study of C IV toward objects with measured values of N(O VI), it would imply that several different ionization mechanisms are controlling the abundance of highly ionized gas in the halo.

Theories for the origin of highly ionized gas in the Milky Way halo must explain the distribution, support, kinematics, and ionization of the gas. For reviews of the models see Spitzer (1990), McKee (1993), and Savage (1995). The ionization of Si IV and C IV is likely either from electron collisions in a hot gas, photoionization, or some combination of both processes. With their high ionization threshold, O VI and N V are more likely to be produced by collisional ionization in hot gas. However, non-equilibrium ionization effects will probably be important because collisionally ionized gas cools very rapidly in the range of temperature from $(1-5) \times 10^5$ K. Such transition temperature gas might occur within the cooling gas of a "Galactic fountain" (Shapiro & Field 1976; Edgar & Chevalier 1986; Shapiro & Benjamin 1991), in the conductively heated interface region between the hot and cool interstellar gas (Ballet, Arnaud, & Rothenflug 1986), in radiatively cooling SN bubbles (Slavin & Cox 1992, 1993) or in turbulent mixing layers (TMLs) where hot gas and warm gas are mixed by turbulence to produce gas with non-equilibrium ionization characteristics (Begelman & Fabian 1990; Slavin, Shull, & Begelman 1993). The heating and ionization of the gas could also occur through magnetic reconnection processes (Raymond 1992; Zimmer, Lesch, & Birk 1997). With such a large number of processes possibly operating, it is difficult to clearly identify the most important processes.

### 10.1. Insights from the Kinematics of O VI

Although the velocities expected for gas in a hot Galactic fountain (Bregman 1980) are large enough to explain the velocities of the Milky Way HVCs, it is now established that some of the HVCs (Complex C and the Magellanic Stream) have low metallicities and are extragalactic in origin (Wakker et al. 1999; Lu et al. 1998; Richter et al. 2001a). Turning to the intermediate velocity gas, we find that velocities of the IVCs are well described by a low temperature Galactic fountain (Houck & Bregman 1990). Furthermore, the IVCs have abundances and z-distances consistent with the idea that they represent gas participating in a Galactic fountain flow (Wakker 2001; Richter et al 2001a, 2001b). Although specific Galactic fountain kinematical predictions have not been made for O VI, it is not obvious that such a prediction would produce the symmetrical distribution of upward and downward O VI cloud motions seen in Figure 6c for sight lines at high latitude with $|b| > 45^\circ$. On the other hand, the high positive velocity wings of O VI absorption extending from

~100 km s$^{-1}$ to ~ 250 km s$^{-1}$ in the spectra of 20 objects could be marking outflowing fountain gas that contains O VI.

Gas in the Galactic halo could be supported through the turbulent outflow of the hot gas associated with a Galactic fountain, by the pressure in a Galactic magnetic field with lines of force mostly parallel to the Galactic plane (Bloemen 1987), or from the pressure provided by the streaming of cosmic rays along magnetic field lines parallel to z (Chevalier & Fransson 1984; Hartquist & Morfill 1986; Breitschwerdt, Voelk, & McKenzie 1991).

Although it appears that the support provided by turbulent pressure can explain the extended H I distribution in the Galaxy (Lockman & Gehman 1991), it is not possible for the turbulence inferred from the C IV absorption line or the O VI line to support the C IV and O VI layers with scale heights estimated to be h(C IV) = 4.4 kpc (Savage et al. 1997) and h(O VI) ~ 2.3 kpc (this paper). For an isothermal atmosphere at temperature T in hydrostatic equilibrium, the scale height is given by h = kT/ <m>g(|z|), where g(|z|) is the gravitational acceleration toward the disk and <m> is the average mass per particle. With the Doppler spread parameter given by b = $(2kT/<m>)^{1/2}$, the scale height is h = $b^2/2g(|z|)$. For |z| in the range of 1 to 10 kpc in the solar neighborhood, the gravitational acceleration toward the disk is relatively constant and approximately $10^{-8}$ cm s$^{-2}$ (Kuijken & Gilmore 1989). This allows us to use the simple expression above assuming constant g for evaluating the extension of the hot (large scale height) component of the gas. To produce gas scale heights of 4.4 kpc and 2.3 kpc therefore requires turbulent velocities, b = $[2h \, g(|z|)]^{1/2} \approx 165$ and 119 km s$^{-1}$ respectively. The average C IV turbulent velocity not including the effects of inflow and outflow averaged over four high latitude lines of sight is 45 km s$^{-1}$ (Savage et al. 1997). That estimate can be compared to the value of <b(O VI)> = 61 km s$^{-1}$ we obtain for high latitude (|b| > 45$^{o}$) lines of sight (see Table 4). The turbulent velocities for C IV and O VI are 3.7 and 1.9 times smaller than necessary for the gas to be supported by turbulent pressure alone. The discrepancy for C IV is large. The discrepancy for O VI would be reduced if the estimate for b(O VI) were to include the high positive-velocity gas seen toward the 20 high latitude objects listed in Table 6.

The observations imply some other support mechanism is required for C IV and may also be required for O VI. The Galactic magnetic field working with the pressure produced by cosmic rays offers an alternate support mechanism (Chevalier & Fransson 1984; Hartquist & Morfill 1986; Bloemen 1987; Breitschwerdt et al. 1991).

### 10.2. Insights from the Amount and Distribution of O VI

The amount and distribution of O VI in the halo can be used to constrain models of the ISM. The amount of O VI through the entire halo compared to other species such as N V, C IV, and Si IV provides information about the ionization mechanisms operating in the halo. The average column density of O VI through the halo can be used to constrain the mass flow rate associated with a Galactic fountain. The observed north-south asymmetry in the distribution of the O VI not only calls into question the assumption of a simple plane-parallel patchy layer for the distribution of the gas but also raises interesting questions regarding the Galactic phenomena that are causing the asymmetry (see §6).

The various theories for the ionization of the gas in the halo provide quite different predictions for ionic ratios among the highly ionized species. The behavior of N(Si IV)/(N V), N(C IV)/N(Si IV), and N(C IV)/N(N V) has been discussed by Savage et al. (1997), and the various model predictions are given by Sembach et al. (1997). An excellent summary of the model predictions for N(C IV)/N(O VI) is given by Spitzer (1996). Models of conductive interfaces (Borkowski, Balbus, & Fristrom 1990) and supernova bubbles (Slavin & Cox 1992) yield similar predictions with N(C IV)/N(O VI) in the range from 0.09 to 0.25. The cooling Galactic fountain calculations from Shapiro & Benjamin (1991) that include the effects of the ionization produced by the photons from recombining He$^{+2}$ predict that N(C IV)/N(O VI) should range from 0.3 to 0.5 depending on the assumptions about the nature of the flow. The theory of turbulent mixing layers (Slavin et al. 1993) for hot gas entrainment velocities of 25 and 100 km s$^{-1}$ and an electron temperature $2x10^5$ K in the post-mixed gas yield N(C IV)/N(O VI) in the range from 1.4 to 8. The observed average value and dispersion of N(C IV)/N(OVI) is $0.62\pm0.10$. The observed C IV to O VI ratio for paths through the entire Milky Way halo is most consistent with the cooling Galactic fountain model. However, the same cooling fountain models predict values of N(N V)/N(O VI) in the range from 0.04 to 0.06 while the observed average of $0.18\pm0.05$ is approximately 3 times larger. None of the models discussed above predict ionic ratios of C IV/O VI and N V/O VI that agree with the observed ratios for lines of sight through the entire halo. Line of sight averaging over the absorption associated with multiple production processes could be the explanation.

Edgar & Chevalier (1986) have calculated the expected column densities for various highly ionized species in a Galactic fountain flow for gas cooling from temperatures $> 10^6$ K. The expected column density of O VI is related to the mass flow rate out of the disk, $\dot{M}$, where $\dot{M} = 1.4m_H n_H \Omega \, \dot{N} / n_H$ for He/H = 0.1. Here $1.4m_H$ is the mean mass per atom, $\Omega$ is the surface area of the flow region, $n_H$ is the initial ionized hydrogen density of the cooling gas, and $\dot{N}$ is the flux of cooling gas (ionized hydrogen atoms cm$^{-2}$ s$^{-1}$). The value of ($\dot{N} / n_H$), which by conservation of mass is the flow velocity, can be estimated from the amount of O VI in the cooling flow. If the cooling gas flow is isobaric Edgar & Chevalier (1986) find that $\dot{N} / n_H = 2.6x10^6$ [N(O VI)/ $10^{14}$ cm$^{-2}$] cm s$^{-1}$. With N(O VI) = $10^{14.09}$ cm$^{-2}$, the average value of N(O VI)sin|b| in the south Galactic polar region, $\dot{N} / n_H = 3.2x10^6$ cm s$^{-1}$ and $\dot{M} / \Omega = 1.1x10^{-3}$ ($n_H/10^{-3}$ cm$^{-3}$) $M_O$ yr$^{-1}$ kpc$^{-2}$. If the fountain activity extends over a region 20 kpc in radius then $\dot{M} = 1.4( n_H/10^{-3}$ cm$^{-3}$) $M_O$ yr$^{-1}$ on each side of the Galactic disk. This mass flow rate is similar to that estimated for the IVCs, but is much larger than that estimated for the HVCs (Savage 1995). This gives some support to the idea that the IVCs rather than the HVCs are related to the cooling gas of a fountain. In this case, the fountain kinematics are best described by a low temperature rather than a high temperature Galactic fountain (Houck & Bregman 1990). The flow rate numbers given here decrease by 1.7 times if the cooling flow is mostly isochoric rather than isobaric.

The constraints provided by the observations of highly ionized gas in the Milky Way have resulted in the development of several hybrid models for gas in the halo. Ito & Ikeuchi (1988) suggest a model that includes a fountain flow to produce hot collisionally ionized gas in the disk and low halo and photoionization by the extragalactic EUV background radiation and Galactic EUV sources to enhance the production of Si IV and C IV at large |z|. In this model the support of gas at large |z| is due to the fountain and chimney activity, while the ionization of the gas is affected by both collisional ionization and photoionization. A Galactic "chimney" refers to vigorous fountain activity in the region of an OB association where correlated supernovae create superbubbles of hot gas which break through the gas of the Galactic plane through Rayleigh Taylor instabilities (Bruhweiler et al. 1980; Cowie, McKee, & Ostriker 1981; McCray & Kafatos 1987; MacLow, McCray, & Norman 1989). It is difficult to produce Galactic O VI by the extragalactic EUV background at the observed column density levels because very low physical densities and extremely long path lengths are required (Sembach et al. 2002). The relatively free flow of EUV photons into the halo from hot stars in OB associations at the base of Galactic chimneys is an additional source of ionizing radiation that could produce enhancements in the amount of Si IV and C IV but not N V and O VI. Ionization may also be provided by hot white dwarf stars. Si IV and C IV can be produced by the He rich white dwarfs while N V and O VI can also be produced by the H rich white dwarfs (Dupree & Raymond 1983). The larger scale heights of the white dwarfs compared to the nondegenerate OB stars allows their ionizing radiation to more easily escape from the Galactic disk into the halo

A second example of a hybrid model is that of Shull & Slavin (1994), which combines cooling SN bubbles at low |z| with the turbulent mixing layers associated with Galactic chimneys at large |z|. These authors were specifically trying to produce an inhomogeneous model in which the ISM changes character at large z in order to explain the different high ion scale heights. In this model the Galactic chimneys produce unusual ionization characteristics at large |z| because non-equilibrium conditions exist in the walls of the chimneys where hot and cool gases are intermixed.

Photoionization by EUV radiation produced by the extragalactic background or by the hot stars in the underlying OB association is another possible source of ionization in the chimney walls. Such hybrid models have a number of adjustable parameters for which it is possible to find values that provide acceptable fits to the observed ionic ratios and their observed changes with |z|.

The simulations of supernova remnant evolution in the lower halo (Shelton 1998) reveal that C IV, N V, and O VI can survive for very long periods of time ($\sim 10^7$ years) in isolated supernova remnants occurring in low density halo environments. The simulations suggest that an ensemble of supernova remnants could produce a time-averaged O VI column density perpendicular to the Galactic plane ranging from 4 to $7 \times 10^{13}$ cm$^{-2}$, which is a reasonable fraction (25 to 45%) of the total amount of O VI observed along the typical halo line of sight with N(O VI)sin|b| = $1.6 \times 10^{14}$ cm$^{-2}$. However, the amount of C IV predicted to occur in such remnants is substantially less than the amount observed by factors of 5-10. Shelton's simulation illustrates that careful considerations of all the possible processes for creating hot gas will ultimately be required to determine the relative importance of each one.

The high degree of irregularity of the distribution of the O VI that extends from sub-degree scales (Howk et al 2002b) to the larger angular scales sampled in this survey imply a great deal of complexity in the distribution of Galactic O VI. The small angular scale variability toward the LMC and SMC (Howk et al. 2002b) shows that the O VI is evidently found in small-scale complex cloud-like or filament-like structures rather than in a more diffuse medium filling a substantial fraction of the volume of the space. However, these small scale structures must be common in order to explain the high frequency of occurrence of large columns of thick disk O VI toward the extragalactic sources observed in our survey program. Evidence for a small filling factor for the O VI follows from the interpretations of O VI emission/absorption observations (Shelton et al. 2001; Dixon et al. 2001). The implications for substantial complexity of the distribution of OVI suggests that experiments designed to image O VI emission will produce a wealth of information about the small and large scale phenomena responsible for the origin of Galactic O VI.

### 10.3. Insights from Observations of Other Galaxies

Measurements of O VI absorption along complete paths through galaxy halos now exist for the Milky Way, LMC (Howk et al. 2002a, 2000b) , and SMC (Hoopes et al. 2002). The results are summarized in Table 10, where we list the average values of log [N(O VI)sin|b|], the average value of b (km s$^{-1}$), and the oxygen abundance. The table also contains estimates of Galactic fountain flow velocities and mass flow rates (see below) and estimates of the star formation rate per unit area in each galaxy.

The values of log [N(O VI)sin|b|] for the three galaxies increase by only a factor of 2.6 in going from the southern Milky Way to the SMC. In contrast the oxygen abundance decreases by a factor of 6.2 times while the star formation rate per unit area of the disk increases by 3 times. Edgar & Chevalier (1986) have calculated the column densities of highly ionized gas cooling from T > $10^6$ K with solar abundances. The expected column density of O VI in a column of cooling gas is proportional to $A_O \dot{N} t_{cool}$ where $A_O$ is the abundance of oxygen with respect to hydrogen, $\dot{N}$ is the flux of cooling gas (hydrogen ions cm$^{-2}$s$^{-1}$), and $t_{cool}$ is the cooling time of the gas. It is often assumed for departures from solar abundances that $t_{cool} \propto n_H^{-1} A_Z^{-1}$, where $n_H$ is the initial ionized hydrogen density in the cooling gas and $A_Z$ is the mean metallicity of the dominant coolants in the gas. This relation assumes that the cooling is dominated by the radiative emission produced by the heavy elements. Since oxygen is the most important coolant we take $A_Z = A_O$. Therefore, with these assumptions, N(O VI) $\propto \dot{N} / n_H$ which is independent of the oxygen abundance. The calculations of Sutherland & Dopita (1993) in their Figure 8 show that for a decrease of elemental abundances from solar to 0.1 solar, the cooling function for a hot plasma with T $\sim 3 \times 10^5$ K , $\Lambda$ (ergs cm$^3$ s$^{-1}$) decreases by only a factor of 3. This is probably because Sutherland & Dopita (1993) allowed O/Fe to increase by a factor of 3.2, while Fe/H and the other heavy elements were decreased in abundance by a factor of 10. Using the recent cooling function calculations of Benjamin et al. (2001), the reduction factor is 7 when all the elemental abundances are reduced by a factor of 10.

The dependence of the cooling time on abundance and density therefore scales more like $t_{cool} \propto n_H^{-1} A_Z^{-\beta}$ with $\beta = 0.85$ if we use the Benjamin et al. (2001) cooling function and maintain the elemental abundances in the solar ratios. Therefore, the dependence of the O VI column density on the various parameters should scale as N(O VI) $\propto$ $[A_O / A_O(solar)]^{1-\beta} \dot{N} n_H^{-1}$, where we have assumed the abundance of oxygen can be taken as a measure of the abundance of the important coolants.

For solar abundances, Edgar & Chevalier (1986) estimate that N(O VI) cm$^{-2}$ = $3.8 \times 10^{14}$ cm$^{-2}$ [v / 100 km s$^{-1}$]. This assumes the gas is cooling isobarically from $10^6$ K. If the gas cools from higher temperatures or isochorically the estimate of N(O VI) increases by factors of several. To allow for sub-solar abundances in the cooling gas we modify this expression to be N(O VI) = $(3.8 \times 10^{14}$ cm$^{-2})$ [v / 100 km s$^{-1}$] $[A_O / A_O(solar)]^{1-\beta}$. Therefore, from the observed value of N(O VI) or N(O VI)sin|b| we can estimate the cooling gas flow velocity, v = $\dot{N}$ $n_H^{-1}$, provided we know the oxygen abundance. From the flow velocity we can derive the mass flow rate per unit area from $\dot{M} / \Omega = 1.4 m_H n_H \dot{N} n_H^{-1}$ for He/H = 0.1.

Values of v (km s$^{-1}$) = $\dot{N}$ $n_H^{-1}$ and $\dot{M} / \Omega$ $(n_H / 10^{-3}$ cm$^{-3})$ $(M_O$ yr$^{-1}$ kpc$^{-2})$ for each galaxy are given in Table 10 assuming $\beta = 0.85$. The values $\dot{M} / \Omega$ scale with the ionized gas density in the cooling flow, $n_H$, assumed to be $10^{-3}$ cm$^{-3}$. The derived values of v = $\dot{N}$ $n_H^{-1}$ and $\dot{M} / \Omega$ $(n_H / 10^{-3}$ cm$^{-3})$ increase from the Milky Way, to the LMC, and SMC by about a factor of 2-3, which is similar to that found for the star formation rate per unit area and should be a measure of the hot gas heating rate. While this trend appears consistent with the expectations, we note there is not a corresponding observed increase in the O VI line breadth, b, from the Milky Way to the Magellanic Clouds. Such an increase would be expected if the line width is a rough measure of the cooling gas flow velocity.

While Hoopes et al. (2002) found that the observed values of N(O VI) in the SMC roughly correlated with measures of the local star formation rates as inferred from H$\alpha$ and X-ray emission, Howk et al. (2002a) found no such relationship for the LMC. This result might be explained simply from the density dependence of the derived values of $\dot{M} / \Omega$. Eventually understanding the origins of the relationships (or lack thereof) among various ISM observables in galaxies is a topic that will require much additional work in the future. However, even in the absence of that work, it is interesting to see that the integrated values of N(O VI) through the halo regions of three very different galaxies are very similar and that the estimated values of $\dot{M} / \Omega$ are roughly similar to the observed values of the star formation rates per unit surface area in the three galaxies.

# 11. SUMMARY

FUSE far-UV spectra of 100 extragalactic objects and two halo stars are used to obtain measures of O VI $\lambda\lambda$1031.93, 1037.62 absorption along paths through the Milky Way halo. The entire set of observations is found in the O VI catalog paper (Wakker et al. 2002). High velocity O VI absorption with |v| > 100 km s$^{-1}$ is seen along ~60% of the sight lines and is considered by Sembach et al. (2002). In this paper we study the Milky Way thick-disk O VI absorption that covers the approximate velocity range from -100 to 100 km s$^{-1}$. Our results are summarized as follows:

1. We find a widespread but highly irregular distribution of O VI in the Galactic thick disk, implying the existence of substantial amounts of hot gas with T ~3x$10^5$ K. The integrated O VI logarithmic column density, log N(O VI), ranges from 13.85 to 14.78 with an average of 14.38 while the logarithmic values of the O VI column density perpendicular to the Galactic plane, log [N(O VI)sin|b|], range from 13.44 to 14.68 with an average of 14.21.

2. Averages of log [N(O VI)sin|b|] over Galactic latitude reveal a ~0.25 dex excess of O VI to the north Galactic polar region compared to the rest of the Galaxy.

3. The standard deviations for the observed values of log N(O VI) and log [N(O VI)sin|b|] are 0.18 and 0.23. The large irregularities in the distribution of the absorbing gas are found to be similar over angular scales extending from less than one to 180 degrees. The small angular scale absorbing structures must be common to explain the large observed sky covering factor of O VI toward extragalactic sources.

4. The observations are not well described by a simple symmetrical plane-parallel patchy distribution of O VI absorbing structures. The simplest departure from such a model that provides a better fit to the observations is a plane-parallel patchy absorbing layer with an average O VI mid-plane density of $n_o$(O VI) = $1.7 \times 10^{-8}$ atoms cm$^{-3}$ and a scale height of ~2.3 kpc combined with a ~0.25 dex excess of O VI absorbing gas in the northern Galactic polar region.

5. The O VI is poorly correlated with other ISM tracers of gas in the halo, including low and intermediate velocity H I, Hα emission from the warm ionized medium, and hot 0.25 keV X-ray emitting gas.

6. The O VI profiles range in velocity dispersion, b, from 30 to 99 km s$^{-1}$, with an average value of 61 km s$^{-1}$ and standard deviation of 15 km s$^{-1}$. The thermal Doppler component of the broadening cannot explain the large observed profile widths since gas at T ~ $3 \times 10^5$ K, the temperature at which O VI is expected to peak in abundance in collisional ionization equilibrium, has b(O VI) = 17.7 km s$^{-1}$. Inflow, outflow, Galactic rotation, and turbulence therefore also affect the profiles.

7. In directions where the Galactic rotation effects are expected to be large, the imprint of Galactic rotation is visible in the O VI absorption.

8. The O VI average absorption velocities for thick disk gas toward high latitude objects (|b| > 45°) range from -37 to 82 km s$^{-1}$, with a high latitude sample average of 0 km s$^{-1}$ and a standard deviation of 21 km s$^{-1}$. Thick disk O VI is observed to be moving both toward and away from the plane with roughly equal frequency.

9. The broad high positive velocity O VI absorption wings extending from ~100 to ~250 km s$^{-1}$ seen in the spectra of 22 objects may be tracing the outflow of gas into the halo, although we can not rule out a more distant origin.

10. A combination of models involving the radiative cooling of hot gas in a Galactic fountain flow, the cooling of hot gas in halo supernova bubbles, and the turbulent mixing of warm and hot halo gases appears to be required to explain the highly ionized atoms found in the halo. If the origin of the O VI is dominated by a fountain flow, a mass flow rate of approximately 1.4 $M_O$ yr$^{-1}$ to each side of the Galactic disk for cooling hot gas with an average density of $10^{-3}$ cm$^{-3}$ is required to explain the average value of log [N(O VI)sin|b|] found in the southern Galactic hemisphere. Such a flow rate is similar to that estimated for the Galactic IVCs but much larger than estimated for the Galactic HVCs. The preferential venting of hot gas into the northern Galactic polar region may explain the ~0.25 dex excess in the values of log [N(O VI)sin|b|] in that direction.

11. The observed values of log[N(O VI)sin|b|] and b (km s$^{-1}$) for gas in the halos of the Milky Way, LMC, and SMC are quite similar even though the three galaxies have very different oxygen abundances and star formation rates. A similarity in the values of log[N(O VI)sin|b|] is an expected aspect of a Galactic fountain flow where the hot gas cooling rate varies by factors of 2 to 3 and the oxygen abundance changes by factors of 6.

We thank W. Sanders for discussions regarding the soft X-ray background. We thank L. M. Haffner for providing the values of I(Hα) in the direction of each survey object based on diffuse Hα emission observations with the Wisconsin Hα Mapper Experiment. R. Benjamin provided helpful comments regarding the non-equilibrium radiative cooling of hot plasmas. This work is based on data obtained for the Guaranteed Time Team by NASA-CNES-CSA FUSE mission operated by Johns Hopkins University. Financial support to U.S. participants has been provided by NASA contract NAS5-32985. K.R.S. acknowledges additional financial support through NASA Long Term Space Astrophysics grant NAG5-3485. B.P. W. acknowledges additional support from NASA grants NAG5-9179, NAG5-9024, and NGG5-8967. B.D.S. acknowledges the opportunity to visit the Institut d'Astrophysique Paris, France during the fall of 2001 and the Center for Astrophysics and Space Astronomy in Boulder, CO during the spring of 2002. Much of the preparation of this paper occurred at these institutions during a sabbatical leave from the University of Wisconsin-Madison.

**TABLE 1**
**THE HIGH STATES OF IONIZATION ACCESSIBLE IN THE UV**

| ION | $\lambda^a$ (Å) | $\log (\lambda f)^a$ | $IP^b$ ($X^{i-1} \rightarrow X^i$) (eV) | $IP^b$ ($X^i \rightarrow X^{i+1}$) (eV) | $T_{MAX}^c$ (K) | Reference disk | halo |
|-----|------|------|------|------|------|------|------|
| O VI | 1031.93 | 2.136 | 113.9 | 138.1 | $3 \times 10^5$ | 1, 2 | 3,4,5 |
|  | 1037.62 | 1.835 |  |  |  |  |  |
| N V | 1238.82 | 2.288 | 77.5 | 97.9 | $2 \times 10^5$ | 6,7,8,9,10 | 6,7,8,9,10 |
|  | 1242.80 | 1.986 |  |  |  |  |  |
| C IV | 1548.20 | 2.470 | 47.9 | 64.5 | $1 \times 10^5$ | 6,7,8,9,10 | 6,7,8,9,10 |
|  | 1550.77 | 2.169 |  |  |  |  |  |
| Si IV | 1393.76 | 2.855 | 33.5 | 45.1 | $0.8 \times 10^5$ | 6,7,8,9,10 | 6,7,8,9,10 |
|  | 1402.77 | 2.554 |  |  |  |  |  |

[a] Values of $\lambda$ and $\log(\lambda f)$ are from Morton (1991).

[b] The energy in eV required to create and destroy the ion $X^i$ are listed.

[c] The temperature at which the species peaks in abundance under conditions of collisional ionization equilibrium (Sutherland & Dopita 1993).

References to papers reporting observations of the highly ionized species in the disk and halo include: (1) Jenkins 1978a; (2) Jenkins 1978b; (3) Widmann et al. 1998; (4) Hurwitz et al. 1998; (5) Savage et al. 2000; (6) Savage & Massa 1987; (7) Sembach & Savage 1992; (8) Savage, Sembach, & Lu 1997; (9) Savage, Meade, & Sembach 2001; (10) Sembach, Savage, & Tripp 1997.

## TABLE 2
## RESULTS FOR O VI IN THE THICK DISK OF THE GALAXY[a]

| OBJECT (1) | $l$ (deg) (2) | $b$ (deg) (3) | z/v (4) | Q (5) | $v_-$ (km s⁻¹) (6) | $v_+$ (km s⁻¹) (7) | $\bar{v}_{exp}$ (km s⁻¹) (8) | $\bar{v}_{obs}$ (km s⁻¹) (9) | $\sigma_v$ (10) | $b$ (km s⁻¹) (11) | $\sigma_b$ (12) | log N(O VI) (cm⁻²) (13) | $\sigma_c$ (14) | $\sigma_{sys}$ (15) | logNsin\|b\| (cm⁻²) (16) |
|---|---|---|---|---|---|---|---|---|---|---|---|---|---|---|---|
| PG 1352+183 | 4.37 | 72.87 | 0.1519 | 1 | -80 | 100 | 1 | 14 | 10 | 59 | 6 | 14.39 | 0.11 | 0.02 | 14.37 |
| PKS 2155-304 | 17.73 | -52.25 | 0.116 | 4 | -85: | 120 | 14 | 24 | 10 | 67: | 3 | 14.34 | 0.01 | 0.05 | 14.24 |
| PG 1404+226 | 21.48 | 72.37 | 0.098 | 1 | -95 | 75 | 3 | -16 | 10 | 59 | 5 | 14.33 | 0.12 | 0.03 | 14.31 |
| NGC 5548 | 31.96 | 70.50 | 5149 | 2 | -120 | 60 | 4 | -32 | 10 | 59 | 4 | 14.50 | 0.05 | 0.02 | 14.47 |
| PG 1402+261 | 32.96 | 73.46 | 0.164 | 2 | -120 | 115 | 3 | -11 | 10 | 73 | 4 | 14.53 | 0.05 | 0.02 | 14.51 |
| Mrk 509 | 35.97 | -29.86 | 0.0343 | 4 | -100 | 100 | 38 | 21 | 10 | 64 | 3 | 14.66 | 0.01 | 0.02 | 14.46 |
| vZ 1128 | 42.50 | 78.68 | ... | 4 | -130 | 80 | 2 | -25 | 10 | 54 | 3 | 14.48 | 0.01 | 0.02 | 14.47 |
| MRC 2251-178 | 46.20 | -61.33 | 0.066 | 3 | -65 | 70 | 10 | -1 | 11 | 51 | 4 | 14.10 | 0.07 | 0.10 | 14.04 |
| PHL 1811 | 47.46 | -44.81 | 0.1919 | 3 | -65 | 100 | 22 | 24 | 10 | 53 | 3 | 14.45 | 0.04 | 0.03 | 14.30 |
| Mrk 506 | 54.18 | 31.50 | 0.043 | 1 | -100: | 20 | 21 | -36 | 10 | 37: | 2 | 14.47 | 0.07 | 0.03 | 14.19 |
| Mrk 829 | 58.76 | 63.25 | 1215 | 2 | -100 | 40 | 7 | -27 | 10 | 48 | 3 | 14.28 | 0.09 | 0.04 | 14.23 |
| Mrk 478 | 59.24 | 65.03 | 0.079 | 3 | -115 | 80 | 6 | -19 | 10 | 65 | 4 | 14.47 | 0.04 | 0.03 | 14.43 |
| 3C 382.0 | 61.31 | 17.45 | 0.0578 | 1 | -130 | 110 | -5 | -14 | 11 | 99 | 4 | 14.55 | 0.07 | 0.03 | 14.03 |
| Mrk 501 | 63.60 | 38.86 | 0.0336 | 2 | -100: | 20 | 11 | -38 | 11 | 49: | 3 | 14.27 | 0.07 | 0.07 | 14.07 |
| Mrk 1513 | 63.67 | -29.07 | 0.0629 | 3 | -75 | 80 | 7 | 6 | 10 | 50 | 4 | 14.34 | 0.05 | 0.02 | 14.03 |
| Mrk 926 | 64.09 | -58.76 | 0.0473 | 1 | -65: | 100: | 7 | ... | .. | ... | ... | <14.23 | ... | ... | <14.16 |
| PG 1444+407 | 69.90 | 62.72 | 0.2673 | 2 | -100 | 100 | 4 | -22 | 11 | 74 | 4 | 14.50 | 0.07 | 0.03 | 14.45 |
| Mrk 304 | 75.99 | -34.22 | 0.0657 | 3 | -40 | 75 | -4 | 9 | 10 | 42 | 4 | 14.20 | 0.08 | 0.05 | 13.95 |
| SBS 1415+437 | 81.96 | 66.20 | 609 | 2 | -50 | 100 | 0 | 27 | 10 | 54 | 4 | 14.14 | 0.12 | 0.03 | 14.10 |
| NGC 7469 | 83.10 | -45.47 | 4892 | 3 | -65 | 45 | -5 | -13 | 10 | 38 | 4 | 13.96 | 0.06 | 0.06 | 13.81 |
| PG 1411+442 | 83.83 | 66.35 | 0.0895 | 2 | -95 | 90 | 0 | -5 | 10 | 58 | 4 | 14.57 | 0.05 | 0.03 | 14.53 |
| PG 1626+554 | 84.51 | 42.19 | 0.1329 | 2 | -65: | 85: | -7 | 7 | 10 | 47: | 4 | 14.25 | 0.09 | 0.08 | 14.08 |
| PG 1415+451 | 84.72 | 65.32 | 0.1139 | 1 | -80 | 100 | 0 | 13 | 11 | 52 | 5 | 14.32 | 0.10 | 0.04 | 14.28 |
| Mrk 487 | 87.84 | 49.03 | 665 | 2 | -55 | 80 | -6 | 24 | 11 | 73 | 4 | 14.25 | 0.11 | 0.04 | 14.13 |
| NGC 7714 | 88.22 | -55.56 | 2798 | 2 | -60 | 45 | -3 | -2 | 10 | 35 | 2 | 13.85 | 0.14 | 0.05 | 13.77 |
| Mrk 290 | 91.49 | 47.95 | 0.0295 | 1 | -70: | 165 | -8 | 24 | 15 | 91: | 10 | 14.21 | 0.12 | 0.05 | 14.08 |
| PG 2349-014 | 91.66 | -60.36 | 0.174 | 1 | -70: | 70: | -3 | ... | .. | ... | ... | <13.97 | ... | ... | <13.91 |
| UGC 12163 | 92.14 | -25.34 | 0.0246 | 2 | -100 | 100 | -32 | 1 | 12 | 81 | 7 | 14.00 | 0.12 | 0.06 | 13.63 |
| H 1821+643 | 94.00 | 27.42 | 0.2844 | 4 | -160 | 115 | -30 | -46 | 10 | 89 | 3 | 14.48 | 0.02 | 0.02 | 14.14 |
| Mrk 876 | 98.27 | 40.38 | 0.1289 | 4 | -100: | 85 | -18 | -12 | 11 | 64: | 6 | 14.43 | 0.02 | 0.05 | 14.24 |
| Mrk 817 | 100.30 | 53.48 | 0.0314 | 4 | -90: | 60 | -9 | -30 | 11 | 52: | 5 | 14.38[b] | 0.01 | 0.06 | 14.29 |
| Mrk 335 | 108.76 | -41.42 | 0.0256 | 4 | -75: | 90 | -21 | 5 | 10 | 48: | 3 | 14.07 | 0.03 | 0.04 | 13.89 |
| Mrk 59 | 111.54 | 82.12 | 1089 | 3 | -75 | 75 | 0 | -11 | 10 | 45 | 4 | 14.29 | 0.04 | 0.02 | 14.28 |
| PG 1351+640 | 111.89 | 52.02 | 0.0882 | 3 | -100: | 100: | -13 | -11 | 10 | 81: | 4 | 14.38 | 0.04 | 0.06 | 14.28 |
| Mrk 279 | 115.04 | 46.86 | 0.032 | 4 | -115: | 100 | -18 | -16 | 10 | 64: | 3 | 14.41 | 0.02 | 0.02 | 14.27 |
| PG 1259+593 | 120.56 | 58.05 | 0.4778 | 4 | -80: | 100 | -10 | 5 | 10 | 62: | 3 | 14.22[b] | 0.02 | 0.05 | 14.15 |
| Mrk 1502 | 123.75 | -50.18 | 0.0611 | 2 | -35 | 65 | -16 | 21 | 10 | 33 | 3 | 14.00 | 0.10 | 0.03 | 13.89 |
| PG 0052+251 | 123.91 | -37.44 | 0.155 | 2 | -70: | 70: | -26 | ... | .. | ... | ... | <13.97 | ... | ... | <13.75 |
| Mrk 352 | 125.03 | -31.01 | 4456 | 2 | -115 | 115 | -33 | -9 | 10 | 85 | 4 | 14.34 | 0.09 | 0.03 | 14.08 |
| Mrk 205 | 125.45 | 41.67 | 0.0708 | 1 | -100: | 100: | -22 | ... | .. | ... | ... | <14.19 | ... | ... | <14.01 |
| 3C 249.1 | 130.39 | 38.55 | 0.3114 | 2 | -75 | 65 | -25 | -17 | 11 | 51 | 4 | 14.32 | 0.07 | 0.04 | 14.11 |
| Mrk 357 | 132.20 | -39.14 | 0.0516 | 2 | -115 | 115 | -24 | 15 | 11 | 99 | 5 | 14.40 | 0.08 | 0.05 | 14.20 |
| NGC 588 | 133.34 | -31.39 | -174 | 2 | -125: | 100: | -30 | ... | .. | ... | ... | <14.07 | ... | ... | <13.79 |
| NGC 595 | 133.53 | -31.31 | -174 | 2 | -80: | 80: | -31 | ... | .. | ... | ... | <14.00 | ... | ... | <13.72 |
| Mrk 209 | 134.15 | 68.08 | 281 | 3 | -115 | 130 | -5 | 0 | 10 | 78 | 5 | 14.43 | 0.04 | 0.02 | 14.40 |
| PG 0804+761 | 138.28 | 31.03 | 0.1020 | 4 | -120 | 80 | -30 | -18 | 10 | 62 | 4 | 14.51 | 0.02 | 0.02 | 14.22 |
| Ton S180 | 139.00 | -85.07 | 0.0619 | 3 | -90 | 90 | 0 | -5 | 11 | 55 | 5 | 14.37 | 0.04 | 0.03 | 14.37 |
| HS 0624+6907 | 145.71 | 23.35 | 0.3699 | 2 | -85 | 90 | -33 | -6 | 11 | 69 | 4 | 14.38 | 0.06 | 0.04 | 13.98 |
| PG 0832+675 | 147.75 | 35.01 | ... | 3 | -60 | 85 | -21 | 4 | 11 | 51 | 4 | 14.26 | 0.06 | 0.03 | 14.02 |
| VIIZw 118 | 151.36 | 25.99 | 0.0796 | 4 | -95 | 50 | -26 | -19 | 10 | 54 | 3 | 14.13 | 0.05 | 0.05 | 13.77 |
| MS 0700.7+6338 | 152.47 | 25.63 | 0.1529 | 1 | -65 | 85 | -26 | 9 | 10 | 41 | 4 | 14.16 | 0.12 | 0.03 | 13.80 |
| NGC 4151 | 155.08 | 75.06 | 995 | 3 | -35 | 100 | -1 | 21 | 11 | 42 | 4 | 14.24 | 0.04 | 0.03 | 14.23 |
| NGC 3310 | 156.60 | 54.06 | 993 | 4 | -135 | 85 | -7 | -26 | 10 | 78 | 4 | 14.56 | 0.03 | 0.02 | 14.47 |
| Mrk 9 | 158.36 | 28.75 | 0.0398 | 3 | -95 | 85 | -19 | -9 | 10 | 58 | 5 | 14.37 | 0.04 | 0.02 | 14.05 |
| Mrk 116 | 160.53 | 44.84 | 751 | 3 | -125 | 110 | -9 | 9 | 10 | 83 | 4 | 14.21 | 0.05 | 0.04 | 14.06 |
| Mrk 106 | 161.14 | 42.88 | 0.1234 | 1 | -100: | 100: | -10 | 1 | 10 | 70: | 5 | 14.45 | 0.08 | 0.04 | 14.28 |
| Mrk 79 | 168.60 | 28.38 | 0.0221 | 2 | -40 | 100 | -10 | 21 | 10 | 44 | 3 | 14.41 | 0.08 | 0.04 | 14.09 |
| NGC 1068 | 172.10 | -51.93 | 1137 | 4 | -115 | 60 | -2 | -8 | 11 | 57 | 7 | 13.98 | 0.04 | 0.03 | 13.87 |
| PG 0953+414 | 179.79 | 51.71 | 0.234 | 4 | -125 | 100: | 0 | 5 | 10 | 77: | 4 | 14.45 | 0.03 | 0.04 | 14.34 |
| Mrk 421 | 179.83 | 65.03 | 0.03 | 4 | -130 | 100 | 0 | 0 | 10 | 79: | 4 | 14.39[b] | 0.03 | 0.03 | 14.35 |
| PG 0985 | 180.84 | -59.49 | 0.0431 | 3 | -95 | 90 | 0 | 4 | 10 | 62 | 4 | 14.34 | 0.04 | 0.02 | 14.28 |
| PG 0947+396 | 182.85 | 50.75 | 0.2059 | 2 | -65 | 100: | 1 | 26 | 10 | 58: | 4 | 14.54 | 0.07 | 0.04 | 14.43 |
| NGC 3991 | 185.68 | 77.20 | 3192 | 1 | -95 | 115 | 0 | -3 | 11 | 71 | 5 | 14.47 | 0.11 | 0.02 | 14.46 |
| Ton 1187 | 188.33 | 55.38 | 0.0699 | 2 | -90 | 65 | 3 | -4 | 10 | 51 | 5 | 14.35 | 0.08 | 0.02 | 14.27 |
| HS 1102+3441 | 188.56 | 66.22 | 0.5099 | 1 | -140 | 95: | 2 | -21 | 10 | 74: | 5 | 14.71 | 0.07 | 0.03 | 14.67 |
| PG 0844+349 | 188.57 | 37.97 | 0.064 | 4 | -85 | 120 | 6 | 29 | 11 | 70 | 4 | 14.32 | 0.04 | 0.02 | 14.30 |
| SBS 0335-052 | 191.34 | -44.69 | 4043 | 2 | -70: | 70: | 6 | ... | .. | ... | ... | <13.99 | ... | ... | <13.84 |
| PG 0832+251 | 199.49 | 33.15 | 0.3309 | 1 | -90 | 120 | 16 | 4 | 11 | 77 | 4 | 14.51 | 0.07 | 0.03 | 14.25 |
| PG 1001+291 | 200.09 | 53.20 | 0.3297 | 1 | -70 | 100: | 8 | 25 | 10 | 59: | 4 | 14.50 | 0.06 | 0.05 | 14.48 |
| HE 0238-1904 | 200.48 | -63.63 | 0.6309 | 1 | -120 | 110 | 5 | 0 | 11 | 66 | 4 | 14.33 | 0.09 | 0.02 | 14.28 |
| Mrk 1095 | 201.69 | -21.13 | 0.0322 | 3 | -35 | 50 | 25 | 5 | 10 | 31 | 3 | 13.89 | 0.07 | 0.08 | 13.45 |
| Mrk 36 | 201.76 | 66.49 | 614 | 1 | -120 | 110 | 4 | 19 | 13 | 90 | 6 | 14.36 | 0.11 | 0.03 | 14.32 |
| NGC 3504 | 204.60 | 66.04 | 1534 | 1 | -70: | 70: | 5 | ... | .. | ... | ... | <14.19 | ... | ... | <14.15 |

| (1) | (2) | (3) | (4) | (5) | (6) | (7) | (8) | (9) | (10) | (11) | (12) | (13) | (14) | (15) | (16) |
|---|---|---|---|---|---|---|---|---|---|---|---|---|---|---|---|
| PKS 0405-12 | 204.93 | -41.76 | 0.5725 | 4 | -80 | 100 | 14 | 12 | 10 | 60 | 5 | 14.14 | 0.04 | 0.04 | 13.96 |
| Mrk 618 | 206.72 | -34.66 | 0.0355 | 2 | -50 | 45 | 20 | -2 | 10 | 30 | 3 | 14.27 | 0.07 | 0.04 | 14.02 |
| NGC 4670 | 212.69 | 88.63 | 1069 | 3 | -100 | 105 | 0 | 5 | 10 | 59 | 3 | 14.51 | 0.03 | 0.01 | 14.51 |
| PG 1116+215 | 223.36 | 68.21 | 0.1764 | 4 | -55 | 115: | 6 | 32 | 10 | 53: | 3 | 14.26 | 0.03 | 0.03 | 14.23 |
| Ton S210 | 224.97 | -83.16 | 0.116 | 4 | -120 | 95 | 1 | -25 | 10 | 73 | 4 | 14.60 | 0.02 | 0.02 | 14.60 |
| PG 1004+130 | 225.12 | 49.12 | 0.2399 | 3 | -130 | 100 | 17 | -12 | 10 | 83 | 3 | 14.63 | 0.03 | 0.02 | 14.51 |
| HE 0450-2958 | 231.13 | -37.59 | 0.286 | 2 | -85: | 85: | 27 | ... | ... | ... | ... | <14.00 | ... | ... | <13.79 |
| NGC 1399 | 236.72 | -53.63 | 1425 | 2 | -50 | 70 | 14 | 0 | 10 | 38 | 6 | 14.10 | 0.11 | 0.06 | 14.01 |
| Mrk 734 | 244.75 | 63.94 | 0.0502 | 2 | -35 | 140: | 7 | 82 | 11 | 60: | 3 | 14.55 | 0.06 | 0.05 | 14.50 |
| HE 0226-4110 | 253.94 | -65.77 | 0.4949 | 4 | -140 | 75 | 5 | -33 | 10 | 76 | 3 | 14.25 | 0.03 | 0.02 | 14.21 |
| PKS 0558-504 | 257.96 | -28.57 | 0.137 | 3 | -115 | 135 | 34 | 11 | 10 | 76 | 5 | 14.42 | 0.03 | 0.02 | 14.10 |
| NGC 1705 | 261.08 | -38.74 | 628 | 4 | -120 | 120 | 21 | 29 | 11 | 70 | 5 | 14.39 | 0.02 | 0.09 | 14.19 |
| PG 1211+143 | 267.55 | 74.32 | 0.0816 | 3 | -100 | 100 | 1 | 10 | 10 | 51 | 5 | 14.17 | 0.06 | 0.02 | 14.15 |
| Mrk 771 | 269.44 | 81.74 | 0.063 | 1 | -10 | 120 | 0 | 41 | 10 | 39 | 7 | 14.19 | 0.16 | 0.07 | 14.18 |
| HE 1115-1735 | 273.65 | 39.64 | 0.217 | 1 | -30: | 90: | 12 | ... | ... | ... | ... | <13.92 | ... | ... | <13.72 |
| IRASF 11431-1810 | 281.85 | 41.71 | 0.0329 | 3 | -75 | 100: | 3 | 38 | 11 | 57: | 4 | 14.56 | 0.04 | 0.05 | 14.38 |
| ESO 265-G23 | 285.91 | 16.59 | 0.0559 | 1 | -20 | 100 | 30 | 48 | 10 | 43 | 4 | 14.32 | 0.11 | 0.05 | 13.78 |
| ESO 572-G34 | 286.12 | 42.12 | 1114 | 3 | -100 | 100: | 0 | 22 | 11 | 72: | 5 | 14.47 | 0.04 | 0.06 | 14.30 |
| 3C 273.0 | 289.95 | 64.36 | 0.1583 | 4 | -110 | 105 | -2 | 6 | 10 | 58 | 3 | 14.73[b] | 0.01 | 0.01 | 14.68 |
| Fairall 9 | 295.07 | -57.83 | 0.047 | 2 | -110 | 100 | -6 | 24 | 11 | 79: | 4 | 14.38 | 0.07 | 0.06 | 14.31 |
| NGC 4649 | 295.88 | 74.34 | 1117 | 1 | -105 | 70 | -1 | -37 | 12 | 71 | 5 | 14.31 | 0.13 | 0.04 | 14.29 |
| Tol 1247-232 | 302.60 | 39.30 | 0.0479 | 2 | -50 | 100: | -16 | 37 | 10 | 52: | 4 | 14.40 | 0.08 | 0.05 | 14.20 |
| PG 1302-102 | 308.59 | 52.16 | 0.2783 | 4 | -65 | 70 | -14 | 7 | 10 | 40 | 3 | 14.23 | 0.04 | 0.03 | 14.13 |
| PG 1307+085 | 316.79 | 70.71 | 0.155 | 3 | -115 | 80 | -4 | -19 | 10 | 60 | 3 | 14.45 | 0.05 | 0.02 | 14.42 |
| HE 1326-0516 | 320.07 | 56.07 | 0.5779 | 1 | -100: | 100: | -13 | ... | ... | ... | ... | <14.14 | ... | ... | <14.06 |
| ESO 141-G55 | 338.18 | -26.71 | 0.036 | 4 | -100 | 85 | -41 | 3 | 10 | 52 | 3 | 14.50 | 0.02 | 0.02 | 14.15 |
| Mrk 1383 | 349.22 | 55.12 | 0.0864 | 4 | -100 | 100: | -7 | -3 | 10 | 55: | 3 | 14.58[b] | 0.02 | 0.01 | 14.49 |
| PKS 2005-489 | 350.37 | -32.60 | 0.0709 | 3 | -100 | 120: | -28 | 38 | 11 | 67: | 4 | 14.78 | 0.02 | 0.03 | 14.51 |
| Tol 1924-416 | 356.94 | -24.10 | 2884 | 2 | -70 | 95 | -20 | 10 | 10 | 50 | 4 | 14.62 | 0.05 | 0.02 | 14.23 |
| <LMC>[c] | 278.66 | -33.33 | 270 | 4 | -36 | 175 | ... | ... | ... | ... | ... | 14.51[c] | 0.13[c] | ... | 14.25 |
| <SMC>[c] | 302.10 | -44.43 | 150 | 4 | -45 | 63 | ... | ... | ... | ... | ... | 14.10[c] | 0.17[c] | ... | 13.94 |

[a] The various entries in this table include:(1) object name; (2) Galactic latitude; (3) Galactic longitude; (4) redshift, z, or radial velocity, v (km s$^{-1}$); (5) data quality (see §2); (6) negative velocity integration limit; (7) positive velocity integration limit; (8) expected average velocity of the O VI absorption assuming a co-rotating halo with an exponential scale height of 2.4 kpc and a turbulent velocity of 60 km s$^{-1}$; (9) average observed LSR velocity of the O VI absorption; (10) 1σ random error in the observed velocity, the systematic velocity error is ~ 10 km s$^{-1}$; (11) observed O VI velocity dispersion (see Eqn. 4); (12) 1σ random error in the velocity dispersion; (13) log of O VI column density for the velocity range v. to v$_*$based on the strong line of the doublet; (14) 1σ combined statistical and continuum placement errors in log N(O VI); (15) systematic error in log N(O VI); (16) log of projected O VI column density.

[b] Wakker et al. (2002) summarize the results of measures of N(O VI)$_W$/N(O VI)$_S$ for objects with S/N > 6.5 (Q = 3 or 4) for which the continuum near O VI $\lambda$1037.62 is reliable. Here, N(O VI)$_W$ and N(O VI)$_S$ refer to values of N$_a$(O VI) derived from the weak and strong component of the O VI doublet, respectively. The integrations extend over a limited velocity range to avoid the contamination problems affecting the O VI $\lambda$1037.62 line. If there is little or no unresolved line saturation the value of N(O VI)$_W$/N(O VI)$_S$ should be close to 1.0 (Savage & Sembach 1991). Wakker et al. (2002) find that for 15 of 20 objects N(O VI)$_W$/N(O VI)$_S$ is within 1σ of 1.0 implying little or no line saturation. For 5 objects the values of N(O VI)$_W$/N(O VI)$_S$ suggest some saturation. The results for those six cases are listed below.

| Object | N(O VI)$_W$/N(O VI)$_S$ | Δlog N(corr) |
|---|---|---|
| Mrk 421 | 1.60±0.18 | 0.57 |
| PG 1259+593 | 1.29±0.11 | 0.24 |
| Mrk 817 | 1.28±0.07 | 0.23 |
| Mrk 1383 | 1.23±0.08 | 0.19 |
| 3C 273 | 1.13±0.04 | 0.11 |

Here, Δlog N(corr), is the saturation correction that must be applied to the value of log N$_S$ in Table 2 to approximately correct for the presence of unresolved saturation. The correction was derived following the technique recommended by Savage & Sembach (1991; see their Table 4) but modified so that the correction can be applied to the strong line value of logN$_a$(O VI). The implied correction is large for Mrk 421 and relatively modest for the other four objects. In two cases (PG 1259+593 and Mrk 1383) the fitted continuua for the O VI $\lambda$1037.62 may be high. Alternate continuum choices would help to remove some, but not all, of the difference between the weak and strong line measures of N(O VI).

[c] Average values of log N(O VI) in the Milky Way halo in the directions of the LMC and SMC are listed along with the standard deviations. The measurements are from Howk et al. (2002a, 2002b) and Hoopes et al. (2002). For the direction to the LMC, values of log N(O VI) toward 12 LMC stars range from 14.22 to 14.67 for integrations extending over the LSR velocity range averaging from –36 to 175 km s$^{-1}$. For a narrower velocity range averaging from –50 to 50 km s$^{-1}$, the values of logN(O VI) range from 13.61 to 14.23 with an average of 14.02 and a standard deviation of 0.18. log N(O VI) in the Milky Way toward 11 SMC stars ranges from 13.77 to 14.39 and the velocity range of the Milky Way halo absorption on average extends from –45 to 63 km s$^{-1}$. The substantial dispersion in the observed values of log N(O VI) of 0.13 and 0.17 dex for Milky Way halo gas toward the LMC and SMC, respectively, shows that the irregularity in the distribution of Milky Way O VI extends to angular scales as small as 5° to 0.05° (Howk et al. 2002b).

**TABLE 3**
**O VI COLUMN DENSITY AVERAGES[a]**

| Quantity | Sample | Number | Median | Average | Stdev |
|----------|--------|--------|--------|---------|-------|
| log N(O VI) | full | 91 | 14.38 | 14.36 | 0.18 |
|  | Yes IVC | 53 | 14.39 | 14.38 | 0.17 |
|  | No IVC | 37 | 14.34 | 14.35 | 0.20 |
|  |  |  |  |  |  |
| log [N(O VI)sin\|b\|] | full | 91 | 14.23 | 14.21 | 0.23 |
|  | \|b\| = 15 to 45 | 38 | 14.09 | 14.08 | 0.21 |
|  | \|b\|= 45 to 60 | 22 | 14.27 | 14.19 | 0.22 |
|  | \|b\| = 60 to 90 | 31 | 14.36 | 14.36 | 0.16 |
|  | \|b\| = 45 to 90 | 53 | 14.29 | 14.29 | 0.20 |
|  | Yes IVC | 53 | 14.22 | 14.21 | 0.21 |
|  | No IVC | 37 | 14.27 | 14.21 | 0.25 |
|  |  |  |  |  |  |
| log [N(O VI)sin\|b\|] | b > 0° | 62 | 14.27 | 14.26 | 0.20 |
| " | b = 15 to 45° | 22 | 14.09 | 14.10 | 0.17 |
| " | b = 45 to 60° | 14 | 14.27 | 14.30 | 0.14 |
| " | b = 60 to 90° | 26 | 14.38 | 14.37 | 0.15 |
|  | b = 45 to 90° | 40 | 14.34 | 14.35 | 0.15 |
|  |  |  |  |  |  |
| log [N(O VI)sin\|b\|] | b < 0° | 29 | 14.09 | 14.09 | 0.25 |
| " | b = -15 to -45° | 16 | 14.09 | 14.06 | 0.26 |
| " | b = -45 to -60° | 8 | 13.94 | 14.02 | 0.22 |
| " | b = -60 to -90° | 5 | 14.28 | 14.29 | 0.20 |
|  | b = -45 to -90° | 13 | 14.20 | 14.12 | 0.25 |

[a] When computing these averages, the values of log N(O VI) for the 11 3σ upper limits were not included.

**TABLE 4**
**O VI KINEMATIC AVERAGES**

| Quantity | Sample | Number | Min | Max | Median | Average | Stdev |
|---|---|---|---|---|---|---|---|
| $\overline{v}_{obs}$ (km s$^{-1}$) | full | 91 | -46 | 82 | 3 | 2 | 21 |
| " | \|b\| = 15 to 45$^o$ | 38 | -46 | 48 | 5 | 5 | 21 |
| " | \|b\| > 45$^o$ | 53 | -37 | 82 | -2 | 0 | 22 |
| " | b > 45$^o$ | 40 | -37 | 82 | -3 | 0 | 23 |
| " | b < -45$^o$ | 13 | -33 | 24 | -1 | -1 | 17 |
| | | | | | | | |
| b (km s$^{-1}$) | full | 91 | 30 | 99 | 59 | 61 | 15 |
| " | b > 15$^o$ | 62 | 37 | 99 | 59 | 62 | 16 |
| " | b < -15$^o$ | 29 | 30 | 98 | 54 | 58 | 18 |
| " | \|b\| = 15 to 45$^o$ | 38 | 30 | 99 | 57 | 60 | 17 |
| " | \|b\| > 45$^o$ | 53 | 33 | 91 | 59 | 61 | 14 |
| " | b > 45$^o$ | 40 | 39 | 91 | 59 | 62 | 13 |
| " | b < -45$^o$ | 13 | 33 | 82 | 57 | 57 | 17 |
| " | Yes IVC | 53 | 30 | 99 | 60 | 62 | 16 |
| " | No IVC | 37 | 31 | 91 | 56 | 58 | 14 |

**TABLE 5**
**O VI ABSORPTION AND H I INTERMEDIATE VELOCITY CLOUDS[a]**

| Object | HI IVC[b] | v(IVC) (km s⁻¹) | Comment |
|--------|-----------|------------------|---------|
| Mrk 209 | IV4 and IV Arch | -100, -55 | OVI component at v(IVC) |
| PG 1351+640 | IV9 and IV19 | -74, -47 | OVI overlaps IVC |
| Mrk 279 | IV9 and LLIV | -76, -40 | OVI overlaps IVC |
| Mrk 487 | IV15 | -85 | No OVI at v(IVC) |
| PG 0953+414 | IV16 | -49 | OVI overlaps IVC |
| NGC 3991 | IV16 | -53 | OVI overlaps IVC |
| PG 0947+396 | IV16 and IV Arch | -66,-46 | OVI partly overlaps IVC |
| PG 1001+291 | IV18 | -28 | OVI overlaps IVC |
| NGC 3504 | IV18 | -52 | No OVI at v(IVC) |
| Mrk 36 | IV18 | -55 | OVI overlaps IVC |
| Mrk 876 | IV20 and Draco | -52, -30 | OVI overlaps IVC |
| Mrk 421 | IV26 | -61 | OVI overlaps IVC |
| NGC 4151 | IV 26 and IV Arch | -41, -29 | OVI partly overlaps IVC |
| HS 1102+3441 | IV Arch | -39 | OVI component at v(IVC) |
| Mrk 817 | IV Arch | -40 | OVI component at v(IVC) |
| Mrk 59 | IV Arch | -44 | OVI overlaps IVC |
| 3C 249.1 | IV Arch | -50 | OVI component at v(IVC) |
| PG 1259+593 | IV Arch | -54 | OVI overlaps IVC |
| NGC 3310 | IV Arch and IV Arch | -65, -47 | OVI overlaps IVC |
| PG 1211+143 | IV spur | -35 | No OVI at v(IVC) |
| Mrk 734 | IV spur | -42 | No OVI at v(IVC) |
| PG 1116+215 | IV spur | -42 | OVI partly overlaps IVC |
| Mrk 116 | LLIV | -39 | OVI overlaps IVC |
| Mrk 9 | LLIV | -40 | OVI overlaps IVC |
| Mrk 106 | LLIV | -40 | OVI overlaps IVC |
| Mrk 205 | LLIV | -48 | No OVI at v(IVC) |
| PG 0832+675 | LLIV | -56 | OVI partly overlaps IVC |
| PG 0804+761 | LLIV | -58 | OVI overlaps IVC |
| Ton 1187 | MIII and IV Arch | -104, -69 | No OVI at v(IVC) |
| Mrk 509 | Complex gp | 60 | OVI overlaps IVC |
| Mrk 478 | Complex K | -60 | OVI overlaps IVC |
| Mrk 501 | Complex K | -83 | OVI component at v(IVC) |
| Mrk 506 | Complex K | -90 | OVI overlaps IVC |
| 3C 273.0 | Other IVC | 25 | OVI overlaps IVC |
| Mrk 335 | Other IVC | -27 | OVI overlaps IVC |
| Mrk 352 | Other IVC | -28 | OVI overlaps IVC |
| Mrk 771 | Other IVC | -29 | No OVI at v(IVC) |
| vZ 1128 | Other IVC | 29 | OVI overlaps IVC |
| Mrk 618 | Other IVC | -37 | OVI partly overlaps IVC |
| NGC 588 | Other IVC | -38 | No OVI at v(IVC) |
| PG 0052+251 | Other IVC | -39 | No OVI at v(IVC) |
| ESO 265-G23 | Other IVC | 40 | OVI component at v(IVC) |
| Mrk 357 | Other IVC | -41 | OVI component at v(IVC) |
| HE 0238-1904 | Other IVC | -42 | OVI overlaps IVC |
| ESO 141-G55 | Other IVC | -45 | OVI overlaps IVC |

| NGC 7714 | Other IVC | -50 | No OVI at v(IVC) |
|---|---|---|---|
| UGC 12163 | Other IVC | -53 | No O VI at v(IVC) |
| Tol 1247-232 | Other IVC | 54 | OVI component at v(IVC) |
| IRAS F11431-1810 | Other IVC | 56 | OVI component at v(IVC) |
| MRC 2251-178 | Other IVC | 59 | OVI partly overlaps IVC |
| HE 1326-0516 | Other IVCs | -61, 39 | No OVI at v(IVC) |
| Mrk 304 | Other IVC | -63 | No OVI at v(IVC) |
| PG 1411+442 | Other IVC | 63 | OVI overlaps IVC |
| PKS 0558-504 | Other IVC | 71 | OVI overlaps IVC |
| NGC 1705 | Other IVC | 88 | OVI overlaps IVC |
| H 1821+643 | Outer Arm | -128 | OVI overlaps IVC |
| 3C 382.0 | Outer Arm | -82 | OVI overlaps IVC |
| HS 0624+6907 | Outer Arm | -92 | No OVI at v(IVC) |
| Mrk 926 | WW534 | -74 | No OVI at v(IVC) |

[a] Additional information about the IVCs toward the O VI survey objects including column densities are found in Wakker et al. (2002) and Wakker (2001).

[b] The IVC name designations listed here are explained more fully in Wakker (2001).

**TABLE 6**
**OBJECTS WITH POSITIVE VELOCITY**
**O VI ABSORPTION WINGS[a]**

| Object | l | b | $v_-$ | $v_+$ | log N(O VI) | $\sigma_{sc}$ | $\sigma_{sys}$ |
|---|---|---|---|---|---|---|---|
| Mrk 509 | 35.97 | -29.86 | 115 | 200 | 13.55 | 0.06 | 0.22 |
| PG 1626+554 | 84.51 | 42.19 | 85 | 210 | 14.08 | 0.11 | 0.07 |
| Mrk 876 | 98.27 | 40.38 | 85 | 140 | 13.23 | 0.13 | 0.31 |
| Mrk 817 | 100.30 | 53.48 | 60 | 140 | 13.30 | 0.08 | 0.21 |
| PG 1351+640 | 111.89 | 52.02 | 100 | 160 | 13.44 | 0.13 | 0.22 |
| PG 1259+593 | 120.56 | 58.05 | 100 | 185 | 13.06 | 0.15 | 0.27 |
| PG 0953+414 | 179.79 | 51.71 | 100 | 235 | 13.83 | 0.06 | 0.13 |
| Mrk 421 | 179.83 | 65.03 | 100 | 185 | 13.51 | 0.11 | 0.19 |
| PG 0947+396 | 182.85 | 50.75 | 100 | 175 | 14.19 | 0.10 | 0.11 |
| PG 0844+349 | 188.56 | 37.97 | 120 | 250 | 13.75 | 0.08 | 0.14 |
| HS 1102+3441 | 188.56 | 66.22 | 95 | 210 | 14.30 | 0.10 | 0.07 |
| PG 1001+291 | 200.09 | 53.20 | 100 | 200 | 13.97 | 0.13 | 0.16 |
| PKS 0405-12 | 204.93 | -41.76 | 100 | 210 | 13.75 | 0.07 | 0.09 |
| Mrk 734 | 244.75 | 63.94 | 140 | 275 | 14.10 | 0.10 | 0.16 |
| NGC 1705 | 261.08 | -38.74 | 120 | 245 | 13.78 | 0.04 | 0.41 |
| HE 1115-1735 | 273.65[b] | 39.64[b] | 90 | 175 | 14.13 | 0.10 | 0.06 |
| IRAS F11431-1810 | 281.85[b] | 41.71[b] | 100 | 210 | 14.25 | 0.06 | 0.12 |
| " | 281.85[b] | 41.71[b] | 210 | 320 | 13.85 | 0.11 | 0.08 |
| ESO 572-G34 | 286.12[b] | 42.12[b] | 100 | 275 | 14.47 | 0.03 | 0.06 |
| 3C 273.0 | 289.95[b] | 64.36[b] | 105 | 160 | 13.17 | 0.11 | 0.33 |
| " | 289.95[b] | 64.36[b] | 160 | 260 | 13.52 | 0.07 | 0.08 |
| Tol 1247-232 | 302.60[b] | 39.30[b] | 100 | 255 | 14.18 | 0.10 | 0.07 |
| Mrk 1383 | 349.22[b] | 55.12[b] | 100 | 160 | 13.42 | 0.09 | 0.18 |
| PKS 2005-489 | 350.37[b] | -32.60[b] | 120 | 225 | 13.92 | 0.06 | 0.17 |

[a] Additional discussions of these wings are found in Sembach et al. (2002). The values of $v_-$, $v_+$, logN(O VI), $\sigma_{sc}$, $\sigma_{sys}$ provide the velocity range of the integration, the observed logarithmic column density, the statistical+ continuum placement error, and the systematic error, respectively ( see §2).
[b] The directions to these objects lie within the boundaries of Radio Loops I and/or IV (see §9.4 and Fig. 9a).

## TABLE 7
## MEASURES OF H I, Hα, AND 0.25 KeV DIFFUSE X-RAYS
## TOWARD EACH SURVEY OBJECT

| name (1) | l (2) | b (3) | log N(O VI) (4) | $\sigma_c$ (5) | N(H I) (6) | σ(H I) (7) | I(Hα) (8) | σ(Hα) (9) | I(X) (10) | σ(IX) (11) |
|---|---|---|---|---|---|---|---|---|---|---|
| 3C 249.1 | 130.39 | 38.55 | 14.32 | 0.07 | 2.83E+20 | 6.00E+17 | 0.76 | 0.048 | 753 | 121 |
| 3C 273.0 | 289.95 | 64.36 | 14.73 | 0.01 | 1.69E+20 | 1.00E+18 | 0.83 | 0.039 | 1688 | 136 |
| 3C 382.0 | 61.31 | 17.45 | 14.55 | 0.07 | 6.22E+20 | 1.40E+18 | 3.27 | 0.04 | 518 | 77 |
| ESO 141-G55 | 338.18 | -26.71 | 14.50 | 0.02 | 6.15E+20 | 1.20E+18 | ... | ... | 604 | 229 |
| ESO 265-G23 | 285.91 | 16.59 | 14.32 | 0.11 | 8.61E+20 | 1.20E+18 | ... | ... | 737 | 115 |
| ESO 572-G34 | 286.12 | 42.12 | 14.47 | 0.04 | 2.75E+20 | 1.40E+18 | 0.75 | 0.031 | 743 | 93 |
| FAIRALL 9 | 295.07 | -57.83 | 14.38 | 0.07 | 2.41E+20 | 8.00E+17 | ... | ... | 929 | 117 |
| H 1821+643 | 94.00 | 27.42 | 14.48 | 0.02 | 3.84E+20 | 5.00E+17 | 1.85 | 0.041 | 691 | 29 |
| HE 0226-4110 | 253.94 | -65.77 | 14.25 | 0.03 | 2.14E+20 | 1.00E+18 | ... | ... | 1263 | 129 |
| HE 0238-1904 | 200.48 | -63.63 | 14.33 | 0.09 | 2.60E+20 | 5.00E+17 | 0.67 | 0.034 | 833 | 119 |
| HE 0450-2958 | 231.13 | -37.59 | <14.00 | ... | 1.22E+20 | 1.30E+18 | 2.58 | 0.037 | 1069 | 135 |
| HE 1115-1735 | 273.65 | 39.64 | <13.92 | ... | 3.71E+20 | 9.00E+17 | 0.91 | 0.038 | 683 | 105 |
| HE 1326-0516 | 320.07 | 56.07 | <14.14 | ... | 2.15E+20 | 1.00E+18 | 1.34 | 0.035 | 751 | 128 |
| HS 0624+6907 | 145.71 | 23.35 | 14.38 | 0.06 | 6.68E+20 | 1.10E+18 | 2.32 | 0.041 | 329 | 86 |
| HS 1102+3441 | 188.56 | 66.22 | 14.71 | 0.07 | 1.62E+20 | 1.20E+18 | 0.87 | 0.036 | 937 | 121 |
| IRAS F11431-1810 | 281.85 | 41.71 | 14.56 | 0.04 | 3.13E+20 | 4.00E+17 | 0.64[c] | ...[c] | 523 | 117 |
| MRC 2251-178 | 46.20 | -61.33 | 14.10 | 0.07 | 2.61E+20 | 1.20E+18 | 0.51 | 0.037 | 630 | 215 |
| Mrk 9 | 158.36 | 28.75 | 14.37 | 0.04 | 4.78E+20 | 5.00E+17 | 0.61 | 0.040 | 448 | 92 |
| Mrk 36 | 201.76 | 66.49 | 14.36 | 0.11 | 2.07E+20 | 1.20E+18 | 0.11 | 0.038 | 933 | 135 |
| Mrk 59 | 111.54 | 82.12 | 14.29 | 0.04 | 1.08E+20 | 8.00E+17 | 0.32 | 0.032 | 1074 | 103 |
| Mrk 79 | 168.60 | 28.38 | 14.41 | 0.08 | 5.29E+20 | 4.00E+17 | ... | ... | 464 | 84 |
| Mrk 106 | 161.14 | 42.88 | 14.45 | 0.08 | 2.44E+20 | 6.00E+17 | 0.69 | 0.040 | 738 | 113 |
| Mrk 116 | 160.53 | 44.84 | 14.21 | 0.05 | 3.02E+20 | 1.10E+18 | 0.65 | 0.041 | 740 | 102 |
| Mrk 205 | 125.45 | 41.67 | <14.19 | ... | 3.06E+20 | 2.00E+17 | 0.76 | 0.039 | 642 | 80 |
| Mrk 209 | 134.15 | 68.08 | 14.43 | 0.04 | 1.11E+20 | 4.00E+17 | 0.52 | 0.031 | 1106 | 109 |
| Mrk 279 | 115.04 | 46.86 | 14.41 | 0.02 | 1.80E+20 | 7.00E+17 | 0.48 | 0.045 | 1058 | 104 |
| Mrk 290 | 91.49 | 47.95 | 14.21 | 0.12 | 1.32E+20 | 2.00E+18 | 0.50 | 0.034 | 1110 | 79 |
| Mrk 304 | 75.99 | -34.22 | 14.20 | 0.08 | 4.94E+20 | 5.00E+17 | 1.80 | 0.037 | 335 | 80 |
| Mrk 335 | 108.76 | -41.42 | 14.07 | 0.03 | 3.64E+20 | 6.00E+17 | 0.78 | 0.045 | 443 | 116 |
| Mrk 352 | 125.03 | -31.01 | 14.37 | 0.09 | 5.19E+20 | 3.00E+17 | 1.13 | 0.043 | 368 | 79 |
| Mrk 357 | 132.20 | -39.14 | 14.40 | 0.08 | 5.06E+20 | 1.40E+18 | 0.98 | 0.036 | 396 | 77 |
| Mrk 421 | 179.83 | 65.03 | 14.39 | 0.03 | 1.51E+20 | 9.00E+17 | 0.58 | 0.035 | 2350 | 278 |
| Mrk 478 | 59.24 | 65.03 | 14.47 | 0.04 | 9.80E+19 | 7.00E+17 | 0.37 | 0.032 | 1145 | 260 |
| Mrk 487 | 87.84 | 49.03 | 14.25 | 0.11 | 1.26E+20 | 4.00E+17 | 0.73 | 0.033 | 1109 | 75 |
| Mrk 501 | 63.60 | 38.86 | 14.27 | 0.07 | 1.84E+20 | 7.00E+17 | 0.99 | 0.037 | 1136 | 125 |
| Mrk 506 | 54.18 | 31.50 | 14.47 | 0.07 | 2.88E+20 | 3.00E+17 | 1.41 | 0.037 | 716 | 98 |
| Mrk 509 | 35.97 | -29.86 | 14.66 | 0.01 | 4.29E+20 | 1.10E+18 | 1.40 | 0.039 | 785 | 117 |
| Mrk 618 | 206.72 | -34.66 | 14.27 | 0.07 | 4.82E+20 | 3.00E+17 | 13.01 | 0.047 | 1306 | 144 |
| Mrk 734 | 244.75 | 63.94 | 14.55 | 0.06 | 2.45E+20 | 6.00E+17 | 0.34 | 0.030 | 767 | 101 |
| Mrk 771 | 269.44 | 81.74 | 14.19 | 0.16 | 1.37E+20 | 9.00E+17 | 0.42 | 0.033 | 837 | 132 |
| Mrk 817 | 100.3 | 53.48 | 14.38 | 0.01 | 9.50E+19 | 9.00E+17 | 0.40 | 0.032 | 1661 | 116 |
| Mrk 829 | 58.76 | 63.25 | 14.28 | 0.09 | 9.80E+19 | 9.00E+17 | 0.50 | 0.034 | 894 | 98 |
| Mrk 876 | 98.27 | 40.38 | 14.43 | 0.02 | 2.40E+20 | 8.00E+17 | 0.96 | 0.038 | 907 | 73 |
| Mrk 926 | 64.09 | -58.76 | <14.23 | … | 2.69E+20 | 2.00E+17 | 0.60 | 0.038 | 513 | 131 |
| Mrk 1095 | 201.69 | -21.13 | 13.89 | 0.07 | 9.38E+20 | 8.00E+17 | 18.3 | 0.049 | 513 | 96 |
| Mrk 1383 | 349.22 | 55.12 | 14.58 | 0.02 | 2.55E+20 | 6.00E+17 | 0.59[c] | …[c] | 823 | 114 |
| Mrk 1502 | 123.75 | -50.18 | 14.00 | 0.10 | 4.59E+20 | 3.00E+17 | 0.74 | 0.035 | 434 | 94 |
| Mrk 1513 | 63.67 | -29.07 | 14.34 | 0.05 | 3.63E+20 | 5.00E+17 | 1.22 | 0.037 | 757 | 127 |
| MS0 700.7+6338 | 152.47 | 25.63 | 14.16 | 0.12 | 4.09E+20 | 1.40E+18 | 0.90 | 0.042 | 593 | 101 |
| NGC 588 | 133.34 | -31.39 | <14.07 | … | 6.84E+20 | 1.30E+18 | 1.64 | 0.040 | 457 | 81 |
| NGC 595 | 133.53 | -31.31 | <14.00 | … | 4.17E+20 | 1.20E+18 | 1.64 | 0.040 | 408 | 79 |
| NGC 985 | 180.84 | -59.49 | 14.34 | 0.04 | 3.49E+20 | 6.00E+17 | 0.79 | 0.031 | 515 | 101 |
| NGC 1068 | 172.10 | -51.93 | 13.98 | 0.06 | 2.73E+20 | 6.00E+17 | 0.81 | 0.035 | 642 | 154 |
| NGC1399 | 236.72 | -53.63 | 14.10 | 0.11 | 1.59E+20 | 5.00E+17 | … | … | 1920 | 290 |
| NGC 1705 | 261.08 | -38.74 | 14.39 | 0.02 | 1.70E+20 | 6.00E+17 | … | … | 868 | 91 |
| NGC3310 | 156.6 | 54.06 | 14.56 | 0.02 | 1.42E+20 | 6.00E+17 | 0.27[c] | …[c] | 833 | 94 |
| NGC 3504 | 204.6 | 66.04 | <14.19 | … | 2.00E+20 | 1.00E+18 | 0.41 | 0.037 | 1033 | 142 |
| NGC 3991 | 185.68 | 77.20 | 14.47 | 0.11 | 1.75E+20 | 1.50E+18 | 0.56 | 0.031 | 910 | 105 |
| NGC 4151 | 155.08 | 75.06 | 14.24 | 0.04 | 2.10E+20 | 6.00E+17 | 0.50 | 0.030 | 1041 | 153 |
| NGC 4649 | 295.88 | 74.34 | 14.31 | 0.13 | 2.04E+20 | 1.10E+18 | 0.30 | 0.035 | 1428 | 130 |
| NGC 4670 | 212.69 | 88.63 | 14.50 | 0.03 | 1.15E+20 | 8.00E+17 | 0.54 | 0.032 | 1008 | 111 |
| NGC 5548 | 31.96 | 70.50 | 14.50 | 0.05 | 1.65E+20 | 5.00E+17 | 0.34 | 0.032 | 1007 | 153 |
| NGC 7469 | 83.10 | -45.47 | 13.96 | 0.06 | 4.42E+20 | 5.00E+17 | 0.91 | 0.035 | 351 | 83 |
| NGC 7714 | 88.22 | -55.56 | 13.85 | 0.14 | 4.71E+20 | 3.00E+17 | 0.59[c] | …[c] | 420 | 81 |
| PG 0052+251 | 123.91 | -37.44 | <13.97 | … | 3.99E+20 | 4.00E+17 | 0.84 | 0.035 | 375 | 81 |
| PG0804+761 | 138.28 | 31.03 | 14.51 | 0.02 | 3.89E+20 | 7.00E+17 | 0.97[c] | …[c] | 560 | 109 |
| PG 0832+251 | 199.49 | 33.15 | 14.51 | 0.07 | 2.46E+20 | 1.30E+18 | 2.16 | 0.038 | 599 | 95 |
| PG 0832+675 | 147.75 | 35.01 | 14.26 | 0.06 | 3.58E+20 | 1.60E+18 | 0.70 | 0.044 | 645 | 125 |
| PG 0844+349 | 188.56 | 37.97 | 14.52 | 0.02 | 3.01E+20 | 7.00E+17 | 1.06 | 0.041 | 662 | 104 |
| PG 0947+396 | 182.85 | 50.75 | 14.54 | 0.07 | 1.21E+20 | 4.00E+17 | 0.55 | 0.039 | 1085 | 124 |
| PG 0953+414 | 179.79 | 51.71 | 14.45 | 0.03 | 1.23E+20 | 9.00E+17 | 0.60[c] | …[c] | 993 | 128 |
| PG 1001+291 | 200.09 | 53.20 | 14.5 | 0.06 | 1.76E+20 | 6.00E+17 | 0.75 | 0.040 | 846 | 102 |
| PG 1004+130 | 225.12 | 49.12 | 14.65 | 0.03 | 3.76E+20 | 7.00E+17 | 1.21 | 0.043 | 553 | 84 |
| PG 1011-040 | 246.50 | 40.75 | 14.15 | 0.04 | 3.66E+20 | 5.00E+17 | 0.83 | 0.040 | 628 | 102 |
| PG 1116+215 | 223.36 | 68.21 | 14.26 | 0.03 | 1.16E+20 | 1.40E+18 | 0.64 | 0.038 | 976 | 139 |
| PG 1211+143 | 267.55 | 74.32 | 14.17 | 0.06 | 2.71E+20 | 7.00E+17 | 0.67 | 0.042 | 951 | 119 |
| PG 1259+593 | 120.56 | 58.05 | 14.22 | 0.02 | 7.60E+19 | 1.70E+18 | 0.35 | 0.031 | 1141 | 114 |
| PG 1302-102 | 308.59 | 52.16 | 14.23 | 0.04 | 3.19E+20 | 7.00E+17 | 2.63 | 0.041 | 858 | 127 |
| PG 1307+085 | 316.79 | 70.71 | 14.45 | 0.05 | 2.21E+20 | 6.00E+17 | 0.55 | 0.037 | 778 | 116 |
| PG 1351+640 | 111.89 | 52.02 | 14.38 | 0.04 | 2.12E+20 | 1.70E+18 | 0.42 | 0.033 | 1092 | 93 |
| PG 1352+183 | 4.37 | 72.87 | 14.39 | 0.11 | 1.69E+20 | 6.00E+17 | 0.43[c] | …[c] | 1281 | 146 |
| PG 1402+261 | 32.96 | 73.46 | 14.53 | 0.05 | 1.50E+20 | 9.00E+17 | 0.41 | 0.032 | 905 | 112 |
| PG 1404+226 | 21.48 | 72.37 | 14.33 | 0.12 | 2.00E+20 | 6.00E+17 | 0.57 | 0.039 | 1064 | 127 |
| PG 1411+442 | 83.83 | 66.35 | 14.57 | 0.05 | 7.70E+19 | 9.00E+17 | 0.28 | 0.033 | 1384 | 95 |
| PG 1415+451 | 84.72 | 65.32 | 14.32 | 0.10 | 7.50E+19 | 6.00E+17 | 0.34 | 0.035 | 1410 | 102 |
| PG 1444+407 | 69.90 | 62.72 | 14.50 | 0.07 | 1.08E+20 | 4.00E+17 | 0.53 | 0.032 | 1101 | 99 |
| PG 1626+554 | 84.51 | 42.19 | 14.25 | 0.09 | 9.00E+19 | 1.20E+18 | 0.48[c] | …[c] | 1400 | 101 |
| PG 2349-014 | 91.66 | -60.36 | <13.97 | … | 3.00E+20 | 2.00E+17 | 0.53 | 0.037 | 591 | 106 |
| PHL 1811 | 47.46 | -44.81 | 14.45 | 0.04 | 3.97E+20 | 9.00E+17 | 0.78 | 0.038 | 481 | 89 |
| PKS 0405-12 | 204.93 | -41.76 | 14.14 | 0.04 | 3.56E+20 | 8.00E+17 | 7.46 | 0.043 | 1190 | 131 |
| PKS 0558-504 | 257.96 | -28.57 | 14.42 | 0.03 | 4.54E+20 | 9.00E+17 | … | … | 576 | 76 |
| PKS 2005-489 | 350.37 | -32.6 | 14.78 | 0.02 | 4.74E+20 | 1.30E+18 | … | … | 668 | 127 |

| | l | b | log N(O VI) | error | log N(H I) | error | I(Hα) | error | I(X) | error |
|---|---|---|---|---|---|---|---|---|---|---|
| PKS 2155-304 | 17.73 | -52.25 | 14.34 | 0.01 | 1.35E+20 | 8.00E+17 | ... | ... | 1278 | 138 |
| SBS 0335-052 | 191.34 | -44.69 | <13.99 | ... | 3.63E+20 | 9.00E+17 | 8.37[c] | ...[c] | 1133 | 120 |
| SBS 1415+437 | 81.96 | 66.2 | 14.14 | 0.12 | 9.80E+19 | 1.10E+18 | 0.29 | 0.033 | 1171 | 106 |
| TOL 1247-232 | 302.6 | 39.3 | 14.40 | 0.08 | 6.51E+20 | 1.10E+18 | 1.60 | 0.037 | 455 | 101 |
| TOL 1924-416 | 356.94 | -24.10 | 14.62 | 0.05 | 5.84E+20 | 1.10E+18 | ... | ... | 721 | 104 |
| TON 1187 | 188.33 | 55.38 | 14.35 | 0.08 | 1.00E+20 | 5.00E+17 | 0.51 | 0.045 | 1334 | 135 |
| TON S180 | 139.00 | -85.07 | 14.37 | 0.04 | 1.24E+20 | 3.00E+17 | 0.32 | 0.034 | 784 | 133 |
| TON S210 | 224.97 | -83.16 | 14.60 | 0.02 | 1.80E+20 | 1.20E+18 | 0.48 | 0.036 | 626 | 92 |
| UGC 12163 | 92.14 | -25.34 | 14.00 | 0.12 | 5.03E+20 | 1.30E+18 | 1.69[c] | ...[c] | 457 | 84 |
| VIIZw 118 | 151.36 | 25.99 | 14.13 | 0.05 | 3.66E+20 | 3.00E+17 | 1.11 | 0.044 | 452 | 95 |
| VZ 1128 | 42.50 | 78.68 | 14.49 | 0.01 | 1.11E+20 | 4.00E+17 | 0.17 | 0.033 | 1200 | 142 |

[a]  This table lists: (1) object name; (2) l; (3) b; (4) log N(O VI); (5) random error in log N(O VI); (6) Values of log N(H I) integrated over the  O VI velocity range from $v_-$ to $v_+$ using the H I observations displayed in Wakker et al. (2002); (7) error in N(H I); (8) the intensity of Hα emission, I(Hα), in the  one degree field containing the survey object from the Wisconsin Hα survey (Haffner et al. 2002).  I(Hα) is in Rayleighs ($10^6$ /$4\pi$ photons cm$^{-2}$ s$^{-1}$ sr$^{-1}$); (9) error in  I(Hα); (10)  I(X), the intensity of the  ROSAT  (R1+R2 band) 0.25 keV soft X-ray diffuse background ( counts s$^{-1}$ arcminute$^{-2}$)  averaged over a 36x36 arc minute region centered on the survey object; (11) error in I(X).

[b] The H I and O VI velocity ranges over which the various quantities have been calculated are the same for each object with the following exceptions where we list  object(O VI $v_-$ and $v_+$; H I $v_-$ and  $v_+$)  with the velocities in km s$^{-1}$.

ESO 265-G23 (-20, 100; -100, 80)      Mrk 279 (-115, 100; -100, 100)
Mrk 304 (-40, 75; -100, 75)              Mrk 506 (-100, 20; -100, 50)
Mrk 734 (-35, 140; -100, 140)          NGC 4151(-35, 100; -100, 100)
NGC 7714 (-60, 45; -80, 45)            PG 0832+675 (-60, 85; -85, 85)
PG 0947+396 (-65, 100; -30,100).

[c]  The Hα  measurement is very uncertain because of starlight contamination of the observation for the survey direction.  The listed intensity is an average over the survey measurements for adjacent directions (see Haffner et al. 2002).  Errors are not listed in these cases.

**TABLE  8**
**EXTRAGALACTIC  SIGHT  LINES**
**WITH  GALACTIC  O VI,  N  V,   AND  C  IV  COLUMN  DENSITIES**

| Object | l | b | log N(O VI) | logN(N V) | log N(C IV) | N(N V)/N(O VI) | N(C IV)/N(O VI) | Ref. |
|--------|---|---|-------------|-----------|-------------|----------------|-----------------|------|
| Mrk 509 | 35.97 | -29.86 | 14.66±0.01 | 13.87±0.07 | 14.45±0.04 | 0.16±0.03 | 0.62±0.06 | 1 |
| 3C 273 | 289.95 | 64.36 | 14.73±0.01 | 13.86±0.06 | 14.49±0.03 | 0.13±0.02 | 0.58±0.04 | 2 |
| PKS 2155-304 | 17.73 | -52.25 | 14.34±0.01 | 13.72±0.13 | 14.11±0.04 | 0.24±0.08 | 0.59±0.06 | 1 |
| ESO 141-55 | 338.18 | -26.71 | 14.50±0.02 | 13.82±0.10 | 14.74±0.10 | 0.21±0.05 | 1.74±0.46 | 3 |
| Fairall 9 | 295.07 | -57.83 | 14.38±0.07 | <13.71(3σ) | 14.21±0.08 | <0.21(3σ) | 0.68±0.19 | 1 |
| NGC 4151 | 156.60 | 54.06 | 14.24±0.04 | ... | 13.93±0.06 | ... | 0.49±0.09 | 1 |
| NGC 5548 | 31.96 | 70.50 | 14.50±0.05 | 13.64±0.10 | 14.40±0.04 | 0.14±0.04 | 0.79±0.13 | 1 |
| Mrk 205 | 125.45 | 41.67 | <14.19(3σ) | ... | 13.94±0.06 | ... | <0.56 (3σ) | 4 |

References for N V and C IV measurements: (1) Savage et al. 1997; (2) Sembach et al. 1997; (3) Sembach et al. 1999; (4) Bowen & Blades 1993.

**TABLE 9**
**GALACTIC SCALE HEIGHT ESTIMATES**

| ION | $n_0(cm^{-3})$ | $\sigma_p$ (dex)[a] | $N_\perp(cm^{-2})$ | h(kpc) | Reference |
|------|------|------|------|------|------|
| H I | 0.16±0.01 | 0.11 | $(1.9\pm0.2)\times10^{20}$ | 0.39±0.3 | 1[b] |
| H$^+$ | 0.036 | 0.22 | $1.0(+0.23,-0.19)\times10^{20}$ | 0.89(+0.25,-0.18) | 2[c] |
| Al III | $2.2\times10^{-9}$ | 0.20 | $6.9(+2.8,-1.0)\times10^{12}$ | 1.02(+0.36,-0.24) | 2 |
| Si IV | $(2.3\pm0.2)\times10^{-9}$ | 0.31 | $(3.6\pm0.6)\times10^{13}$ | 5.1±0.7 | 1[b] |
| C IV | $(9.2\pm0.8)\times10^{-9}$ | 0.30 | $(1.2\pm0.2)\times10^{14}$ | 4.4±0.6 | 1[b] |
| N V | $(2.6\pm0.3)\times10^{-9}$ | 0.24 | $(2.7\pm0.5)\times10^{13}$ | 3.3±0.5 | 1[b] |
| O VI | $1.7\times10^{-8}$ | ~0.25 | $\sim1.2\times10^{14}$ | ~2.3 | 3[d] |

[a] $\sigma_p$ (dex) is a logarithmic patchiness parameter included in the parameter estimate to allow for the irregular distribution of the gas (see Savage et al. 1990 and Savage et al. 1997). The value for O VI simply represents the standard deviation in log [N(O VI)sin|b|] for objects in the southern Galactic hemisphere.

[b] The same sample of disk and halo stars was used to derive $n_0$, $N_\perp(cm^{-2})$, and h(kpc) for H I, Si IV, C IV, and N V. Therefore, the parameters listed for these four species are affected by the same object sample bias. For the other ions the sample bias will be different which makes a cross-comparison of results from ion to ion more uncertain. The value of $n_0(cm^{-3}) = 0.16\pm0.11$ cm$^{-3}$ listed for H I indicates that the object sample studied by Savage et al. (1997) for stars at low z is dominated by paths through the lower density warm neutral medium of the Galaxy. Sight lines through pronounced H II regions were avoided to reduce the possibility of contaminating the disk sample with low |z| lines of sight through regions of photoionized Si IV and C IV. However, this selection process also removes from the sample those regions of higher than normal gas density and hot star density where one might expect to have an excess of collisionally ionized gas. These scale heights are more uncertain than the formal error bars suggest. The number of extragalactic objects leading to the estimate for C IV is 13. If the distribution of C IV in the sky is as complex as that of O VI, the Galactic positions of those thirteen objects will influence the derived scale height. To reliably compare the extensions of these different high ions away from the Galactic plane it will be necessary to obtain measures of N V, C IV, and Si IV for a number of those objects where measurements of O VI now exist.

[c] The scale height estimate for free electrons given in Savage et al. (1990) used the pulsar dispersion data from Reynolds (1989, Table 1) along with dispersion measures for pulsars in the globular clusters M13 and M53. The technique for estimating the various parameters is identical to that used for Al III.

[d] The scale height estimate for O VI is uncertain because a simple plane parallel patchy cloud distribution does not provide a good fit to the observations (see §6). The estimate here assumes the O VI distribution is given by a patchy plane parallel layer with a 0.25 dex enhancement to log N(O VI) in the northern Galactic hemisphere. The value of $n_0$(O VI) is from the FUSE survey of O VI in the Galactic Disk (Jenkins et al. 2001).

References. (1) Savage et al. 1997; (2) Savage et al. 1990; (3) this paper.

TABLE 10
RESULTS FOR MILKY WAY, LMC, AND SMC

| Galaxy | $<$[N(O VI)sin\|b\|]$>$[a] | $<b>$[a] | $A_O$[b] | $\dot{N}$ / $n_H$[c] | $\dot{M}$ / $\Omega$[d] | SFR/$\Omega$[e] |
|---|---|---|---|---|---|---|
| | $(10^{14}$ cm$^{-2})$ | (km s$^{-1}$) | | (km s$^{-1}$) | $(M_O$ yr$^{-1}$ kpc$^{-2})$ | $(M_O$ yr$^{-1}$ kpc$^{-2})$ |
| Milky Way South | 1.2 | 60 | $6.61$x$10^{-4}$ | 32 | $1.1$x$10^{-3}$ | $3$x$10^{-3}$ |
| Milky Way North | 1.8 | 62 | " | 47 | $1.7$x$10^{-3}$ | " |
| LMC | 1.9 | 55 | $2.24$x$10^{-4}$ | 59 | $2.1$x$10^{-3}$ | $8$x$10^{-3}$ |
| SMC | 3.4 | 56 | $1.07$x$10^{-4}$ | 118 | $4.2$x$10^{-3}$ | $6$x$10^{-3}$ |

[a] Average values of N(O VI)sin\|b\| and the line widths, b(km s$^{-1}$), are from this paper for the Milky Way, from Howk et al. (2002a) for the LMC, and from Hoopes et al. (2002) for the SMC.

[b] The oxygen abundances, $A_O$ = O/ H are from Allen (1973) for the Milky Way, from Russell & Dopita (1992) for the LMC and the SMC. We list the Allen (1973) abundances because they were adopted for the Edgar & Chevalier (1986) fountain calculations. A current estimate from Holweger (2001) for the Sun is $A_O$ = $5.4$x$10^{-4}$.

[c] The cooling gas fountain flow velocity v = $\dot{N}$ / $n_H$ is estimated from the observed value of the O VI column density and
N(O VI) cm$^{-2}$ = $3.8$x$10^{14}$ [v / 100 km s$^{-1}$] [ $A_O$ / $A_O$(Solar) ]$^{1-\beta}$ cm$^{-2}$, with $\beta$ =0.85 see §10.3.

[d] The cooling Galactic fountain gas mass flow rate per unit area of the disk is estimated from $\dot{M}$ / $\Omega$ = $1.4 m_H n_H$ $\dot{N}$ $n_H^{-1}$, see §10.3. Note that the values of $\dot{M}$ / $\Omega$ scale with the initial ionized hydrogen density of the cooling gas, $n_H$, here assumed to be $10^{-3}$ cm$^{-3}$ which seems reasonable for the Milky Way. For the LMC and SMC the value of $n_H$ may be substantially larger producing a corresponding increase in $\dot{M}$ / $\Omega$.

[e] The value of the star formation rates per unit area of the disk, SFR/$\Omega$ ($M_O$ yr$^{-1}$ kpc$^{-2}$) for the Milky Way is from the McKee & Williams (1997) global estimate of 4 $M_O$ yr$^{-1}$ and a assumed disk radius of 20 kpc. For the LMC and SMC we use the global SFR estimates from Kennicutt et al. (1995) based on H$\alpha$ observations and adopt surface areas of 31 kpc$^2$ and 8.0 kpc$^2$ for the LMC and SMC which are equal to the size of the regions of each galaxy studied.

# FIGURES

**FIG. 1.** – Aitoff projections of the all sky distribution of the objects studied in the O VI survey program. In (a) the Galactic center is at the center and Galactic longitude increases to the left. In (b) the Galactic anticenter is at the center of the figure. Object names are shortened to avoid clutter in the figure. Although more objects have been observed in the northern Galactic hemisphere, the measurements suffer from a sample bias since few objects appear in the region from $b = 30°$ to $60°$ and $l = 320°$ to $50°$. This is the region of high interstellar obscuration produced by dust in the Sco-Oph association. The average directions for stars in the LMC and SMC toward which Milky Way halo gas absorption has been studied by Howk et al. (2002a, 2002b) and Hoopes et al. (2002) are also shown.

**FIG. 2.** – Portions of the FUSE LiF1A spectra of Mrk 421, Mrk 1095, and PKS 2005-489 covering the wavelength range from 1015 to 1040 Å. Galactic atomic absorption lines are identified. $H_2$ lines are marked with the extra tick marks for Mrk 1095 and PKS 2005-489. $H_2$ absorption toward Mrk 421 is not detected between 1015 and 1040 Å. The weaker O VI $\lambda 1037.62$ line lies near strong absorption by C II* $\lambda 1037.02$ and the $H_2$ (5-0) R(1) and P(1) lines at 1037.15 and 1038.16 Å. The stronger O VI $\lambda 1031.93$ line is usually relatively free of blending since the nearby $H_2$ (6-0) P(3) and R(4) lines at 1031.19 and 1032.36 Å are generally weak. Terrestrial emission from H I Lyβ $\lambda 1025.72$ and lines of O I are identified with the circled plus signs.

**FIG. 3.** – A sample of the O VI $\lambda 1031.93$ absorption line profiles from Wakker et al. (2002) is displayed. Flux ($10^{-14}$ erg cm$^{-2}$ s$^{-1}$ Å$^{-1}$) is plotted against LSR velocity with the solid line showing the continuum placement including an estimate of the contaminating $H_2$ (6-0) R(3) and R(4) absorption when necessary. In the case of H 1821+643 a strong absorption component in O VI associated with the planetary nebula K-16 is shown at ~ 13 km s$^{-1}$ (see §8.6).

**FIG. 4.** – Examples of the analysis process for investigating saturation in the O VI profiles for PKS 2005-489 and Mrk 279. For each object the upper two panels show absorption line profiles for the O VI $\lambda\lambda 1031.93$ and 1037.62 lines plotted against LSR velocity. For the O VI $\lambda 1031.93$ absorption the smooth solid line shows the continuum and $H_2$ (6-0) R(3) and R(4) absorption correction when needed. For the O VI $\lambda 1037.62$ absorption, the smooth line only shows the continuum, since a correction for the $H_2$ (5-0) R(1) and P(2) and C II* absorption has not been attempted. The lower panel for each object shows the derived apparent column density profiles, $N_a(v)$ from the strong (heavy line) and weak (light line) O VI absorption. The units for $N_a(v)$ are atoms cm$^{-2}$ (km s$^{-1}$)$^{-1}$. For PKS 2005-489 these two curves agree well from –40 to 130 km s$^{-1}$. For Mrk 279 the agreement is good from –50 to 100 km s$^{-1}$. The good agreement for $N_a(v)$ for the weak and strong O VI lines implies that unresolved saturation is not affecting the O VI profiles. At velocities outside the indicated ranges the disagreement between the $N_a(v)$ curves for the strong and weak O VI lines is due to the various contaminating ISM lines spanning the region of the weak O VI $\lambda 1037.62$ absorption.

**FIG. 5.** – The overall kinematical behavior of the O VI-bearing gas in the Milky Way halo. The horizontal bars display the LSR velocity ranges of the O VI absorption. The width of each bar is a measure of the value of the column density in each velocity range. The observations are grouped according to l and b with l increasing from 0 to 360° from the bottom to the top of the figure. The four separate panels divide observations for ranges of latitude: $-90° \leq b \leq -50°$; $-50° < b \leq 0°$ ; $0° < b \leq 50°$ ; and $50° < b \leq 90°$. O VI absorption with $|v| \geq 120$ km s$^{-1}$ is detected along a large percentage (65%) of the lines of sight. The absorption we associate with the thick disk of the Milky Way is mostly confined to the velocity range from -100 to 100 km

s$^{-1}$. The separation between the thick disk absorption and the high velocity absorption is clear in some cases and ambiguous in other cases (see §4).

**FIG. 6.** – Number distribution of log N(O VI), log [N(O VI)sin|b|], $\bar{v}_{obs}$ (km s$^{-1}$) , and b (km s$^{-1}$) for O VI absorption associated with the thick disk of the Milky Way are shown in (a), (b), (c), and (d), respectively. Ranges, medians, averages, and standard deviations of these quantities for various subsamples of the objects are listed in Tables 3 and 4. In the histograms for log N(O VI) and log [N(O VI)sin|b|] upper limits are indicated as detections. In the various histograms the upper histogram is for the full object sample while the heavy line histogram is for Galactic polar objects with |b| > 45$^{o}$.

**FIG. 7.** – The distribution of log N(O VI) on the sky is plotted in these Aitoff projections with the Galactic center at the center in (a) and the Galactic anticenter at the center in (b). Galactic longitude increases to the left in both figures. The values of log N(O VI) are coded as color circles with 12$^{o}$ radius surrounding the direction to each object indicated with the small data point. The bar at the bottom of each panel indicates the conversion between log N(O VI) and color. Upper limits are indicated by small colored circles with a 2.5$^{o}$ radius. The object identifications can be determined with reference to Figures 1a and 1b. When two objects lie with separations < 12$^{o}$ the color area extends to a distance half way between each object. The mottled appearance on some of the colored circles indicates measurements with lower reliability. The multiple data points for the directions to the LMC and SMC indicate the large number of sight lines probed by stars in these two galaxies (see Howk et al. 2002a, 2002b, Hoopes et al. 2002)

**FIG. 8.** – The distribution of log N(O VI) on the sky is plotted in these equal area polar projections with the north Galactic pole center at the center in (a) and the south Galactic pole at the center in (b). Galactic longitude increases in the clockwise sense in (a) and in the counter clockwise sense in (b). Measurements and limits are indicated as colored circles with a 12$^{o}$ or 2.5$^{o}$ radius, respectively. When two objects lie with separations < 12$^{o}$ the color area extends to a distance half way between each object. The color coding for the values of log N(O VI) is the same as in Figures 7a and 7b. The mottled appearance on some of the colored circles indicates measurements with lower reliability. The multiple data points for the directions to the LMC and SMC indicate the large number sight lines probed by stars in these two galaxies (see Howk et al. 2002a, 2000b; Hoopes et al. 2002)

**FIG. 9.** – Values of log N(O VI) for the thick disk of the Milky Way are represented as circles displayed in these Aitoff projections of the sky with the Galactic center at the center of the figures and Galactic longitude increasing to the left. The circle size is proportional to log N(O VI) according to the code shown. Upper limits to log N(O VI) are denoted with triangles with a size proportional to the limit. In (a) the values of log N(O VI) are displayed as filled symbols and the contours of non-thermal Galactic Radio Loops as defined by Berkhuijsen et al. (1971) are shown as the solid lines. In (b) the values of log N(O VI) are displayed with the open symbols and the grey scale shows the 0.25 keV X-ray sky diffuse background count rate as measured by the ROSAT Satellite (Snowden et al. 1997).

**FIG. 10.** – Values of log N(O VI) and log N(O VI)sin|b| for the Milky Way thick-disk absorption are plotted against l and b. Measurements reported as 3σ upper limits are plotted with the downward pointing arrows. The appearance of this figure probably is affected by sample bias (see Fig. 1). For example, dust obscuration in the Sco-Oph region results in a deficiency of data points in the region l = 320$^{o}$ to 50$^{o}$ and b = 20$^{o}$ to 50$^{0}$. This is a direction in the sky where the values of log N(O VI) are likely to be large. In (e) the cross hatched region indicates the expected positions for measurements described by a simple plane parallel distribution for O VI absorbing structures with a scale height of 2.3 kpc, average mid-plane

density of $1.7 \times 10^{-8} cm^{-3}$, implying log [N(O VI)sin|b|] = 14.09, and a degree of irregularity with a standard deviation of 0.25 dex. While this simple model roughly describes the observations for the southern Galactic hemisphere and for b < 45° in the north, it underestimates the values of log N(O VI) from b = 45° to 90°.

**FIG. 11.** – $\Delta N(O VI)_{ij} = |N(O VI)_i - N(O VI)_j|$ for ij pairs of observations with $\sigma[\Delta N(O VI)]$ ≤ 0.25 is displayed in the upper panel as a function of the angular separation between the ij pairs, $\Delta\theta_{ij}$ (degrees). The different symbols indicate north to north Galactic hemisphere pairs (small solid circles), north to south Galactic hemisphere pairs (x symbol), and south to south Galactic hemisphere pairs (small open circles). In the upper panel $\Delta N(O VI)_{ij} / N(O VI)_{ij}$ is displayed where $N(O VI)_{ij} = (N(O VI)_i + N(O VI)_j)/2$. In the middle panel averages (heavy dashed line) and dispersions (light dashed line) of $\Delta N(O VI)_{ij} / N(O VI)_{ij}$ are shown averaged over ranges in $\Delta\theta_{ij}$ of 15° and plotted every 5°. The 90, 75, 50, 25, and 10% levels of $\Delta N(O VI)_{ij} / N(O VI)_{ij}$ measured over 15° wide intervals in $\Delta\theta_{ij}$ are also shown. Note that the level of irregularity in the distribution of O VI hardly changes as $\Delta\theta_{ij}$ ranges from several to 170°. For results over small angular scales ranging from 0.05° to 5° in the directions to the LMC and SMC see Howk et al. (2002b). In the lower panel similar variability results averaged over 15° ranges in $\Delta\theta_{ij}$ are shown for H I. $<\Delta N(H I)_{ij} / N(H I)_{ij}>$ is plotted versus $<\Delta\theta_{ij}>$, where the individual values of $N(H I)_i$ and $N(H I)_j$ are obtained from the 21 cm emission line observations in the direction to each program object. The values of N(H I) for each object direction are listed in Table 7.

**FIG. 12.** – log [N(O VI)sin|b|] for gas in the Milky Way disk and thick disk is plotted against log |z(kpc)| for stars from the Copernicus satellite disk sample (open squares), for Milky Way absorption toward stars in the LMC and SMC (open upward pointing triangles, and for the 100 extragalactic and two Galactic stellar lines of sight listed in Table 2 (open circles). For all data samples the 3σ upper limits are plotted as the downward pointing triangles. The extragalactic observations from Table 2 are plotted on the right hand side of the figure. The data points plotted in the two groupings are organized as for Fig. 10e where the left-most points are for b = -90° and the right-most points are for b = 90°. The three solid lines show the expected behavior of the log N(O VI)sin|b| versus log |z(kpc)| distribution for a smoothly distributed exponentially stratified plane parallel Galactic atmosphere with an O VI mid-plane density $n_o(O VI) = 1.7 \times 10^{-8} cm^{-3}$ and exponential scale heights of 1.0, 2.5, and 10 kpc. The large irregularity in the distribution of O VI and the enhancement in the amount of O VI over the northern Galactic hemisphere introduces serious complications when trying to estimate a Galactic scale height for O VI.

**FIG. 13.** – The average velocity of the O VI thick disk absorption, $\bar{v}_{obs}$, is displayed on an Aitoff projection of the sky with the Galactic anticenter at the center of the figure and longitude increasing to the left. The values of $\bar{v}_{obs}$ are displayed in color with yellow to red indicating positive LSR velocity and green to blue indicating negative LSR velocity according to the color to velocity code shown. For the direction to each object the color circle extends to a radius of 12°. When two objects are closer than 12°, the color coding extends to a distance half way between the two objects. The weak imprint of Galactic rotation is evident; for l ~ 60° to 140° the measurements for |b| < 40° are mostly blue and green, while for l ~ 240° to 320° the colors are mostly yellow and red. Toward the Galactic polar regions, |b| > 45°, where the effects of Galactic rotation are small we see thick disk O VI moving with both positive and negative velocities with equal frequency suggesting both outflow and inflow.

**FIG. 14.** – The average velocity of the O VI thick disk absorption toward each survey object, $\bar{v}_{obs}$, is plotted against the average expected velocity, $\bar{v}_{exp}$, assuming co-rotation of disk and halo gas. The expected velocity was calculated assuming the O VI layer has an exponential

scale height of 2.4 kpc and a turbulent velocity described by $b_{turb} = 60$ km s$^{-1}$. In panels a and c results are shown for high Galactic latitudes (|b| = 45° to 90°) in the north and south, respectively. In panels b and d results are shown for lower latitudes (|b| < 45°) in the sorth and south, respectively. The different symbols denote the reliability of the limits of the integration of the O VI absorption. For the filled circles there were no problems. For the half-filled circles there are possible contamination problems. For the open circles, substantial overlap between Galactic thick disk absorption and high velocity absorption made it necessary to introduce a relatively arbitrary cutoff to that part of the absorption associated with the thick disk. At high latitudes the observations (panels a and c) are dominated by the effects of inflow and outflow. At low latitudes (panels c and d) the effects of Galactic rotation should dominate. Typical $\pm 1\sigma$ errors on $\overline{v}_{obs}$ are ~10 km s$^{-1}$.

**FIG. 15.** – Values of the O VI line width, b(km s$^{-1}$), for the thick disk of the Milky Way are displayed in an Aitoff projection with the Galactic anticenter at the center of the figure and longitude increasing to the left. Increasing symbol size corresponds to increasing b (km s$^{-1}$) according to the code given in the figure. Filled circles are for reliable measures of b(km s$^{-1}$). Open symbols are for cases where blending with high velocity O VI absorption introduces considerable uncertainty in the measurement of b(km s$^{-1}$) for O VI absorption associated with the thick disk.

**FIG. 16.** – The distribution of log N(O VI) on the sky in the velocity range from -30 to -90 km s$^{-1}$ (a) and 30 to 90 km s$^{-1}$ (b) are plotted in these Aitoff projections with l = 120° at the center and longitude increasing to the left. The circle size is proportional to log N(O VI) toward each survey object integrated over the velocity range of the IVCs according to the code shown. Upper limits for log N(O VI) are denoted with triangles with a size proportional to the limit. The grey scale displays the integrated H I column density in intermediate negative (Fig 16a; -30 to -90 km s$^{-1}$) and intermediate positive (Fig 16b; 30 to 90 km s$^{-1}$) velocity gas as determined from the H I 21-cm spectra shown in Figure 1 of Wakker et al. (2002). The presence or absence of an IVC has no apparent effect on the amount of O VI in the IVC velocity range.

**FIG. 17.** – log N(O VI) versus log N(H I) for O VI absorption and H I emission over the velocity range -30 to -90 km s$^{-1}$ (in panel a) and 30 to 90 km s$^{-1}$ (in panel b). When a distinct H I IVC is clearly present the symbols are filled circles. When there is no pronounced H I IVC the symbols are open circles. In these cases the value of log N(H I) refers to weak H I emission in the IVC velocity range. There is no relationship between the column density of O VI in the IVC velocity range and the presence or absence of a H I IVC. The presence of a H I IVC does not effect the distribution of O VI on the sky. This agrees with the impression one gets from a visual inspection of Figs. 16a and 16b.

**FIG.18.** – log N(O VI) is plotted against a measure of the O VI line width, b(km s$^{-1}$) for the Galactic thick disk observations (open circles), the Copernicus satellite O VI disk observations (open squares), the LMC thick disk observations ( upward pointing open triangles), and the SMC thick disk observations (downward pointing open triangles). The trend of increasing column density with increasing line width that holds from the Milky Way disk and halo sample also holds for the LMC and SMC disk+halo sample.

**FIG. 19.** – (a) log N(O VI) for the thick disk of the Milky Way is plotted against N(H I) from 21 cm H I observations. The radio measurements (see Wakker et al. 2002) have been integrated over the same velocity range as for the O VI observations (i.e., from $v_-$ to $v_+$). If the O VI traced a gas phase filling a significant volume of space one might expect to see an anti-correlation between N(O VI) and N(H I). Neither an anti-correlation nor a correlation are evident between O VI and H I.  (b)  log N(O VI) for the thick disk of the Milky Way is plotted against log I(Hα) where I(Hα)  is the intensity of Hα emission in Rayleighs ($10^6 /4\pi$ photons cm$^{-2}$ s$^{-1}$ sr$^{-1}$) from the WHAM survey. No correlation is evident between O VI and Hα.  (c)  log N(O VI) for the thick disk of the Milky Way toward each object in Table 2 is plotted against log I(X) where I(X) is the  ROSAT 0.25 keV (R1+R2 band) X-ray count rate ($10^{-6}$ counts s$^{-1}$ arcmin$^{-2}$) in the direction of each object averaged over eight   12'x12' pixels centered on the object to increase signal-to-noise. The ROSAT pixel closest to the object was omitted from the averaging since many of the AGNs are X-ray bright.  The different symbols denote the value of log N(H I) in each direction. (d) log I(Hα) is plotted against log N(H I) for the O VI survey directions.  The reasonably good correlation is typical of that found when correlating many interstellar quantities. (e) log I(X) is plotted against log I(Hα) for the O VI survey directions.  (f) log N(H I) is plotted against log I(X) for the O VI survey directions. The anti-correlation of  N(HI) and I(X) is a well known property of the 0.25 keV X-ray background.

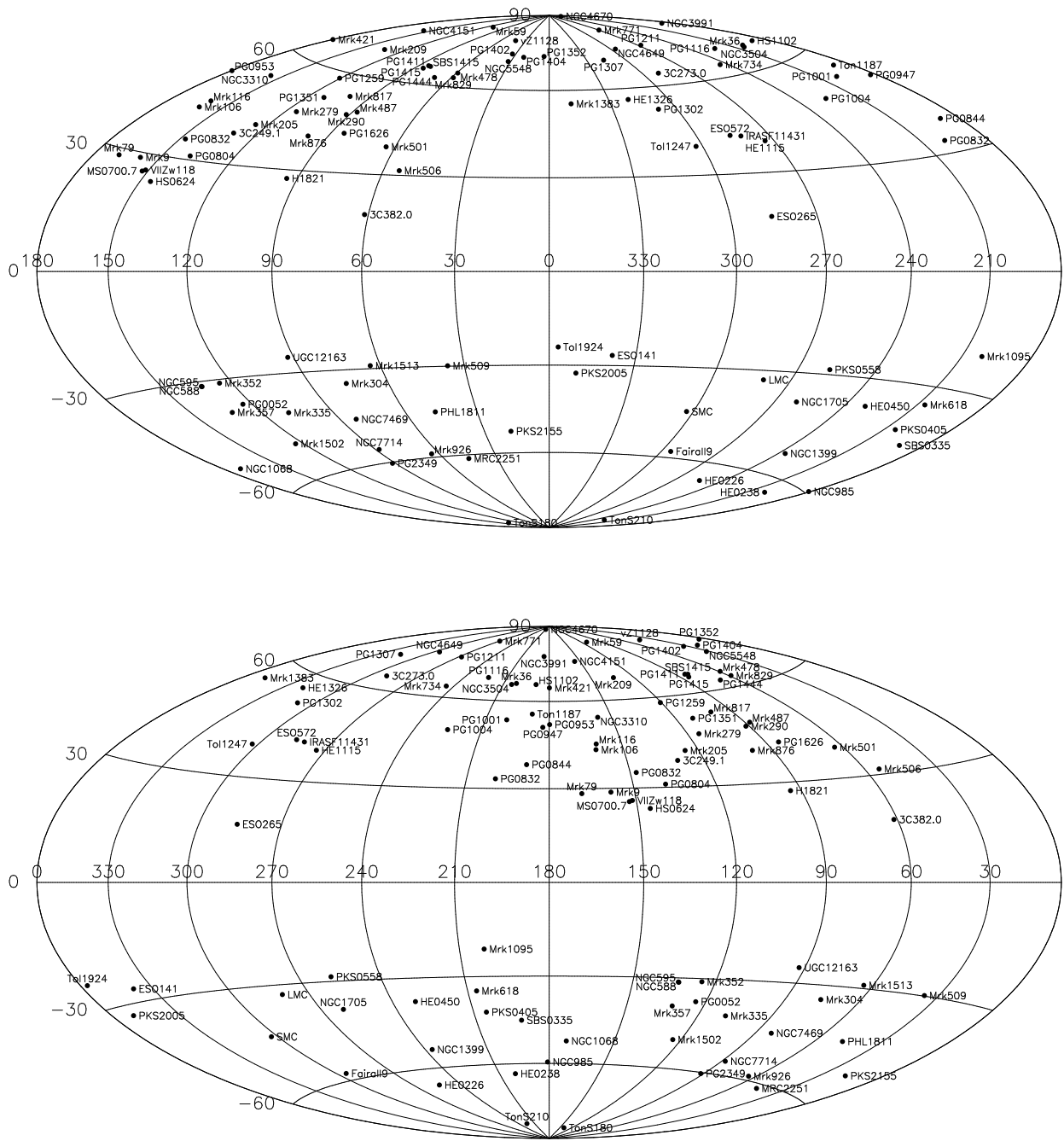

Fig. 1.—

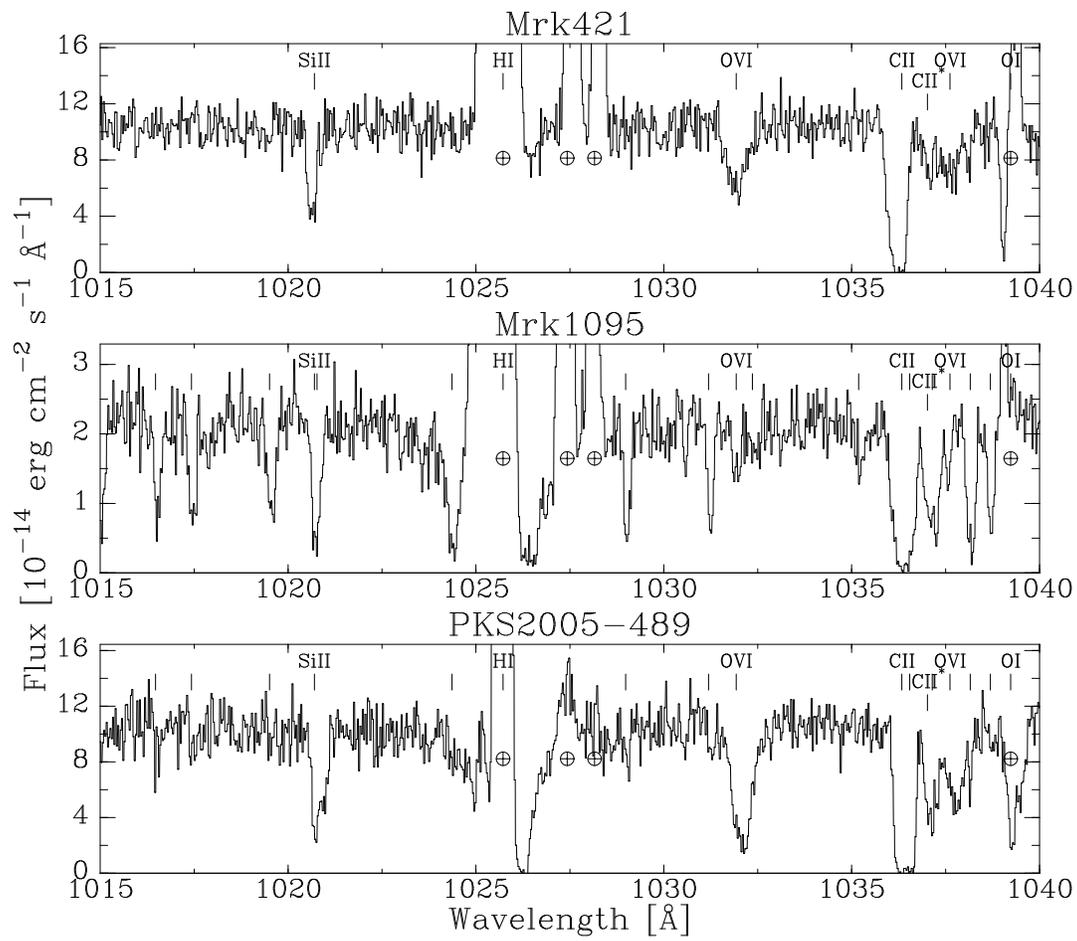

Fig. 2.—

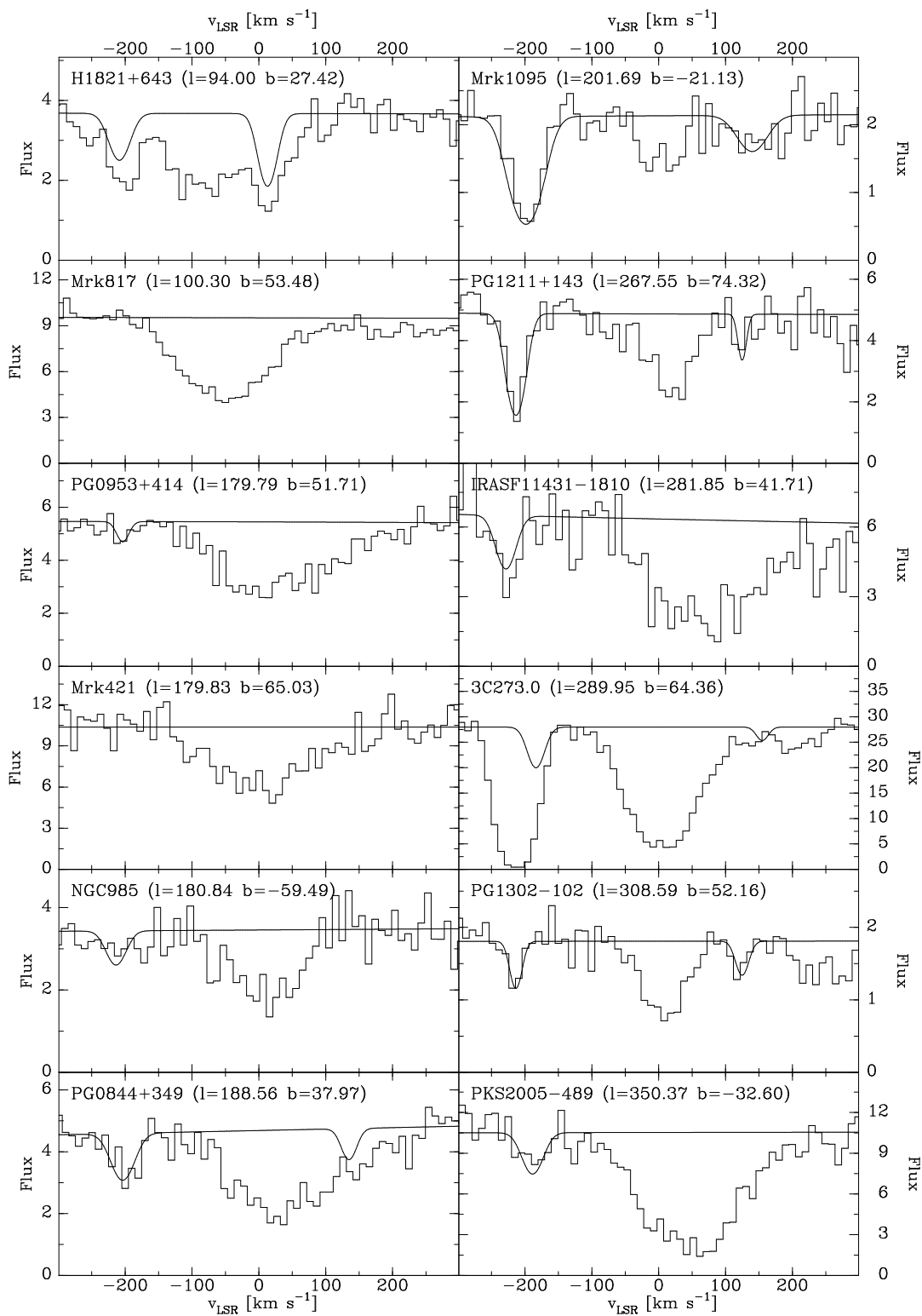

Fig. 3.—

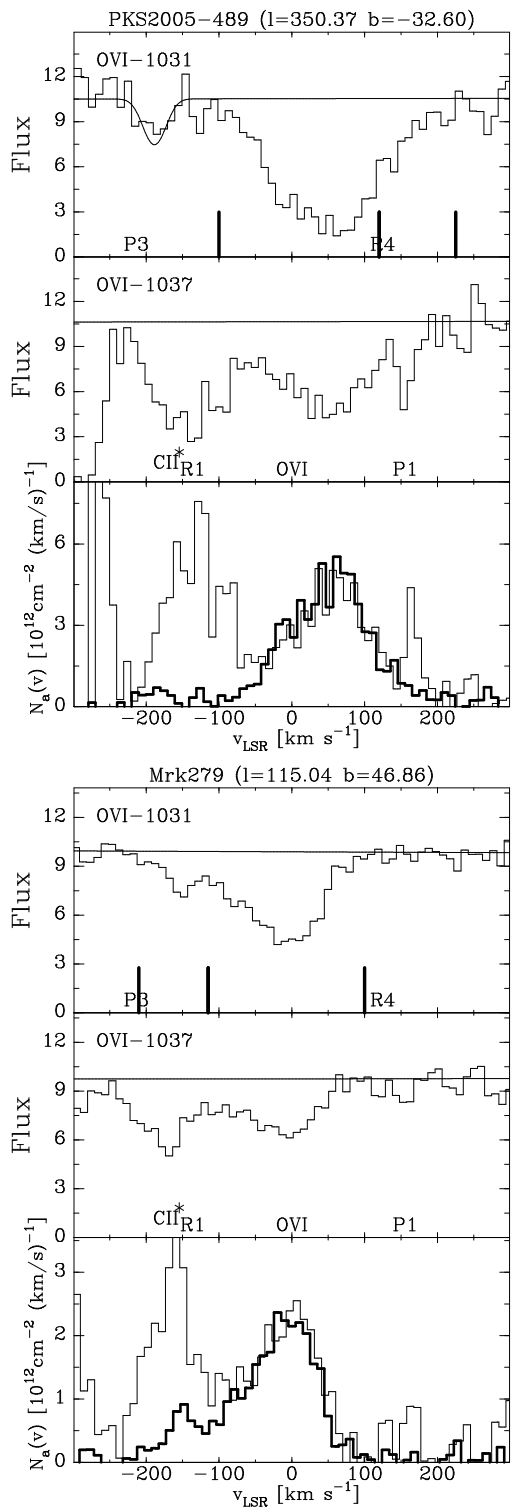

Fig. 4.—

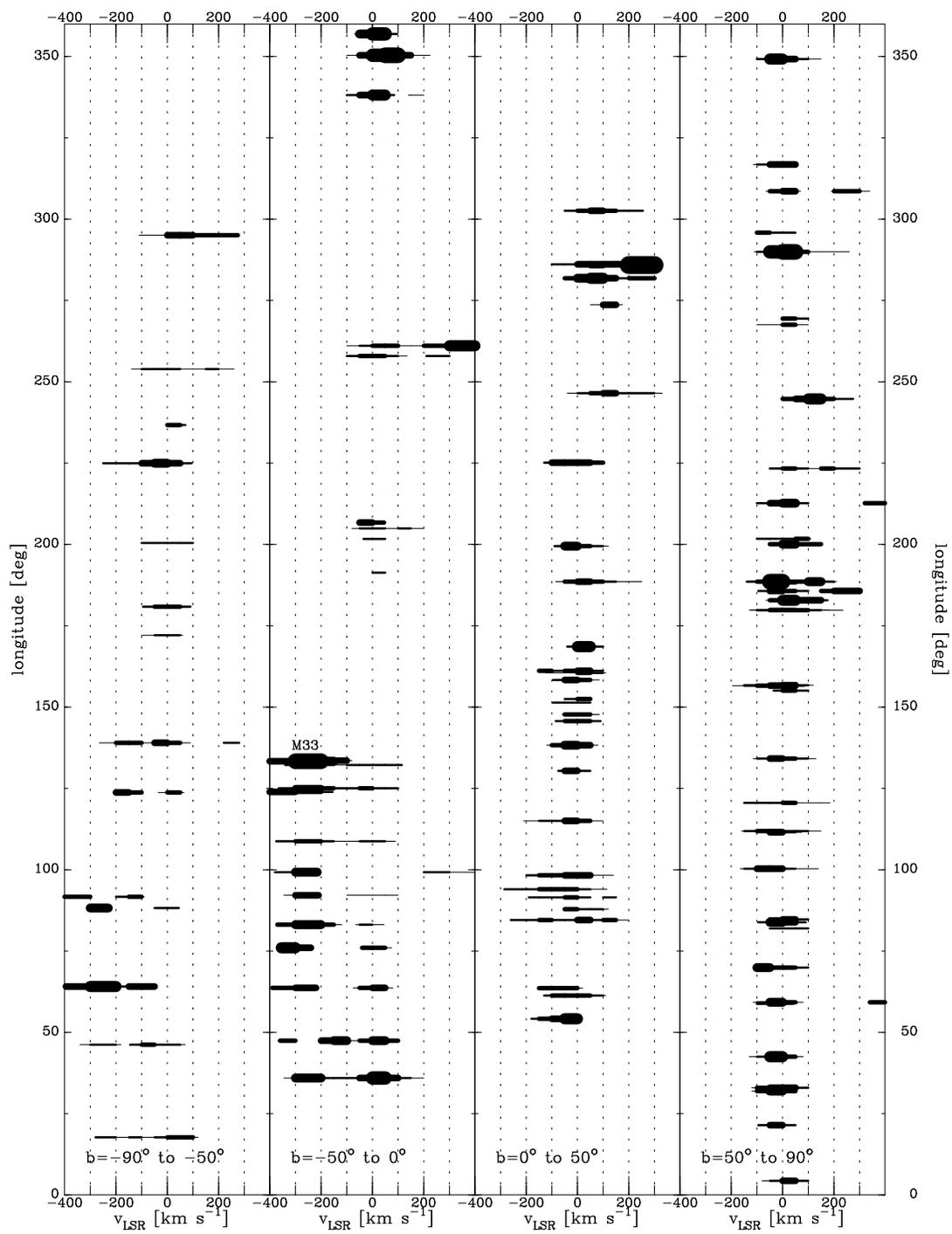

Fig. 5.—

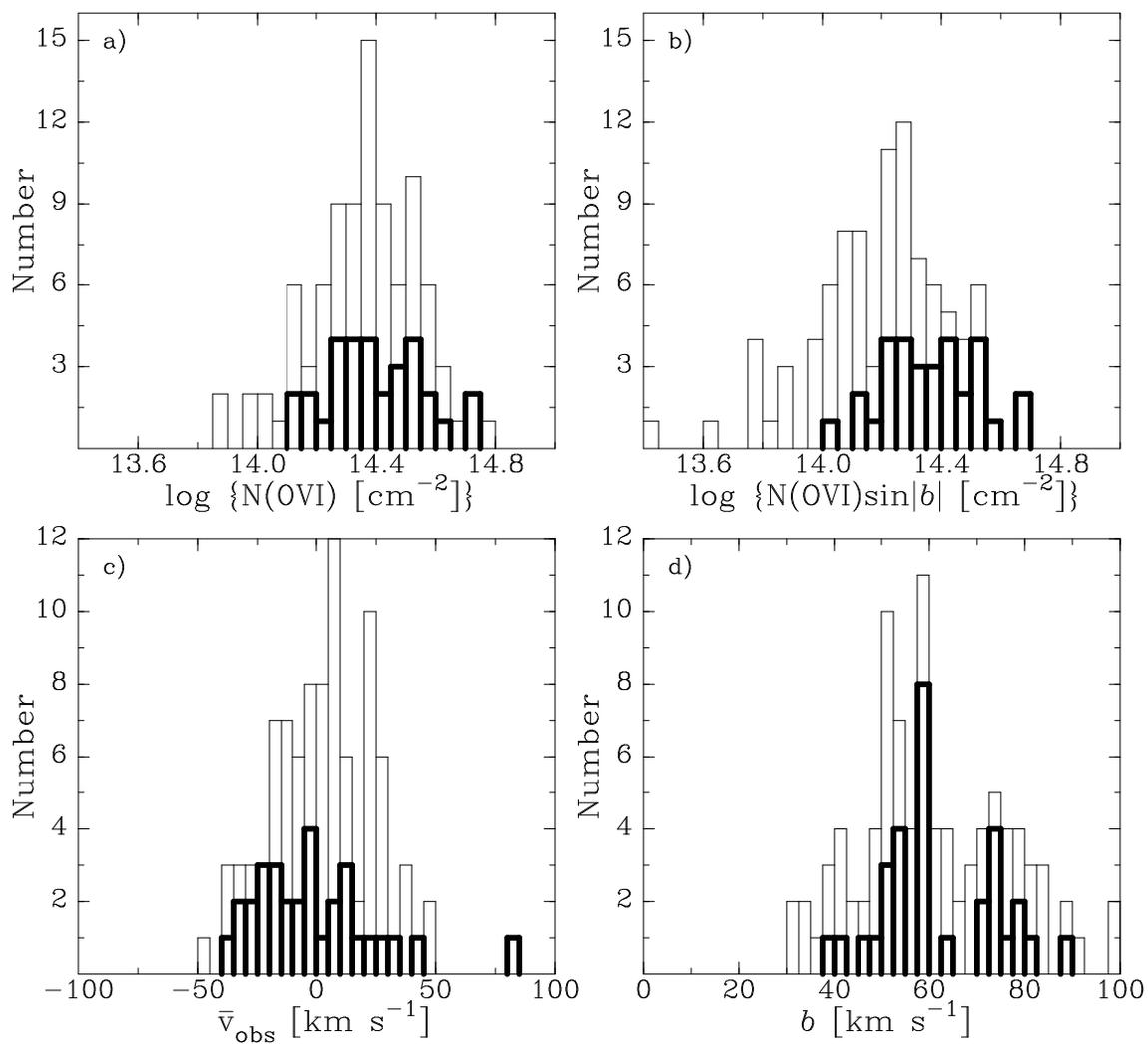

Fig. 6.—

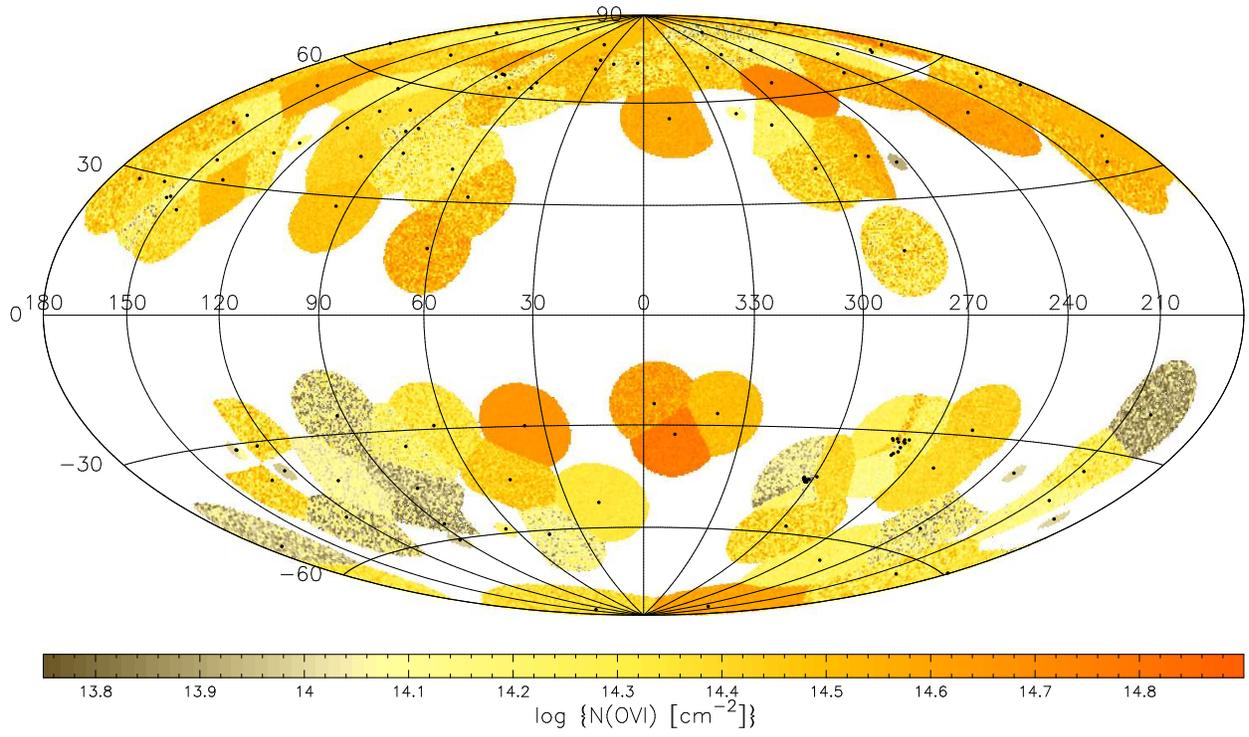

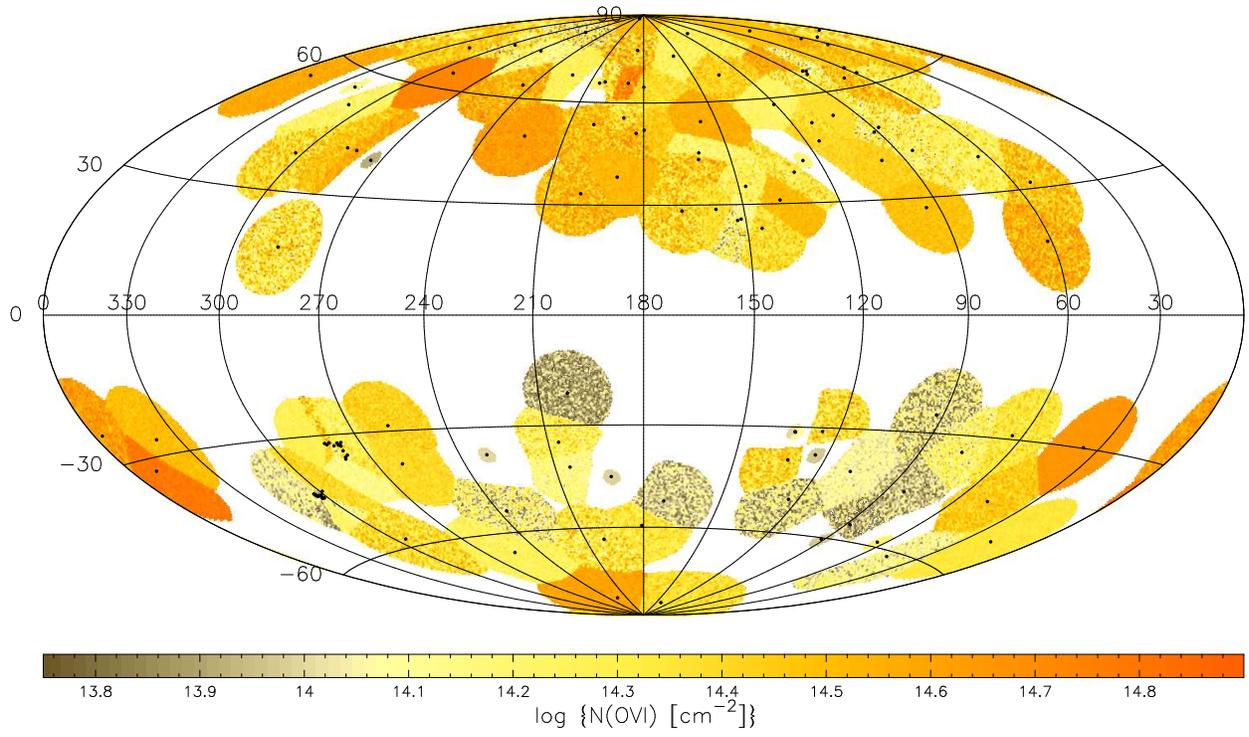

Fig. 7.—

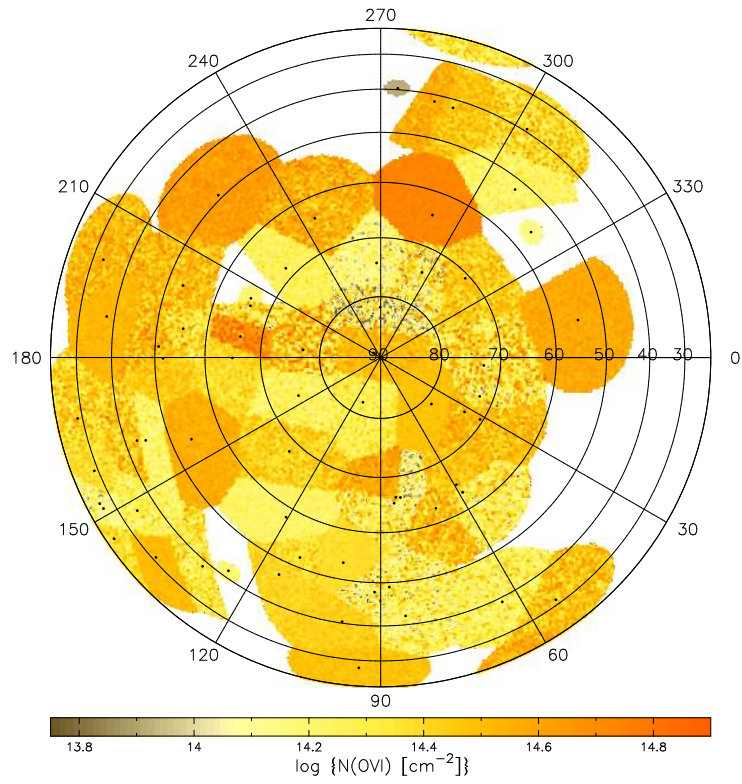

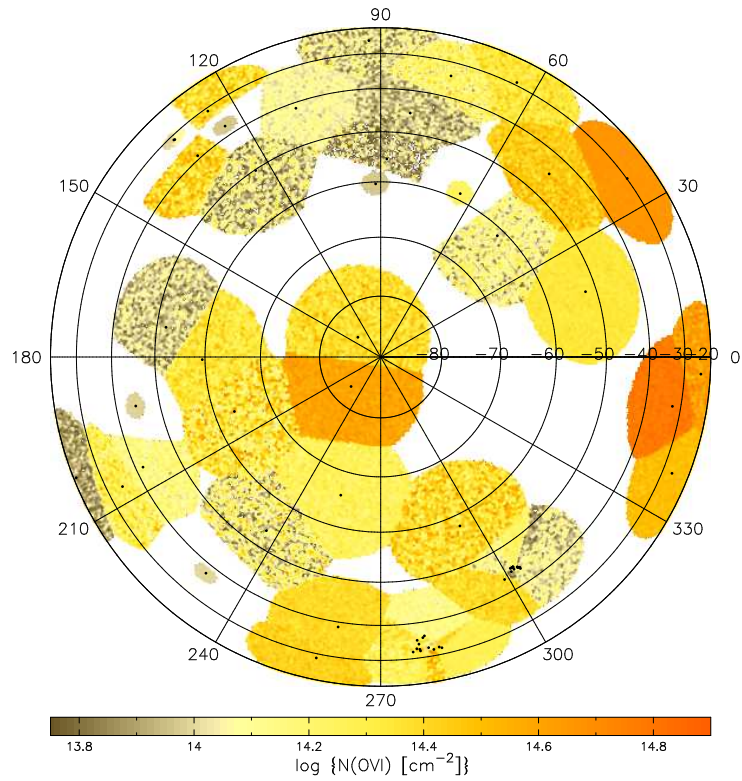

Fig. 8.—

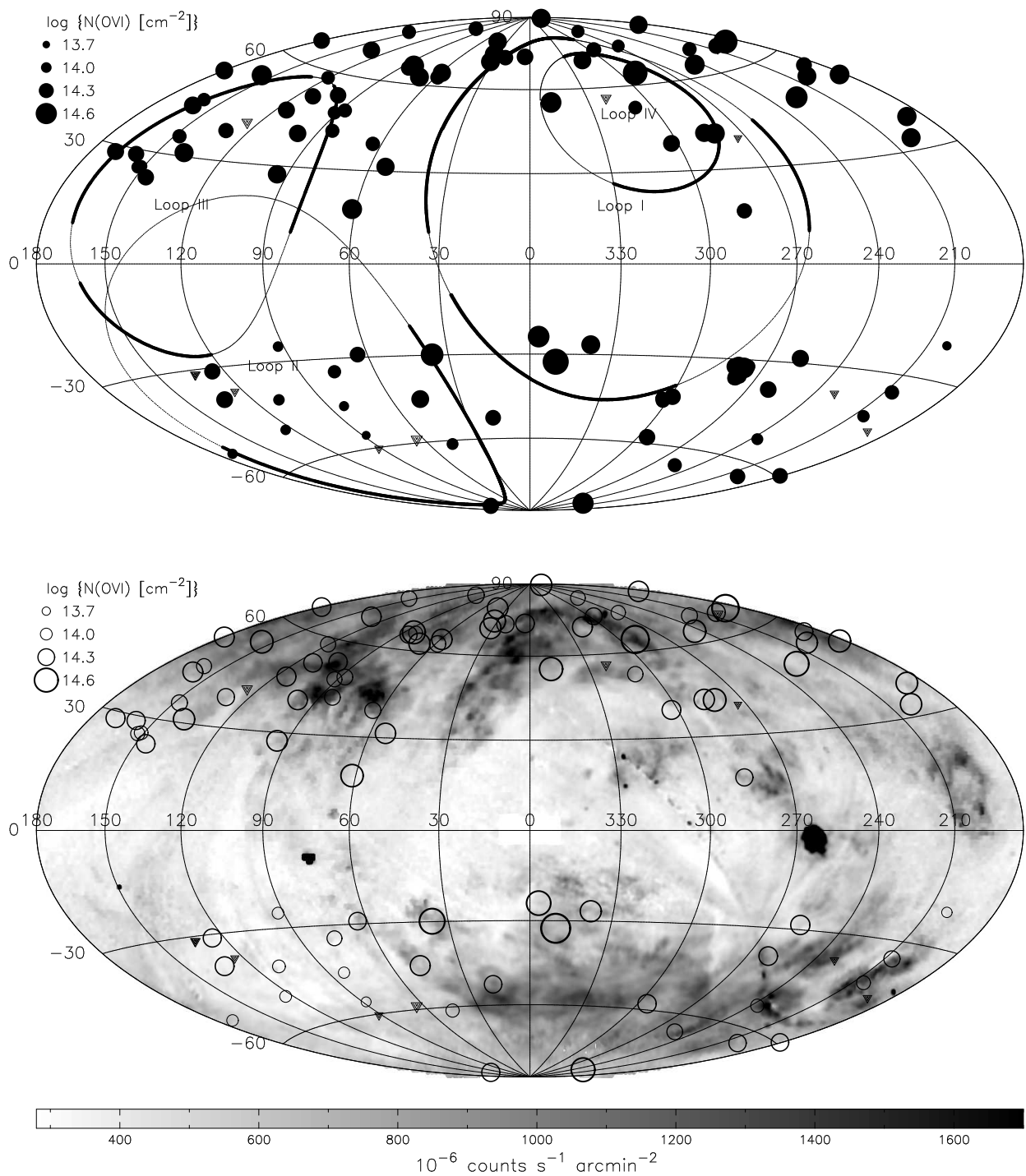

Fig. 9.—

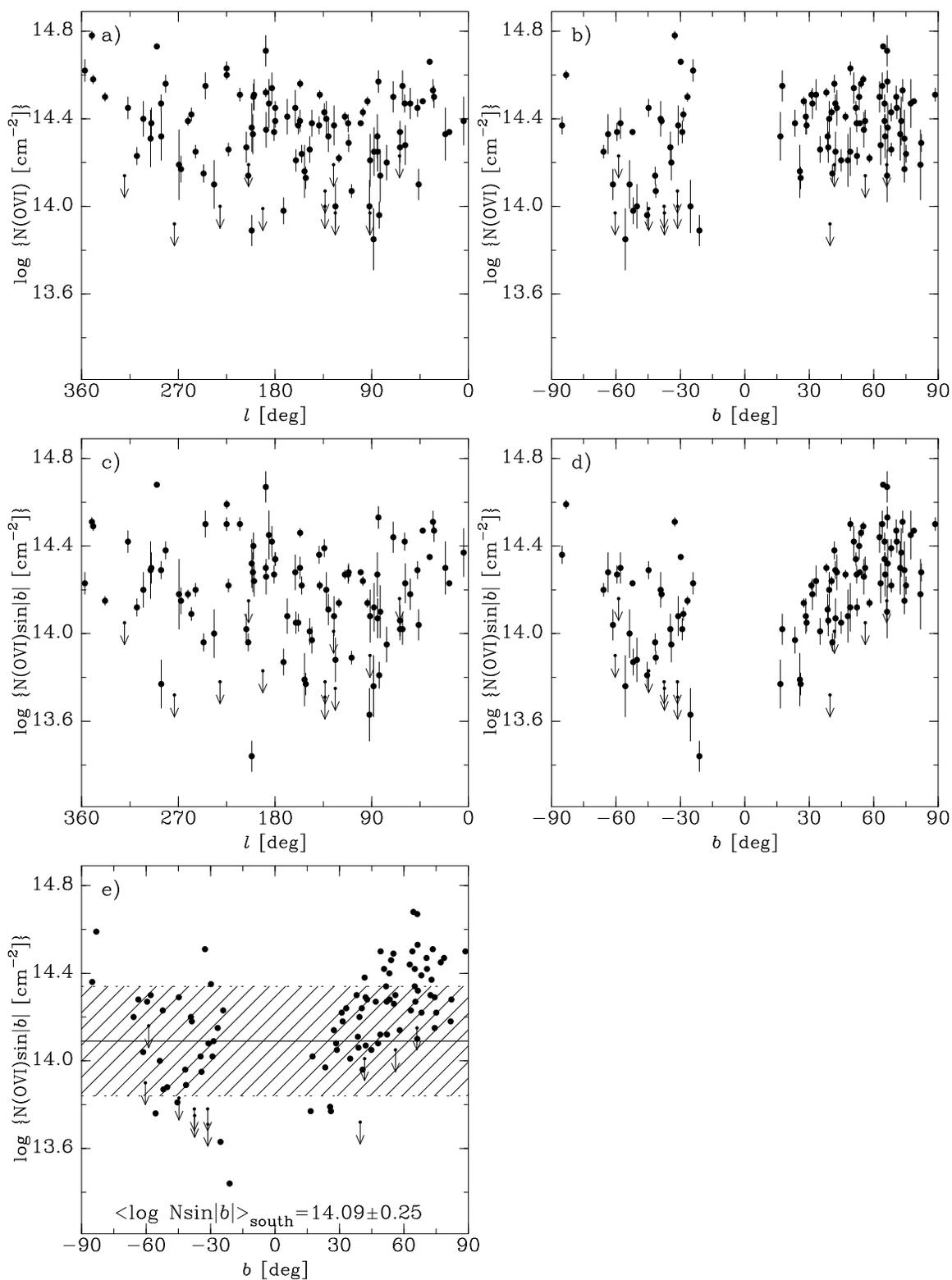

Fig. 10.—

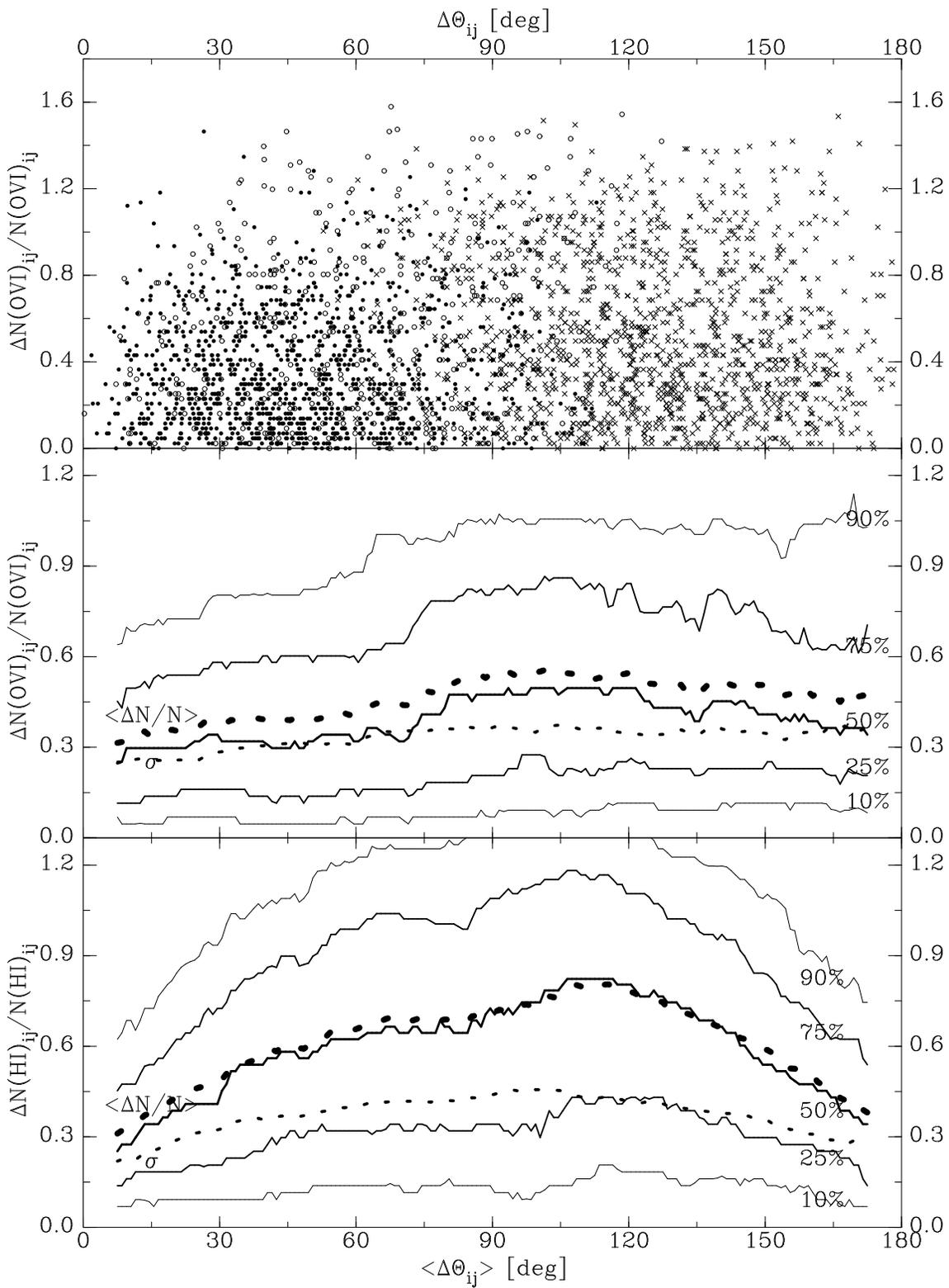

Fig. 11.—

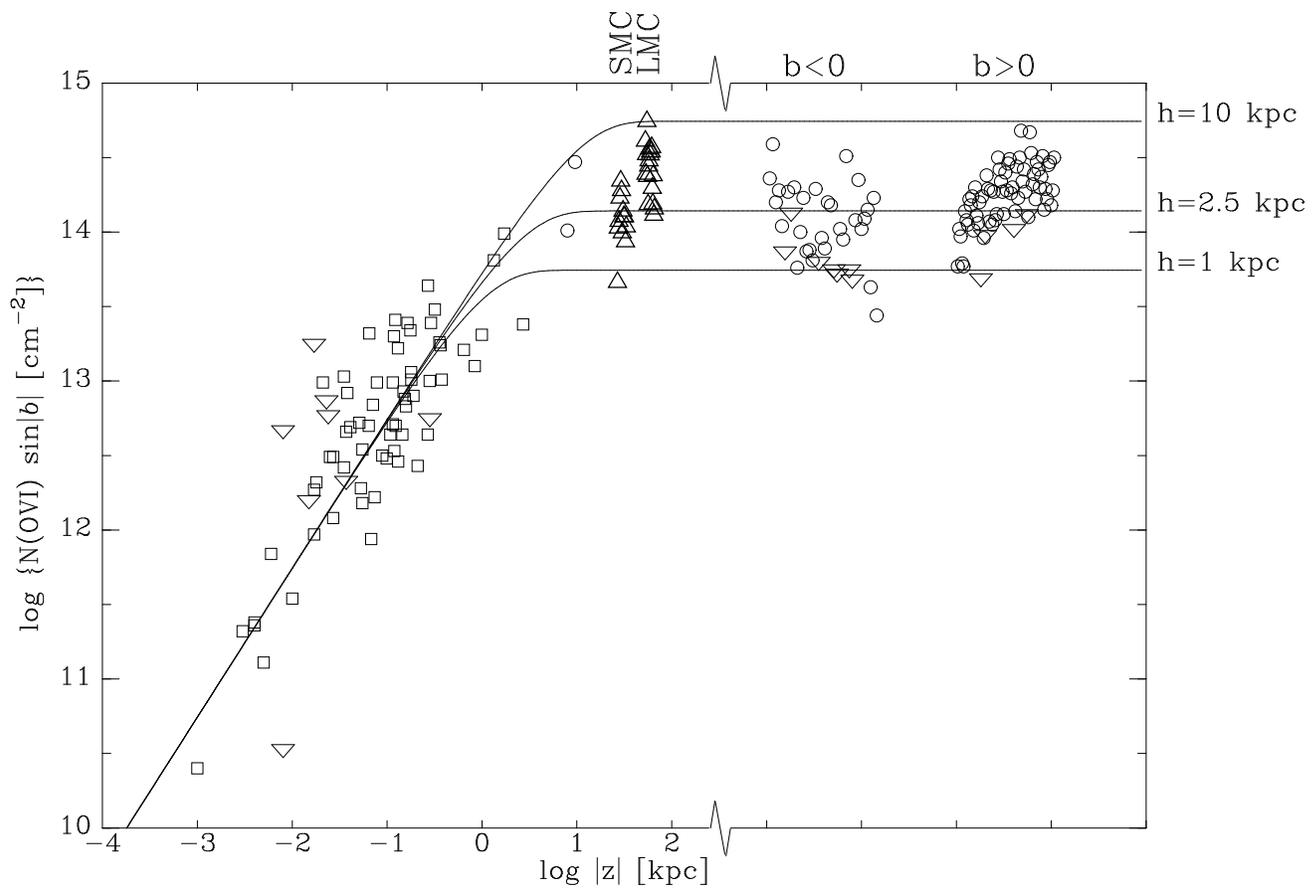

Fig. 12.—

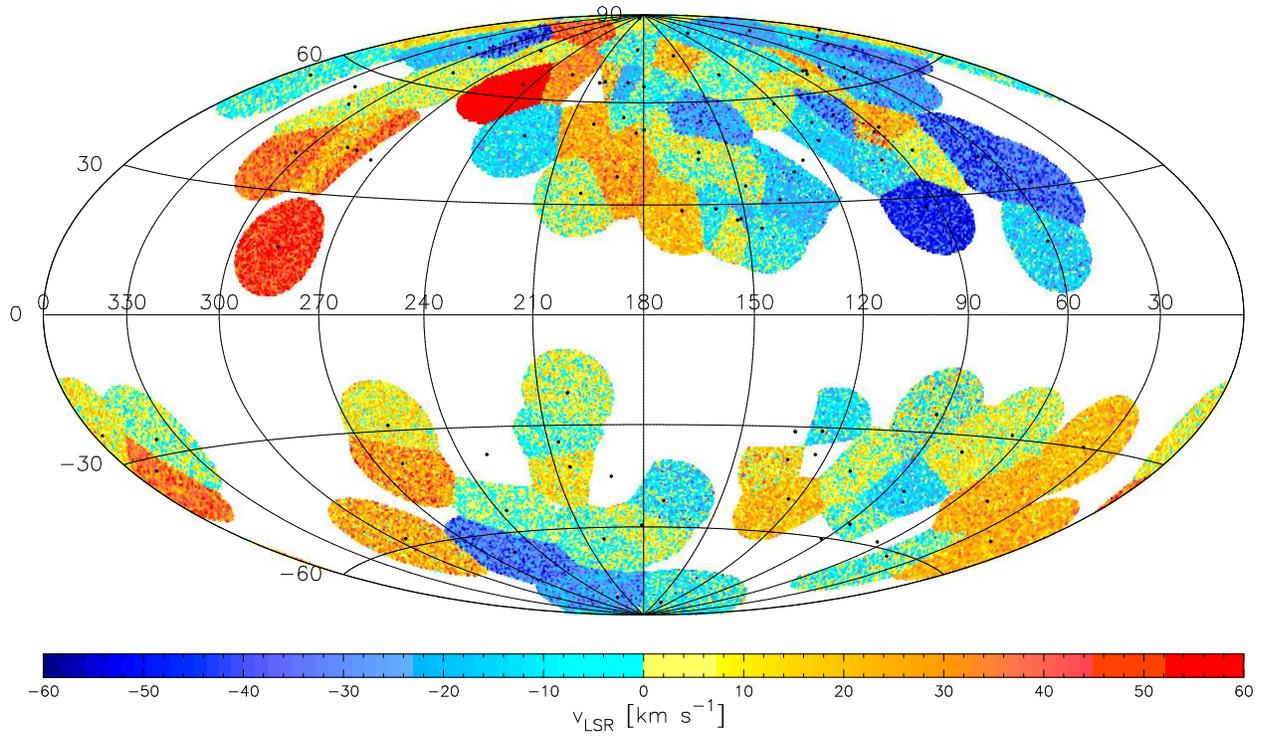

Fig. 13.—

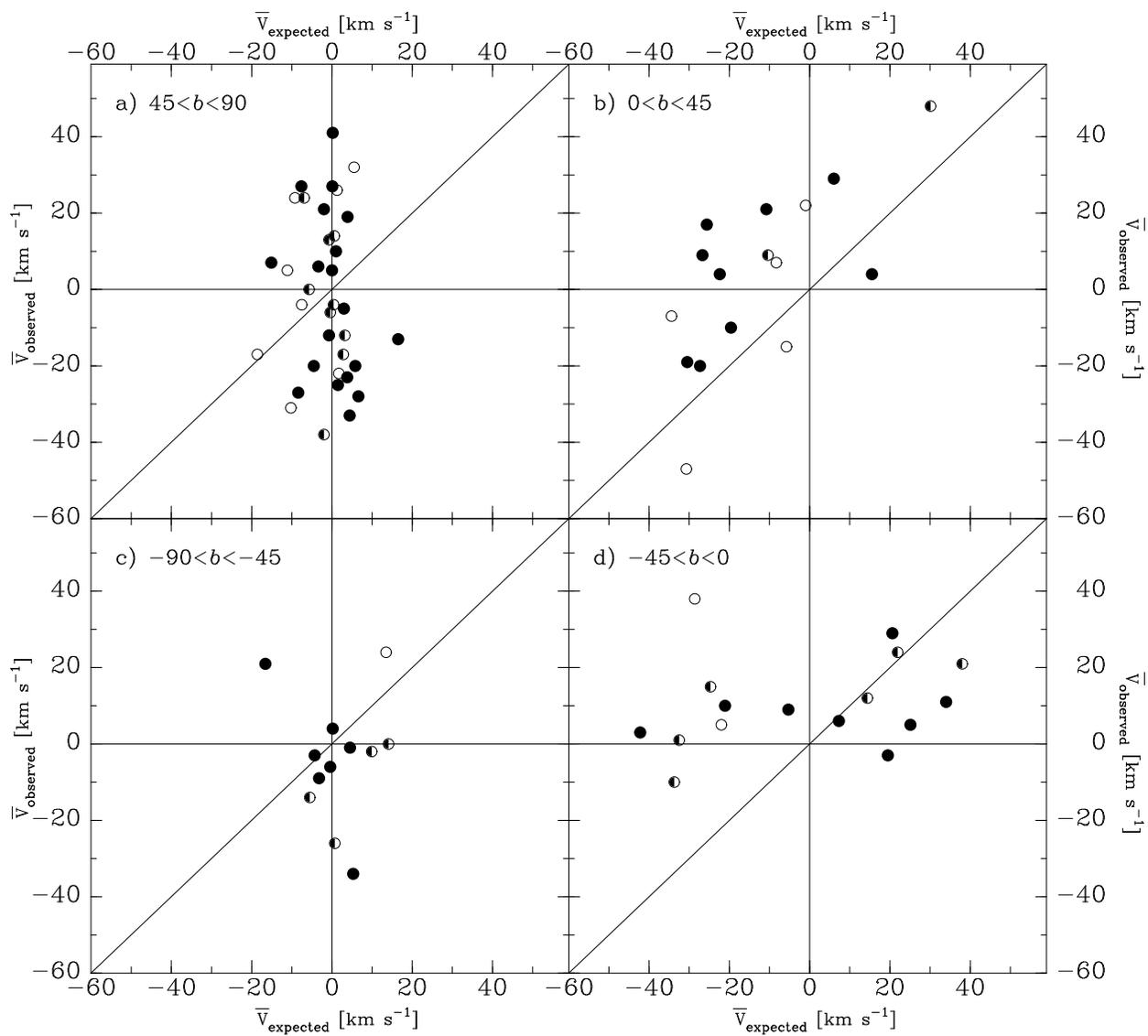

Fig. 14.—

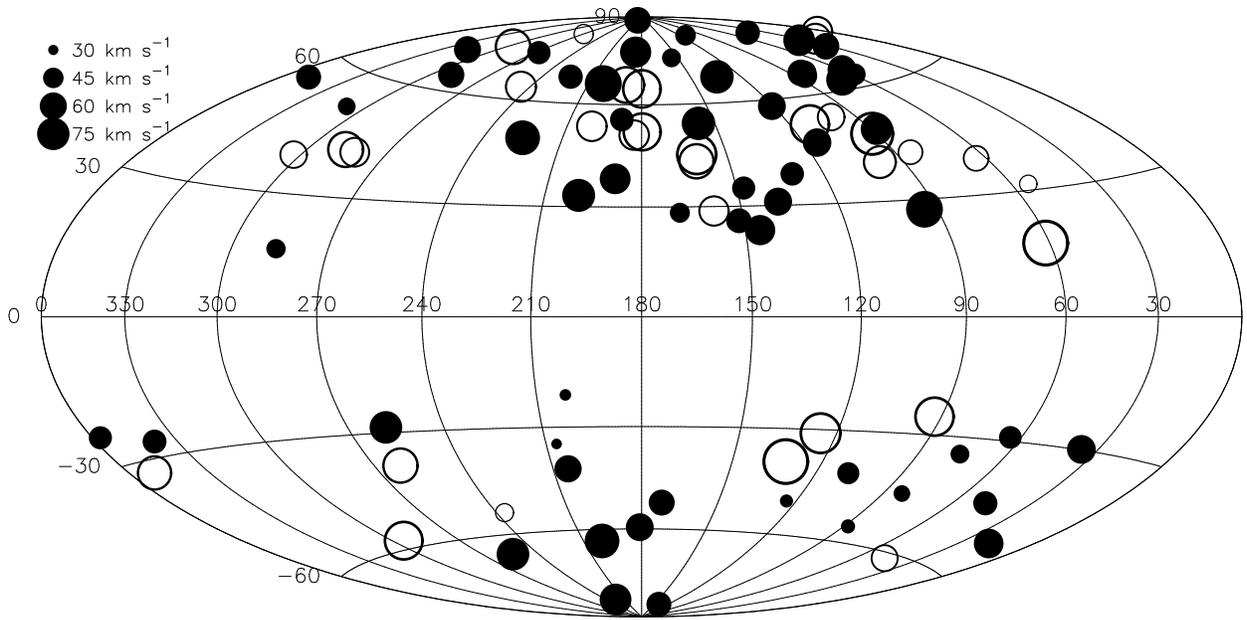

Fig. 15.—

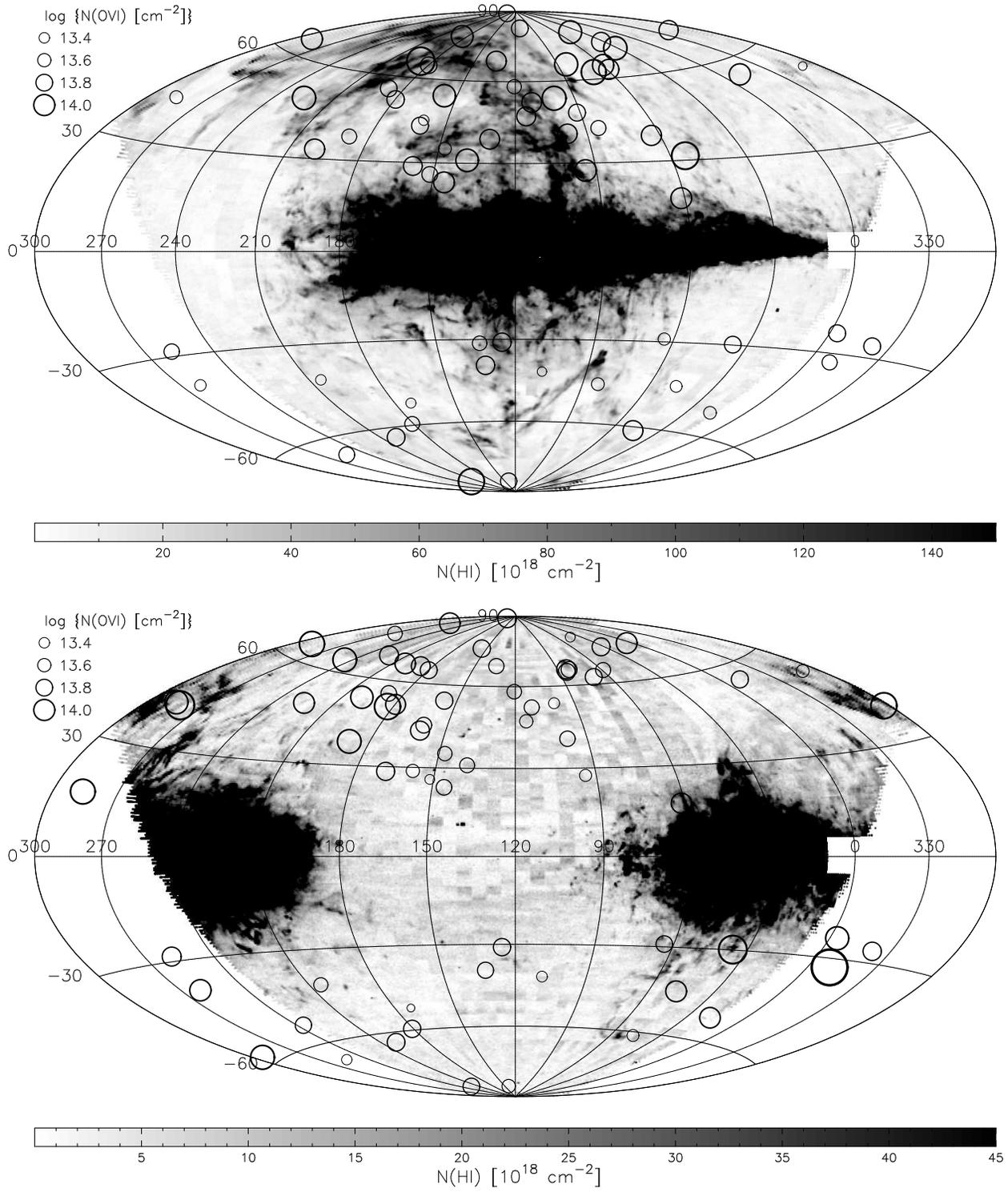

Fig. 16.—

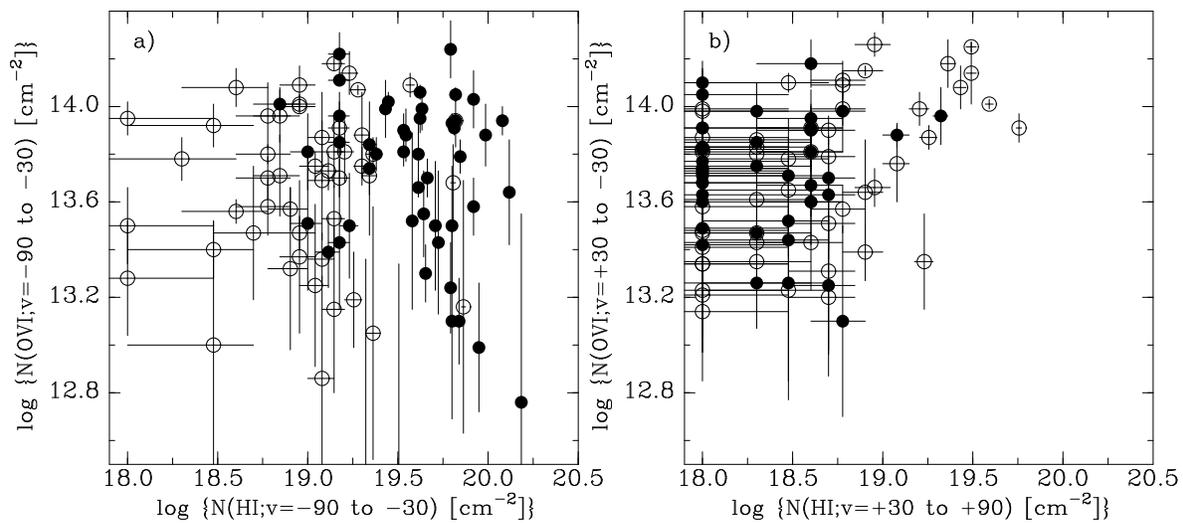

Fig. 17.—

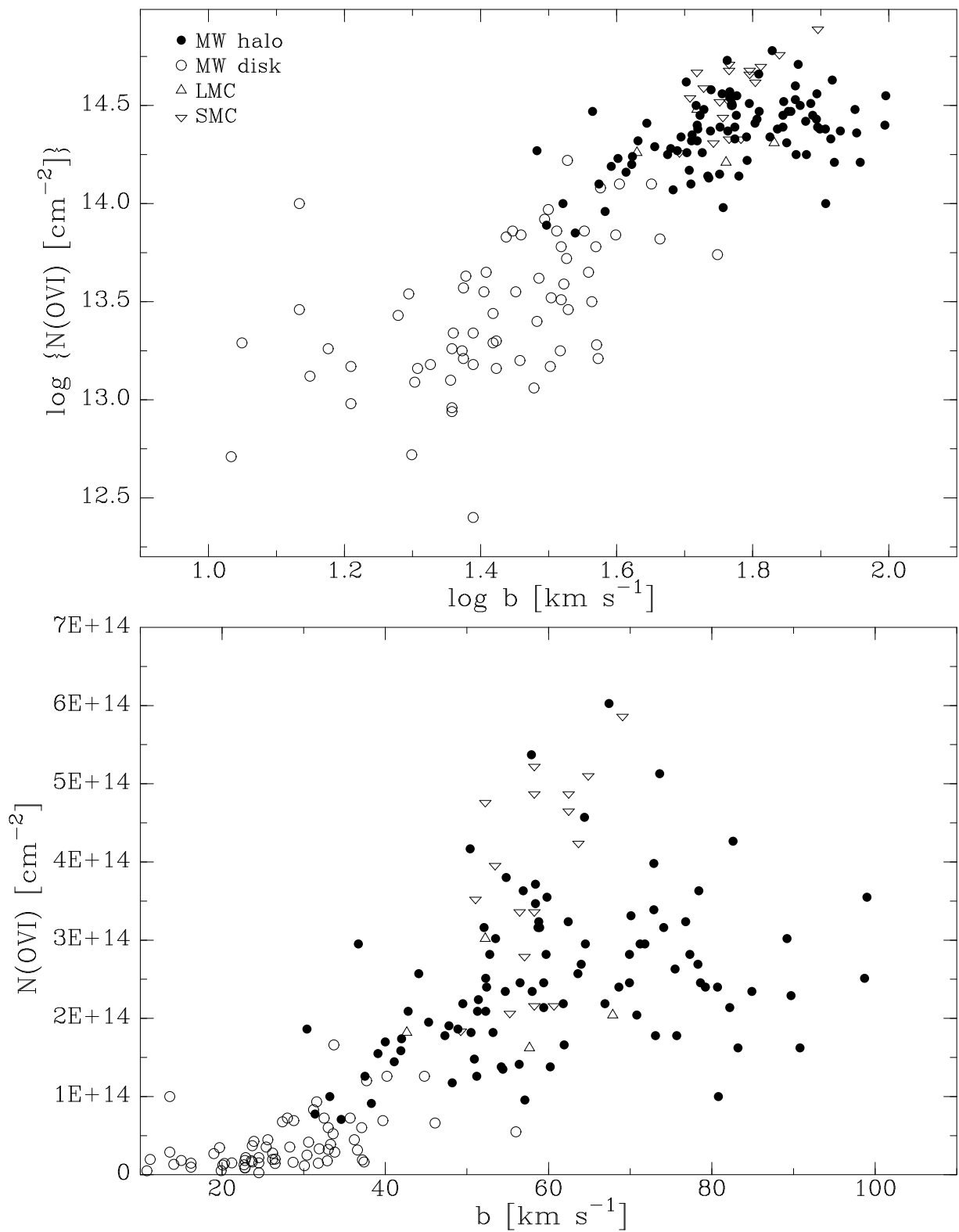

Fig. 18.—

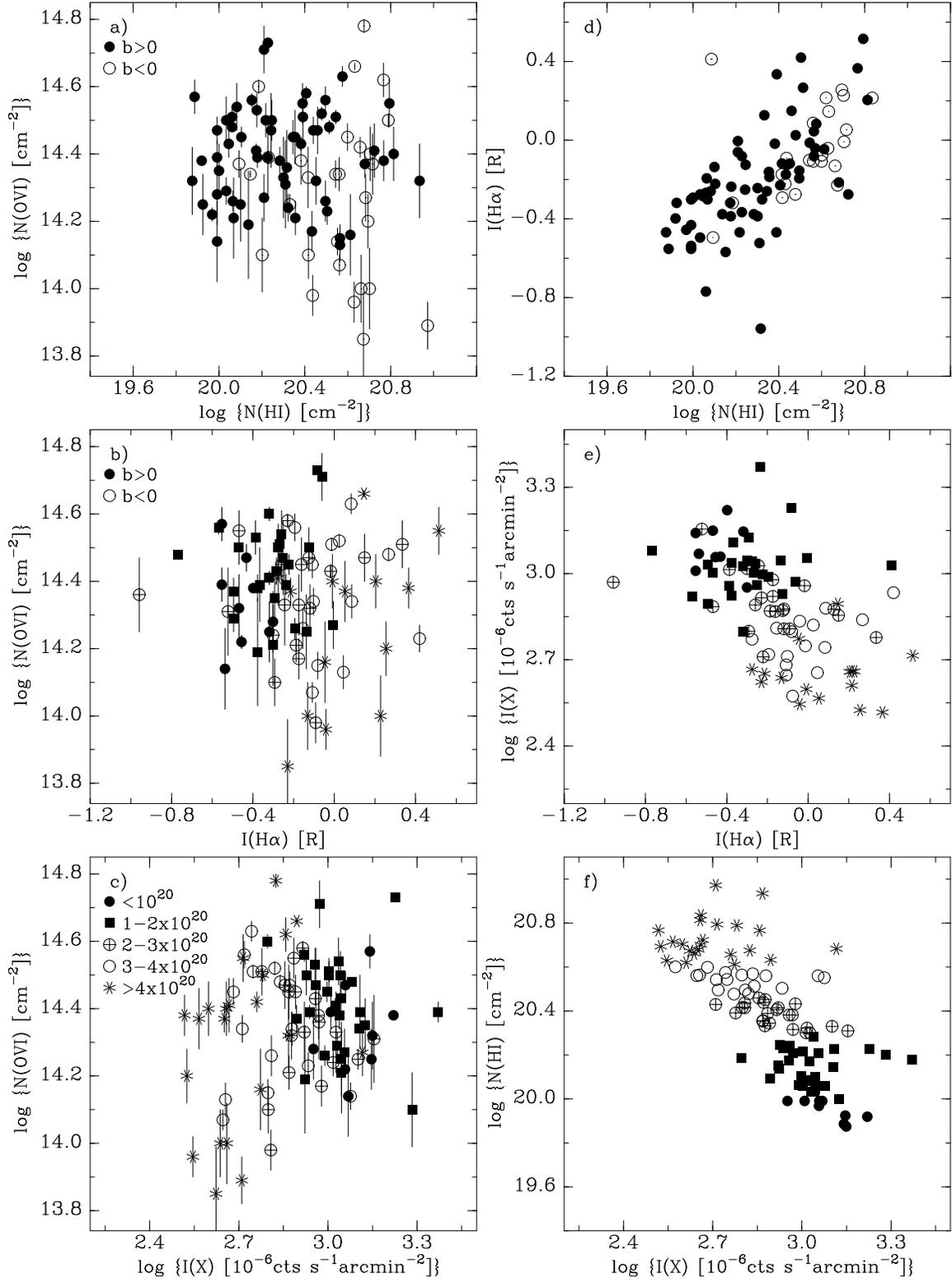

Fig. 19.—